\documentclass[useAMS,usenatbib]{mn2e}
\pdfoutput=1
\usepackage{amsmath,amssymb,pstricks,wasysym,stmaryrd,enumerate,longtable,fancyhdr,calc}
\usepackage{floatflt,graphicx,calc,epsfig,lscape}
\usepackage[T1]{fontenc}
\usepackage{mathptm}
\usepackage[pdftex,colorlinks=true,citecolor=blue]{hyperref}
\newcommand{\rad}{rad~m$^{-2}$}
\newcommand{\dg}{$^{\circ}$}
\newcommand{\DOT}{.}

\newcommand{\da}{$\dagger$}

\newcommand{\kms}{km~s$^{-1}$}
\title[Extragalactic sources towards the Galactic Centre]{\bf Extragalactic 
sources towards the central region of the Galaxy}
\author[Subhashis Roy, A. Pramesh Rao \& Ravi Subrahmanyan]
{Subhashis Roy$^{1}$\thanks{E-mail: roy@ncra.tifr.res.in} A. Pramesh Rao$^1$ \&
Ravi Subrahmanyan$^2$ \\
$^1$ National Centre for Radio Astrophysics (TIFR), \\ 
Pune University Campus, Post Bag No.3, Ganeshkhind, Pune 411 007, India.\\
$^2$ Australia Telescope National Facility, CSIRO, Locked bag 194, Narrabri,
NSW 2390, Australia}
\begin{document}

\date{}
\pagerange{\pageref{firstpage}--\pageref{lastpage}}

\maketitle

\label{firstpage}
\begin{abstract}
We have observed a sample of 64 small diameter sources towards the central
$-$6\dg~$~<~l~<~$6\dg, $-$2\dg~$~<~b~<~$2\dg\ of the Galaxy with the aim of
studying the Faraday rotation measure near the Galactic Centre (GC)
region.  All the sources were observed at 6 and 3.6 cm wavelengths using the
ATCA and the VLA. Fifty nine of these sources are inferred to be extragalactic.
The observations presented here constitute the first systematic study of the
radio polarisation properties of the background sources towards this direction
and increases the number of known extragalactic radio sources in this part of
the sky by almost an order of magnitude. Based on the morphology, spectral
indices and lack of polarised emission, we identify four Galactic HII regions
in the sample.
\end{abstract}
\section{Introduction}

Since extragalactic sources are located outside the Galaxy, the effect of ISM
on the propagation properties of electromagnetic waves from these objects can
be modelled without distance ambiguities as in the cases of pulsars, and
thereby allowing us to observe the integrated effect of the medium along large
($\sim$20 kpc) line of sight distance.
Unfortunately, only a few extragalactic sources have been identified within the
central few degrees of the Galaxy ($-$6\dg~$~<~l~<~$6\dg,
$-$5\dg~$~<~b~<~$5\dg), which limits their usefulness as probes to study the
Galactic Centre (GC) ISM.  High obscuration at optical wavelengths and the
confusion due to the high concentration of stars at infrared wavelengths have
prevented identification of extragalactic sources in this region. High angular
resolution studies at centimetre wavelengths (e.g., \citet{BECKER1994} at 5 GHz
and \citet{ZOONEMATKERMANI1990} at 1.4 GHz) have identified compact radio
sources, but in the presence of a large number of Galactic sources near the
Galactic Centre (GC), identifying the extragalactic sources is non-trivial, and
only about half a dozen extragalactic sources in this region have been
identified \citep{BOWER2001}.

To study the Faraday rotation measure (RM) near the centre of the Galaxy
($-$6\dg~$~<~l~<~$6\dg, $-$2\dg~$~<~b~<~$2\dg), we have selected a sample of 64
small diameter ($< 10''$) sources (see Sect.~1.1) in the region, which we have
studied with high angular resolution at 6~cm (C band) and 3.6~cm (X band) with
the ATCA and the VLA.  All the sources were studied for linear polarisation and
the width of the frequency channels were chosen to avoid bandwidth
depolarisation up to a RM of 15,000 rad m$^{-2}$. Though the NVSS
\citep{CONDON1998} was capable of detecting polarised emission from
sources, in those cases where the RM is high ($>$ 350~rad~m$^{-2}$) its
bandwidth of 50 MHz would cause bandwidth depolarisation.  Our observations,
for the first time, provide reliable measurements of the polarisation properties
of the sources in the region.  These observations have almost an order of
magnitude higher sensitivity (in Stokes I) and up to 3 times higher resolution
as compared to the previous VLA Galactic plane survey (GPS), and this high
sensitivity together with higher resolution has helped to identify the Galactic
sources in the initial sample.

In this paper, we provide information on the morphology, polarisation fraction,
spectral indices and rotation measure of these sources, and in a companion
paper (henceforth Paper II) draw inferences about the magnetic field in the GC
region.

\subsection{Sample selection}

We surveyed the literature and formed a sample of possible extragalactic radio
sources in the central $-$6\dg $ < l < $6\dg, $-$2\dg $ < b < $2\dg\ of the
Galaxy.  These sources were selected on the basis of their small scale
structure ($\le$ 10$^{''}$) and non-thermal spectra ($\alpha \le -0.4$,
S($\nu$) $\propto$ $\nu^ \alpha$). For the sources to have detectable linear
polarisation and so be useful for the RM study, an estimate of the polarisation
fraction of the sources are required. However, in the absence of any reliable
information on their polarisation fraction in the literature, we assumed the
unresolved sources to be polarised at the mean polarisation fraction of
extragalactic small diameter sources of 2.5\% \citep{SAIKIA1988}.  Sources with
measured flux density greater than 10 mJy at 5 GHz were selected.  The source
catalogues used for this selection were VLA images of the GC region at 327 MHz
\citep{LAROSA2000}, the VLA survey of the Galactic plane (GPS) at 1.4 GHz
\citep{ZOONEMATKERMANI1990}, \citep{HELFAND1992} and at 5 GHz
\citep{BECKER1994}. Sources observed by \citet{LAZIO1998a} and the 365 MHz
Texas survey \citep{DOUGLAS1996} were also used for this purpose. In those
cases where no flux density estimates were available at 5 GHz, the flux
densities at this frequency was estimated by extrapolating the 1.4 GHz flux
densities using spectral indices measured between 327 MHz and 1.4 GHz. A total
of 64 sources were found that satisfied all of the above criteria.

\section{Observations and data reduction}

Details of array configurations, frequencies used and the date of radio
observations are in Table~\ref{table.pol.obs}. The ATCA observations were made
using a 6~km array configuration. Twelve sources, G357.435$-$0.519,
G357.865$-$0.996, G358.002$-$0.636, G358.982+0.580, G359.388+0.460,
G359.568+1.146, G359.844$-$1.843, G359.911$-$1.813, G0.537+0.263,
G1.028$-$1.110, G1.035+1.559, G2.143+1.772 were observed on 06 Feb 2000 during
the pilot run of the project. We observed 24 more sources G353.462$-$0.691,
G354.719$-$1.117, G354.740+0.138, G356.000+0.023, G356.161+1.635,
G356.567+0.869, G356.719$-$1.220, G358.591+0.046, G358.917+0.072,
G359.546+0.988, G359.993+1.591, G0.313+1.645, G0.846+1.173, G1.505$-$1.231,
G1.954$-$1.702, G2.423-1.660, G4.005+1.403, G4.188$-$1.680, G4.256$-$0.726,
G4.898+1.292, G5.260$-$0.754, G5.358+0.899, G5.511$-$1.515 and G6.183$-$1.480
on 30th September and 02nd, 06th and 08th October, 2000.  To unwrap possible
n$\pi$ wrap in polarisation angles measured between two frequencies, 4 more
polarised sources from the pilot run G357.865$-$0.996, G359.388+0.460,
G0.537+0.263, G1.028$-$1.110 were re-observed on 06th or 08th Oct 2000. For the
same reason, 4 sources from the pilot run G359.388+0.460, G359.911$-$1.813,
G0.537+0.263, G2.143+1.772 along with G359.993+1.591 were re-observed on 12th
April 2002.  All these observations were made using the multi-channel continuum
observing mode, and the data were acquired in 16 independent frequency channels
covering 128-MHz bands centred at the observing frequencies.  Each target
source was typically observed for a total of 40--50 minutes. Since we used an
E-W array configuration, multiple-snapshot mode was used to get a satisfactory
{\it uv}-coverage and each source was observed $\approx$10 times equally spaced
in hour-angle.  The sources 1741$-$312 and 1748$-$253 were used to calibrate
the antenna based amplitudes and phases (secondary calibrators), and their flux
densities were measured based on the observation of the primary flux density
calibrator PKS B1934$-$638. This source (polarisation fraction $\le$ 0.2\%)
was also used to determine antenna based polarisation leakages.  The
calibration and editing of the ATCA data was performed using {\sc MIRIAD}.
Calibrated data was converted into Stokes I, Q, U, V, and further analysis were
carried out using AIPS. Maps made at different frequencies for a particular
source were convolved to the same resolution. The Polarisation angle ($\phi$)
being given by $\phi$=0.5~tan$^{-1}(U/Q)$ (where, signs of Q and U are
considered separately to unambiguously determine the value of $\phi$), we
divided the Stokes U image by the Q image and measured the polarisation angle
and its error (AIPS task COMB). Finally, the polarisation angle images at
different frequency bands were fitted to the equation, 
\begin{equation}
\label{rm.equation}
\phi = RM.\lambda^2 + \phi_0 +n\pi
\end{equation}
using the AIPS task RM. In this equation, $n$ is an integer, $\lambda$ is the
wavelength, and $\phi_0$ denotes the intrinsic polarisation angle (i.e., when
the observing frequency tends to infinity). If the rms residuals exceed four
times the expected rms noise, the fitted values were rejected. 

Since there were 16 frequency channels per 128 MHz band of the ATCA data, we
tried to measure the RM from the ATCA observations in two ways by using the
AIPS task RM. (i) Since AIPS task RM cannot fit more than four frequency
channels to measure RM, in order to maximise signal to noise ratio as well as
to check for high RM in the data, we divided the central 12 frequency channels
of 4.8 and 5.9 GHz band into 3 equal parts. Polarisation angles were measured
from first and third part of each of these bands, and these were used as input
to the AIPS task RM. This allowed us to measure RM as high as 30,000 \rad.
(ii) If the RM measured by the previous method is not very
high (i.e., $\le$ 2000 rad m$^{-2}$, and was the case for all except one
source), data from each of the 4.8, 5.3, 5.9 and 8.5 GHz band were averaged and
polarisation angle measured from each of these bands. These polarisation angle
images and their error maps were used as the input to the AIPS task RM.

We used the VLA in its BnA configuration to observe the relatively weak sources
in the sample. The default continuum mode, which provides a single frequency
channel of bandwidth 50 MHz in each IF band, was used. Observations were
centred at frequencies 4.63, 4.88, 8.33 and 8.68 GHz. 27 new sources,
G353.410$-$0.360, G354.815+0.775, G355.424$-$0.809, G355.739+0.131,
G356.905+0.082, G358.149$-$1.675, G358.156+0.028, G358.643$-$0.034,
G358.605+1.440, G358.930$-$1.197, G359.392+1.272, G359.2$-$0.8 (Mouse),
G359.604+0.306, G359.717$-$0.036, G359.710$-$0.904, G359.871+0.179,
G0.404+1.061, G1.826+1.070, G2.922+1.028, G3.347$-$0.327, G3.748$-$1.221,
G3.928+0.253, G4.005+0, G4.619+0.288, G4.752+0.255, G5.791+0.794 and
G5.852+1.041 were observed on 11th and 13th February 2001 in two different
bands. Moreover, the source G353.462$-$0.691 from ATCA observations was found
to be almost unpolarised in the 6~cm band, and was re-observed on 11th
February with the VLA. To unwrap possible n$\pi$ wrap in polarisation angles
measured between two frequencies, 5 more sources from the ATCA pilot run,
G358.002$-$0.636, G1.035+1.559, G359.844$-$1.843, G359.911$-$1.813 and
G2.143+1.772 were re-observed on the 2nd day (13th Feb, 2001) of the VLA
observations. From these sources, we re-observed G359.604+0.306 and
G359.717$-$0.036 with the ATCA on 12th April 2002 to uniquely determine their
RM.

The secondary calibrators used for the observation were the same as in the
ATCA observations.  Polarisation calibration was performed using the
unpolarised source 3C84. 3C286 was used as the primary flux density calibrator
and to estimate the instrumental phase difference between the two circularly
polarised (RR and LL) antenna signals. Each source was observed for typically
5 minutes at 2 different hour angles. All the data were calibrated and
processed using the {\sc AIPS} package.

The source G356.905+0.082 has a compact core, which is weakly polarised, and
an extended highly polarised halo.  Due to zero spacing problem, the 8.5 GHz
image of the halo could have missing flux density problems.  Since the
polarised intensity per beam of the core is similar to the contribution by the
halo, which has small scale structures in polarised image and is not properly
sampled at 8.5 GHz, we could not reliably estimate the RM from the core or the
halo.  Therefore, we have not provided its RM in Table~\ref{gc.rm.list}.

\begin{table}
\caption{Journal of observations}
\label{table.pol.obs}
\begin{tabular}{|c c c c c c|}
\hline
Epoch & Telescope & Array & Obs.  & Frequency & No. of \\
     &           &config- & time & (GHz)     & sources  \\
     &           &uration &(hours)&           & observed \\
\hline

06 Feb 2000 & ATCA & 6A & 10 & 4.80 \& 5.95 & 12 \\
30 Sept 2000 & ATCA & 6B & 12 & 4.80 \& 5.95 & 12 \\
02 Oct 2000 & ATCA & 6B & 12 & 4.80 \& 5.95 & 12 \\
06 Oct 2000 & ATCA & 6B & 12 & 5.31 \& 8.51 & 14 \\
08 Oct 2000 & ATCA & 6B & 12 & 5.31 \& 8.51 & 13 \\
11 Feb 2001 & VLA  & BnA & 06 & 4.63 \& 4.88 & 28 \\
13 Feb 2001 & VLA  & BnA & 06 & 8.33 \& 8.68 & 32 \\
12 Apr 2002 & ATCA & 6B & 11 & 5.06 \& 5.70 & 08 \\

\hline
\end{tabular}
\end{table}

\section{Results}

Based on the ATCA data, 24 sources were found to have at least one polarised
component. Images of these sources are shown in
Fig.~\ref{5ghz.atca.larger.maps}.  The FWHM sizes of the beams in the ATCA
images are $\approx$6$^{''} \times 2^{''}$, and the typical RMS noise is 0.23
mJy/beam in Stokes I and about 0.15 mJy/beam in Stokes Q and U. These images
show  7 of the 24 sources to have a single unresolved component and the
remainder are either partially resolved or have multiple components.

Of the sources observed using the VLA, 21 have at least one polarised
component. Images of these sources along with 2 polarised phase calibrators are
shown in Fig.~\ref{5ghz.vla}. The FWHM sizes of the beams in the VLA images are
$\sim$2$^{''} \times 1.5^{''}$ and the RMS noise is typically 75 $\mu$Jy/beam.
In these images, four sources appear as single unresolved components. Since
1741$-$312 and 1748$-$253 have been observed by both ATCA and the VLA, but
1741$-$312 has a faint emission around the compact source as seen in higher
resolution VLA maps, we have presented its properties in Stokes I from the VLA
data. 

Stokes I images of the sources in the sample that were not detected to have
polarised emission are shown in Fig.~\ref{5ghz.atca.unpol.src.maps} and
Fig.~\ref{5ghz.vla.unpol.src.maps} respectively.

The properties of the 24 polarised sources observed with ATCA are presented in
Table~\ref{pol.src.prop.at} and 21 sources observed with VLA are in
Table~\ref{pol.src.prop.vla}. In these tables, the following conventions
are used.  Column 1: the source name using Galactic co-ordinates ($l \pm b$).
Column 2: the component designation; `N' denotes Northern, `S' denotes
Southern, `E' denotes Eastern and `W' denotes Western, `C' denotes central,
`EX' denotes highly extended and `R' denotes ring type.  Columns 3 and 4: Right
ascension (RA) and Declination (DEC) of the radio intensity peaks of the
components in J2000 co-ordinates.  Column 5: the deconvolved size of the
components with their major and minor axes in arc-seconds and the position angle
(PA) in degrees (formatted as major axis $\times$ minor axis, PA). A few
sources that are observed to have multiple resolved components in the 8.5~GHz
images are, however, not well resolved in the 4.8 GHz images. For these
sources, we have measured the size parameters of the components from the 8.5
GHz images and we put a `*' symbol beside these measured parameters.  Columns 6
and 7: the corresponding peak and total flux density of the components at 4.8
GHz in units of mJy beam$^{-1}$ and mJy respectively.  Column 8: total flux
density of the component at 4.8 GHz as measured by the VLA GPS survey.  Column
9: percentage polarisation of the components.  Column 10: spectral indices of
the components measured between 8.5 and 4.8 GHz.  A few of the sources are
extended over several synthesised beam-widths and for these sources we have
convolved the 8.5 GHz images to the resolution at 4.8 GHz and then made
spectral index images. The spectral indices of the individual components are
measured from these images, we put an `s' beside the spectral index for these
extended sources.  Columns 11 and 12: spectral indices between 4.8 and 1.4 and
between 1.4 and 0.3 GHz respectively.  Column 13: the source classification;
`EG' denotes an extragalactic source (based on the morphology), and G denotes a
Galactic source. Several sources show the morphology typical of FR$-$I or
FR$-$II sources and this is noted along with the extragalactic classification.
Sources which appear unresolved (deconvolved source size $<<$ beam size) are
denoted by U, slightly resolved (deconvolved source size $\lesssim$ beam size)
by SR, double by D and T denotes a triple source consisting of a pair of lobes
and a core.  C+E denotes a flat spectrum core with extended emission either in
the form of a lobe or jet.  If there are several objects in the field which
appear to be unrelated, we label the object as M.

For computing spectral indices in Table~\ref{pol.src.prop.at} and
\ref{pol.src.prop.vla}, the 1.4 GHz flux densities of the sources have been
taken from the VLA GPS and the NRAO VLA Sky Survey (NVSS) \citep{CONDON1998}.
If the measured flux density of a component in the GPS differs from that in
the NVSS by more than 20\%, we have put a `$\dagger$' mark beside the computed
spectral indices (column 11).  We have visually examined the NVSS images of
these sources and if we find that the source is not in a confused region of the
image we have used the flux density from NVSS to compute the spectral index; in
such cases, we put `(N)' beside the measured spectral index. For the source
G359.871+0.171, the 1.4 GHz flux density has been assumed to be the same as
that measured by \citet{LAZIO1999} at 1.5 GHz and we put `(L)' beside its
measured spectral index between 4.8 and 1.4 GHz.  For a few sources in the
list, the 1.4 GHz flux density in unknown.  In these cases, we put a `**' in
column 11 and enter the spectral index between 4.8 and 0.3 GHz in column 12
with `(0.3/4.8)' written below. The P-band flux densities of the sources have
been taken from the Texas survey at 365 MHz \citep{DOUGLAS1996}. However, the
Texas survey is known to have large uncertainty in flux densities for sources
near the Galactic plane and which is more near the complex GC region.
Therefore, if any source is detected in the GC image at 330 MHz
\citep{LAROSA2000}, we have used their flux density to compute the spectral
index (column 12 in Table~\ref{pol.src.prop.at}, \ref{pol.src.prop.vla} and
column 11 in Table~\ref{unpol.src.prop.at}, \ref{unpol.src.prop.vla}) and put
`GC' in parenthesis beside the spectral index measured. For 5 sources the flux
densities at 330 MHz have been taken from \citet{ROY2002}, and we put `(GM)'
beside the spectral index in column 12.  We have also taken the flux densities
of 3 sources at 330 MHz from S. Bhatnagar (private communication) and put
`(GM1)' in column 12. Many of the sources resolved at frequencies of 1.4 GHz
and above appear unresolved in the low frequency Texas survey. For these
sources, we only compare their integrated flux densities between 1.4 and 0.3
GHz, and put `(i)' beside the measured spectral index in column 12.

In Table~\ref{unpol.src.prop.at} \& \ref{unpol.src.prop.vla} we present the
properties of the sources which are not detected in polarised emission.  These
tables are similar to Table~\ref{pol.src.prop.at} \& \ref{pol.src.prop.vla},
except that we have omitted the column representing percentage polarisation
(column 9 in Table~\ref{pol.src.prop.at} \& \ref{pol.src.prop.vla}).  For the
four Galactic HII regions we have identified, we write `G$-$HII' in column 12
of this Table.

\begin{figure}
\begin{minipage}{0.45\textwidth}
\centering
\includegraphics[width=\textwidth,clip=true,angle=0]{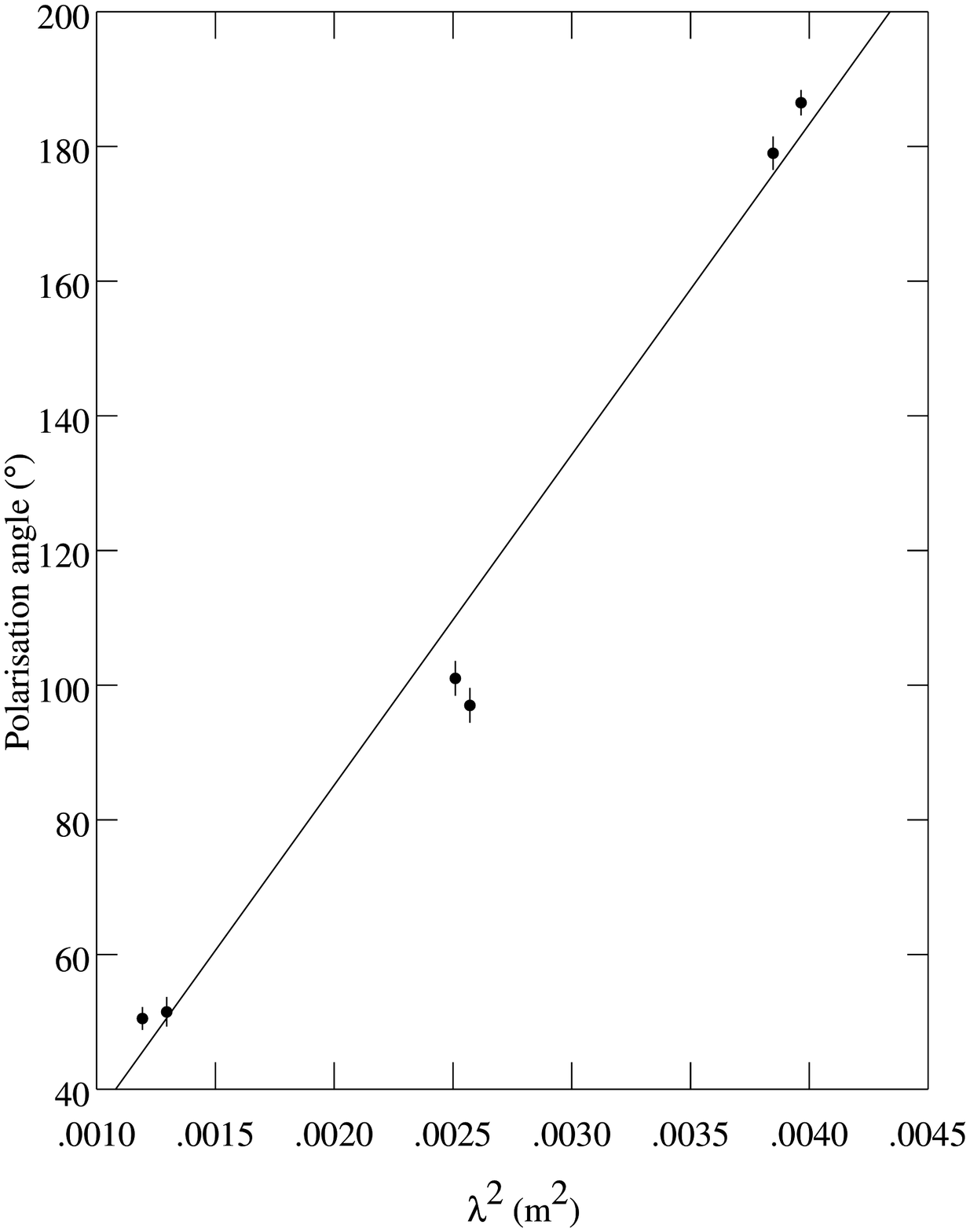}
\caption{An example of bad fit of Equation~\ref{rm.equation} to the measured
polarisation angles vs. square of wavelength plot (reduced $\chi^2$=17). The
polarisation angles are measured towards the source G358.002$-$0.636.}
\label{bad.fit}
\end{minipage}
\hfill
\begin{minipage}{0.45\textwidth}
\centering
\includegraphics[width=\textwidth,clip=true,angle=0]{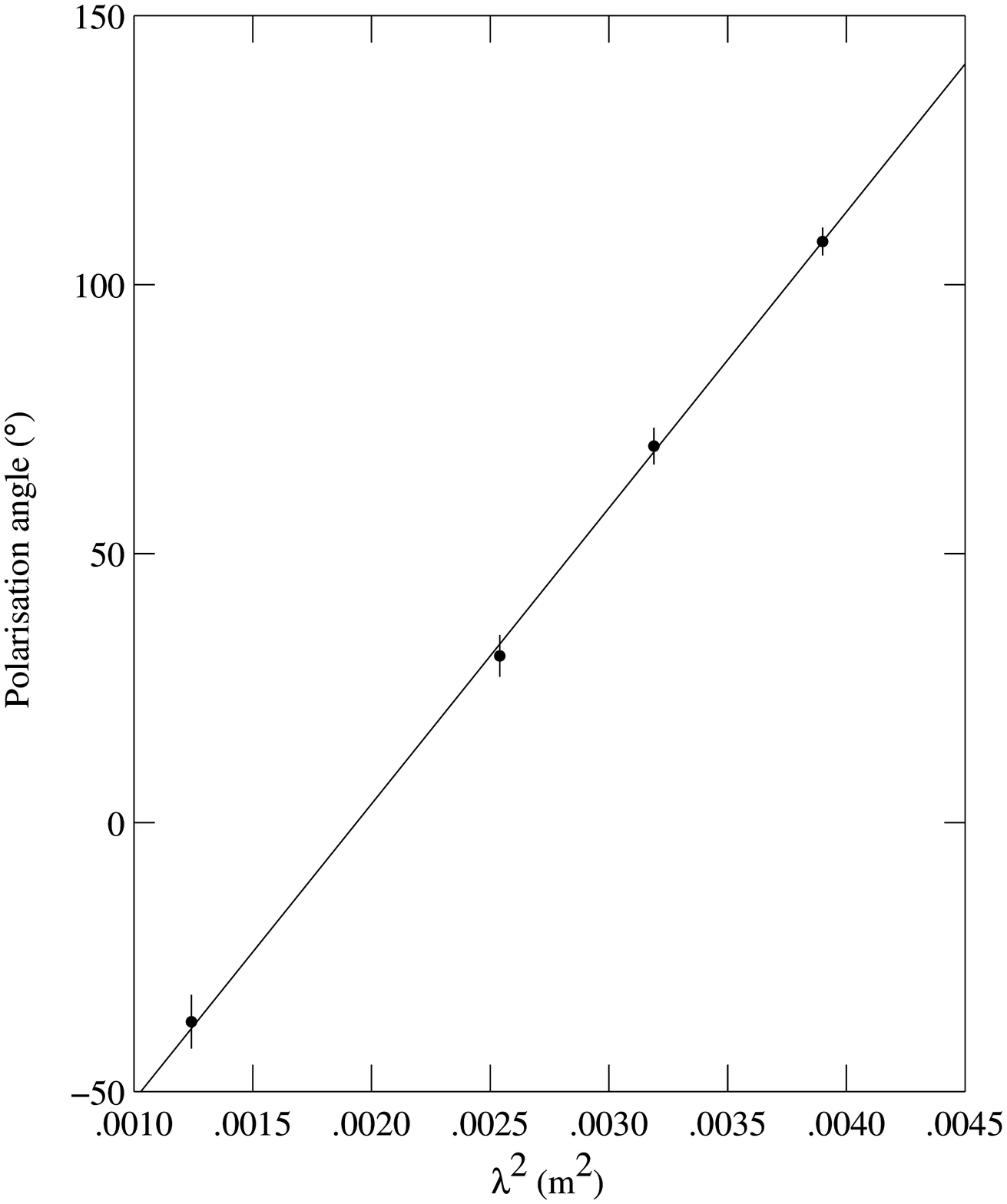}
\caption{An example of good fit of Equation~\ref{rm.equation} to the measured
polarisation angles vs. square of wavelength plot (reduced $\chi^2$=0.2). The
polarisation angles are measured towards the source G356.567+0.869.}
\label{good.fit}
\end{minipage}
\end{figure}

The measured RM towards 44 sources (65 components) and 2 secondary calibrators
are given in Table~\ref{gc.rm.list}, which is arranged as follows:\\ Column 1:
the source name in Galactic co-ordinates (G$l \pm$b).  Column 2 \& 3: RA
(J2000) and DEC (J2000) of the source components. The co-ordinates of these
components are based on the peak in the polarised intensity (if the peak in the
polarised intensity do not coincide with the peak in total intensity, the
component position will be slightly different than what is given in
Table~\ref{pol.src.prop.at} \& \ref{pol.src.prop.vla} based on the peak in
total intensity).  Column 4 \& 5: the measured RM (in rad m$^{-2}$) and the
error in these measurements at the position of the peak in the polarised
emission.  Depolarisation fraction is defined as the ratio of the polarisation
fraction of any component between lower to that at higher frequency, and in
column 6 we provide the depolarisation fraction (D) of the source components
between 4.8 and 8.5 GHz.  Column 7: Percentage error in the depolarisation
fraction ($\Delta$(D) in \%). Assuming that the emission mechanism is
synchrotron, the orientation of the electric field in the radiation has been
used to infer the orientation of the magnetic field in the plasma and we
provide the direction of this magnetic field ($\theta$) and the error in this
measurement ($\Delta\theta$) in columns 8 \& 9 respectively.  Column 10:
Reduced Chi-square ($\chi^2$) of the fit of Equation~\ref{rm.equation} to the
measured polarisation angles.  We show an example of bad fit ($\chi^2$=17) in
Fig.~\ref{bad.fit}, and one example of good fit ($\chi^2$=0.2) in
Fig.~\ref{good.fit}. Column 11: Measured polarisation angles at different
frequencies. The observed frequencies (in MHz) and the measured polarisation
angles (in degrees) are tabulated in pairs, and each frequency, polarisation
angle pair are separated by commas.

The source G359.2$-$0.8 (Mouse) is a known Galactic non-thermal source. Since
this source is known to be within 5 kpc from the Sun \citep{UCHIDA1992}, the RM
towards this object is expected to be quite small. To check if observations
confirm its low RM, we observed this source. From Table~\ref{gc.rm.list}, its
RM is indeed measured to be very small ($-$5 $\pm$18 \rad). However, our
samples are selected to measure the RM introduced by the GC region, but this
source only samples the local ISM, and its RM is not used in any further
analysis.

\subsection{Accuracy of the spectral index measurements}

The errors in the derived spectral indices depend on the accuracies of the
flux densities at different frequencies. One of the drawbacks of a Fourier
synthesis array is that if a source is resolved on the shortest interferometer
baseline, its flux density will most likely be underestimated in the image.
In the GC region, where the sky density of sources is high and emission at
various size scales may co-exist, confusion could be a significant source of
errors in imaging and hence in flux density estimates.  Because our radio
observations are performed at relatively high frequencies, the contribution
from extended Galactic synchrotron background is negligible. Additionally, most
of the sources in our sample have small angular sizes and, therefore, the
problem of missing flux density should be minimal.  The ATCA and VLA
observations were both made using a single array at both the frequencies.
Since problem of any missing flux density increases with increase in observing
frequency, this might result in an underestimation of the source spectral
index.  We have detected extended emission of up to $\sim$10$'$ scale in our
4.8 GHz images of G353.410$-$0.360, G355.424$-$0.809, G355.739+0.131,
G356.905+0.082, G358.149$-$1.675, G358.643$-$0.034, G359.2$-$0.8 (Mouse),
G359.717$-$0.036, G0.313+1.645 and G5.791+0.794.  To examine for any missing
flux density at 8.5 GHz, we first restricted the Fourier Transform of the CLEAN
components of the 4.8 GHz images to the {\it uv}-range of the 8.5 GHz
visibilities. A comparison of the corresponding images with the original 4.8
GHz images (without the restriction in visibility coverage) shows that except
for G353.410$-$0.360, G355.739+0.131, G356.905+0.082, G358.643$-$0.034,
G358.149$-$1.675, G359.2$-$0.8 (Mouse) and G359.717$-$0.036, the missing flux
density is less than 10\% of the total flux density, and the corrsesponding
error in their spectral indices between 8.5 and 4.8 GHz is less than 0.25.
Among the above 7 sources, 4 have been identified as HII regions.  The spectral
indices between 8.5 and 4.8 GHz of the 2 remaining extragalactic sources
G356.905+0.082 and G358.149$-$1.675 and the Galactic non-thermal source
G359.2$-$0.8 (Mouse) (Table~\ref{pol.src.prop.vla}) have been measured only
for the compact components (e.g., core and hot spots).  For other sources, the
error in the measured flux densities are expected to be about 5\%, and the
corresponding error in their spectral indices between 8.5 and 4.8 GHz is about
0.1.

Owing to extended emission and enhanced source confusion near the GC, images
made from radio interferometers and with sparse spatial frequency coverage
suffer from systematic errors. 
To estimate the errors in the source flux densities in the GPS and Texas
survey, we compared the 1.4 GHz flux densities of our sample sources as
measured by the GPS and the NVSS and found that the differences in flux
density measurements were about 20 per cent. 20\% differences in flux density
contributes an error of about 0.15 in the measured spectral index 
between 4.8 and 1.4 GHz.  We also compared the flux densities of the 12 sources
which are detected in both the Texas survey and in the 330 MHz VLA GC image
\citep{LAROSA2000} and noticed that the difference is almost a factor of two
for six of these sources.  The errors in the flux density measurents from the GC
image are expected to be relatively smaller and, therefore, if the 0.3 GHz flux
density of a source is taken from the Texas survey, an error $\sim$0.5 can
occur in its spectral index measurement between 1.4 and 0.3 GHz.  We note that
other than the missing flux density problem, there are several other systematic
errors which could lead to erroneous measurement of spectral indices. One
of these errors is source variability, which we discuss below.

\subsection{Effect of source variability}
\label{variability}
Estimation of source spectral indices and the RM requires multi-frequency
observations. Most of our multifrequency observations were separated by a
time period, which varies from a few days to an year or two. Any source
variability in these timescale will affect the measured spectral indices and
possibly the measured RM. However, we note that the variability of the
extragalactic sources at frequencies higher than a few GHz is often intrinsic
in nature \citep{WAGNER1995} and in some cases caused by interstellar
scintillation (ISS) \citep{LOVELL2003}. The typical variability timescale for
intrinsic variability is from a few hours to years \citep{WAGNER1995}, and from
hours to months in the case of ISS \citep{LOVELL2003}.  However, only the core
dominated objects with flat spectrum ($\alpha >-0.5$) can show significant
variability in timescale of days or less \citep{WAGNER1995}. Among the sources
observed, only G356.905+0.082, G357.865$-$0.996, G358.002$-$0.636,
G0.537+0.263, G3.928+0.253 falls into this category. Among them, the
multifrequency observations of G356.905+0.082 and G3.928+0.253 were separated
by only 2 days, and the probability of significant variability at this
timescale is believed to be low.  As described below, G357.865$-$0.996 is
variable, G0.537+0.263 does not appear to be variable, and we do not have data
to check for variability of G358.002$-$0.636. We note that variability in
timescale of few hours could not be identified for any of our sources.

We identify 12 sources from Sect.~2, G357.865$-$0.996, G358.002$-$0.636,
G359.388+0.460, G359.604+0.306, G359.717$-$0.036, G359.844$-$1.843,
G359.911$-$1.813, G359.993+1.591, G0.537+0.263, G1.028$-$1.110, G1.035+1.559,
G2.143+1.772, towards which our observations were separated upto 2 years, and
their measured properties could suffer substantially from variability. Among
these sources G1.028$-$1.110, G1.035+1.559, G359.604+0.306, G359.717$-$0.036
and G359.993+1.591 are extended (more than few arc seconds in angular size).
Consequently, their variability timescale will be many years, and they are not
considered to be variable in this paper.  Among the rest of the sources,
G357.865$-$0.996, G359.388+0.460, G0.537+0.263, G359.911$-$1.813, G2.143+1.772
were observed with ATCA on days, where certain frequency band was common to
both the epochs. On comparing the flux densities of these sources in the same
band (with appropriate correction of flux densities from their measured
spectral indices due to change of frequency), from both the observing days, it
was found that only the flux density of the source G357.865$-$0.996 has changed
by more than 10\%.  Unfortunately, such a method could not be applied for the
remaining 2 sources G358.002$-$0.636 and G359.844$-$1.843.  Since,
G359.844$-$1.843 is a steep spectrum source (Table~\ref{pol.src.prop.at}), the
emission from it is likely to be originated from extended region (e.g., from
jets), and the variability timescale is large.  However, G358.002$-$0.636 is a
rather flat spectrum source (core dominated), and could be variable. We note
that the fit of Equation~1 to its polarisation angles is bad (Reduced
$\chi^2$=17, Table~\ref{gc.rm.list}), and one possible reason for this is
source variability. Consequently, the measured RM of G357.865$-$0.996 and
G358.002$-$0.636 is excluded from the sample used to derive statistical
estimates of the RM introduced by the GC region. We do not consider the effect
of variability on the spectral indices of sources estimated from other
catalogues (i.e., other than $\alpha_{8.5/4.8}$).

\subsection{Contribution of the intrinsic Faraday rotation to the measured RM}
\label{subsect_intrinsic}
When the polarised emission from the source reaches the observer, rotation of
the polarisation angle could occur (i) within the source, (ii) in the Inter
Galactic Medium (IGM) or (iii) in the ISM of the Galaxy. Compared to the ISM of
our Galaxy, the electron density of the IGM is very small, and consequently the
Faraday rotation introduced by the IGM is negligible. However, if the
synchrotron electrons are mixed with thermal electrons at the source, or, if
there is an intervening galaxy, the ISM of which introduces RM, or if there
is cluster of galaxies along the line of sight, there can be Faraday
rotation introduced outside our Galaxy.
As discussed by \citet{GARDNER1966, KRONBERG1972, VALLEE1980}, the polarisation
angle could deviate from the $\lambda^2$ law due to the following three
mechanisms. \\ (i) If the synchrotron optical depth becomes significant at some
frequency, the polarisation direction makes a transition from parallel to
perpendicular to the projected magnetic field. (ii) If there are multiple
unresolved emission components with differing spectral indices and polarisation
characteristics, it can cause complex wavelength dependent variations in
polarisation angle. (iii) If there are significant gradients in Faraday
rotation across or through the emission region of the source, then also
polarisation angle can have complex dependence on the observing wavelength.
The polarisation angles of the source G358.002$-$0.636 deviates significantly
from the $\lambda^2$ law, and other than source variability
(Sect.~\ref{variability}), it could have been caused by one of the above
processes.

In the case of Faraday rotation outside the Galaxy, the Faraday screen is
likely to be located several orders of magnitude farther away than the GC ISM.
As a result, emission from different parts of the source is viewed within one
synthesised beam, the linear scale of which is much larger than what is sampled
in our ISM.  At such length scales ($\sim$1 kpc), the magnetic field within the
intervening Faraday screen is likely to be uncorrelated. As a result, there is
differential Faraday rotation within the beam, which gives rise to source
depolarisation. From the data, the RM towards the source G358.917+0.072 is
found to be the highest with a measured value of 4768 \rad, and it shows a high
depolarisation fraction (0.3 between 4.8 and 8.5 GHz), which is likely to be
caused by differential Faraday rotation \citep{KRONBERG1972}.

Following the arguments given above, if the reduced $\chi^2$ of the fit is
greater that 4.6, the probability of occurrence of which is less than 1\%
\citep{BEVINGTON1969} or any of the source component shows a depolarisation
fraction of less than 0.6 (Table~\ref{gc.rm.list}) between 4.8 and 8.5 GHz,
we suspect that there is significant RM introduced outside the Galaxy and the
RM towards those components have not been used any further in this or in
deriving the properties of the Faraday screen in Paper II. We note that the
intrinsic RM of most of the extragalactic sources are quite small $\sim10$
\rad\ \citep{SIMARD-NORMANDIN1980} and the RMs measured from our sample is
quite high ($\sim1000$ \rad), and consequently the RM introduced outside the
Galaxy should have little effect on the RMs of most of the sources in our
sample.
%

%
\onecolumn
\begin{landscape}
\begin{longtable}{c c c c c c c c c c c c c} 
\caption{Measured parameters of the polarised sources from the ATCA data} \\
\label{pol.src.prop.at}\\
\hline
Source & Cmp & RA & DEC & S${maj}$ $\times$ S$_{min}$, PA & S$_p$ & S$_t$ & S$_{5}$ & \% & $\alpha_{8.5/4.8}$ & $\alpha_{4.8/1.4}$ & $\alpha_{1.4/0.3}$ & Notes\\
            &     & (J2000) & (J2000) &      &              & mJy  &  mJy        & Poln. &           &    &  &  \\
\hline
\endfirsthead

\hline
Source & Cmp & RA & DEC & S${maj}$ $\times$ S$_{min}$, PA & S$_p$ & S$_t$ & S$_{5}$ & \% & $\alpha_{8.5/4.8}$ & $\alpha_{4.8/1.4}$ & $\alpha_{1.4/0.3}$ & Notes \\
            &     & (J2000) & (J2000) &       &         & mJy    &  mJy   & Pol. &           &    &  &  \\
\hline
\endhead
354.740+0.138 & N & 17 31 56.99 & $-$33 18 34.1 & 0.8$\times$0.0, 146 & 61 & 67 & 52.6 & 2.6 & $-$1.2 & $-$1.0\da & $-$0.8 (i) & EG--SR \\
              & S & 17 31 57.00 & $-$33 18 41.6 & 4.3$\times$1.6, 0   & 14 & 13 &      & 8   & $-$1.1 &      &       &   \\
              &   &             &             &              &    &    &      &     &      &      &          &  \\
356.000+0.023 & E & 17 35 39.74 & $-$32 18 54.2 & 2.3$\times$1.2, 103* & 30 & 42 & 34.7 & 7  & $-$0.9 & $-$0.8 & $-$1.1 (GM) & EG--T \\
              & C & 17 35 40.15 & $-$32 18 57.0 & 0.9$\times$0.0, 24*  & 29 & 34 &      & 9  & $-$0.6 & $-$0.5 &       &  \\
              & W & 17 35 40.40 & $-$32 18 59.0 & 1.3$\times$1.1, 107* & 36 & 38 & 79.2 & 6  & $-$1.0 & $-$0.6(i) &    &   \\
              &   &             &             &               &    &    &      &      &      &          &      \\
356.161+1.635 & E & 17 29 42.63 & $-$31 18 02.1 & 3.1$\times$1.0, 81  & 21 & 38 &      & 5.6 & $-$1.1 & $-$0.7 & $-$0.9 (i) & EG--D \\
              & W & 17 29 43.80 & $-$31 18 05.6 & 2.4$\times$0.7, 101 & 44 & 69 &      & 11  & $-$1.0 & $-$0.8 &     &     \\
              &   &             &             &              &    &    &      &     &      &      &          &          \\
356.567+0.869 & N & 17 33 44.85 & $-$31 22 33.6 & 6.1$\times$4.4, 119 & 11 & 31 &      &     & $-$1.5 & $-$0.55\da & $-$0.9 (i) & EG--D \\
              & S & 17 33 45.82 & $-$31 22 51.2 & 3.4$\times$1.3, 140 & 21 & 46 &      & 14  & $-$1.6 & $-$0.65 &     &     \\
              &   &             &             &              &    &    &      &     &      &       &            &   \\
356.719$-$1.220 & E & 17 42 27.41 & $-$32 22 08.3 & 3.4$\times$2.1, 141 & 19 & 30 &      & 11  & $-$1.0 & $-$0.8 & $-$0.9 (i) & EG--D \\
              & W & 17 42 27.96 & $-$32 22 15.3 & 2.0$\times$1.8, 102 & 24 & 35 &      & 7   & $-$1.0 & $-$0.6 &    &      \\
              &   &             &             &              &    &    &      &     &      &      &            &  \\
358.002$-$0.636 & C & 17 43 17.87 & $-$30 58 18.7 & 0.4$\times$0.0, 165 & 277 & 283 &     & 1  & $-$0.4 & $-$0.26 & $-$0.3 &  EG--U \\
              &      &          &             &              &    &     &      &     &     &       &       &     \\
358.917+0.072 & C & 17 42 44.00 & $-$29 49 16.0 & 0.3$\times$0.0, 2 & 99 & 101 & 74  & 1.5 & $-$1.4 & $-$1.05 & $-$1.2 (GC) & EG--U \\
              &   &             &             &            &    &     &     &     &      &      &      &    \\
359.388+0.460 & C & 17 42 21.47 & $-$29 13 00.9 & 0.5$\times$0.0, 178 & 39 & 41 &    & 3   & $-$0.9 & $-$0.65\da & $-$0.4 (GC) & EG--U \\
              &   &             &             &              &    &     &    &     &      &      &        \\

359.844$-$1.843 & C & 17 52 30.91 & $-$30 01 06.6 & 1.4$\times$1.0, 74 & 154 & 181 &     & 1.6 & $-$0.6 & $-$0.2 & $-$0.9 (GC) & EG--SR \\
              &   &             &             &              &    &     &     &    &      &      &         &  \\
359.911$-$1.813 & C & 17 52 33.1 & $-$29 56 44.8 & 0.6$\times$0.0, 01 & 132 & 149 &    &  4  & $-$0.9 & $-$0.2    & $-$0.9 (GC) & EG--U \\
              &   &            &             &             &     &     &    &     &      &         &     &        \\
359.993+1.591 & E & 17 39 26.49 & $-$28 06 18.2 & 1.2$\times$0.0, 176 & 13 & 14  &     &    & $-$1.2 & $-$0.9 & $-$0.5 (GC) & EG--D  \\
              & W & 17 39 26.94 & $-$28 06 12.6 & 1.5$\times$0.0,46   & 11 & 12  &     & 10 & $-$0.8 & $-$0.66 &       &    \\
              &   &             &             &              &    &     &     &    &      &      &             &        \\
0.313+1.645 & E & 17 40 00.27 & $-$27 48 11.6 & 3.4$\times$2.1, 144* & 17 & 32 &     & 16 & $-$0.7 & $-$0.6 &     &  EG--T   \\
            & C & 17 40 00.65 & $-$27 48 15.9 & 2.3$\times$0.9, 155* & 07 & 08 &     &    & $-$0.1 &  & $-$0.55 (GC) & \\
            & W & 17 40 01.08 & $-$27 48 22.0 & 2.3$\times$0.0, 131* & 11 & 21 &     &    & $-$1.3 & $-$0.8 &           &  \\
            &   &             &             &               &    &    &     &    &      &      &           &   \\

0.537+0.263 & C & 17  45  52.48 & $-$28  20  26.6 & 2.5$\times$0.0, 175 & 71 & 79 & 121 & 1.6 & 0.64 & 0.44 &   & EG--U \\
            &   &               &               &              &    &     &    &     &      &      &           & \\
1.028$-$1.110 & N & 17 52 22.51 & $-$28 37 34.1 & 2.4$\times$0.9, 153 & 23 & 26 &     & 6   & $-$1.1 & $-$0.8    & $-$1.0 (GC) & EG--D \\
            & S & 17 52 22.81 & $-$28 37 42.1 & 1.3$\times$0.0, 142 & 40 & 45 &     & 6   & $-$1.1 & $-$1.0    &    &      \\
            &   &             &             &              &    &    &     &     &      &         &            &    \\
1.035+1.559 & N & 17 42 01.90 & $-$27 13 10.3 & 2.3$\times$0.0, 17  & 50 & 67 &     & 24  & $-$1.2 & $-$0.73   & $-$0.86 (GC) & EG--T \\
            & C & 17 42 03.42 & $-$27 14 16.3 & 1.7$\times$0.0, 85  & 02 & 02 &     &     & 0.0  &         &      &     \\
            & S & 17 42 05.22 & $-$27 15 10.2 & 3.7$\times$2.6, 119 & 02 & 09 &     &     &      &         &      &     \\
            &   &             &             &              &    &    &     &     &      &         &        &   \\
1.505$-$1.231 & E & 17 53 55.37 & $-$28 15 53.8 & 1.8$\times$0.8, 109 & 30 & 56 &     & 17  & $-$1.7  & $-$0.43 & $-$1.3 (i) & EG--D \\
            & W & 17 53 59.21 & $-$28 17 21.3 & 0.7$\times$0.0, 13  & 23 & 25 &     & 12  & $-$1.0 & $-$0.7    &        &   \\
            &   &             &             &              &    &    &     &     &      &      &   &     \\
2.143+1.772 & C & 17 43 51.24 & $-$26 10 59.6 & 0.8$\times$0.4, 150 & 45 & 48 &     & 4.5 & $-$1.5 & $-$1.25   & $-$1.1 & EG--U    \\
            &   &             &             &              &    &    &     &     &      &         &               &     \\
4.005+1.403 & S & 17 49 31.68 & $-$24 46 58.5 & 1.7$\times$0.0, 163 & 52 & 71 &     & 4   & $-$0.8 &  $-$0.7   & $-$0.8 (i) & EG--D \\
            & N & 17 49 32.30 & $-$24 46 45.7 &              & 01 & 03 &     &     &      &         &             &     \\
            &   &             &             &              &    &    &     &     &      &         &               &   \\
4.188$-$1.680 & S & 18 01 44.34 & $-$26 10 43.6 & 1.8$\times$0.7, 25  & 29 & 44 &     & 13  & $-$1.4 &  $-$0.5   & $-$1.3 (i) & FR-II\\
            & C & 18 01 45.97 & $-$26 10 17.6 & 1.4$\times$0.6, 22  & 06 & 06 &     &     & $-$0.7 &         &      &    \\
            & N & 18 01 47.04 & $-$26 10 02.0 & 6.7$\times$4.2, 16  & 04 & 16 &     &     & $<$-2.0& $-$0.84 &      &    \\
            &   &             &             &              &    &    &     &     &      &         &          &     \\
4.256$-$0.726 & E & 17 58 11.76 & $-$25 38 49.1 & 1.0$\times$0.0, 165 & 08 & 09 &     &     & $-$0.8 & $-$0.8    & $-$0.84 (i) & EG--D  \\
            & W & 17 58 13.00 & $-$25 38 44.7 & 1.7$\times$0.5, 63  & 13 & 17 &     & 18  & $-$1.5 & $-$1.0    &    &           \\
            &   &             &             &              &    &    &     &     &      &         &            &      \\
5.260$-$0.754 & C & 18 00 31.50 & $-$24 47 21.3 & 0.7$\times$0.5, 68*; & 66 & 89 &     & 10  & $-$0.8 & $-$0.8    & $-$0.45 &  EG--U  \\
            &   &             &             & 0.6$\times$0.4, 84*  &    &    &     &     &      &         &          &      \\
            &   &             &             &              &    &    &     &     &      &         &       &        \\
5.358+0.899 & E & 17 54 27.34 & $-$23 52 33.8 & 1.4$\times$0.0, 8*; & 24 & 45 &     &     & $-$0.4 & $-$0.85   & $-$0.95 (i) & EG--T \\
            &   &             &             & 1.7$\times$0.5, 7*  &    &    &     &     &      &         &        &  \\
            & W & 17 54 27.78 & $-$23 52 36.2 & 0.8$\times$0.2, 40* & 31 & 33 &     & 6   & $-$1.2 & $-$0.9    &   &        \\
            &   &             &             &              &    &    &     &     &      &         &            &      \\
5.511$-$1.515 & S & 18 03 59.28 & $-$24 56 45.6 & 5.3$\times$1.0, 01  & 62 & 85 &     & 4   & $-$0.8 & $-$0.6    & $-$1.1 & EG--D   \\
            & N & 18 03 59.43 & $-$24 56 30.8 &              & 02 & 04 &     &     &      & $-$0.8    &             &     \\
            &   &             &             &              &    &    &     &     &      &         &   &       \\
6.183$-$1.480 & N & 18 05 18.04 & $-$24 20 26.7 & 3.6$\times$1.6, 09  & 08 & 10 &     &     & $-$0.8 & $-$1.0\da   & $-$1.2 (i) & EG--D \\
            & S & 18 05 18.04 & $-$24 20 45.5 & 3.6$\times$0.9, 08  & 19 & 24 &     & 14  & $-$1.1 & $-$0.9    &      &     \\
\hline
\multicolumn{12}{l}{Note: In Table 2, 3, 4 and 5 of this paper, the symbols
shown below indicate the following} \\
\multicolumn{12}{l}{`*' -- size of the object estimated from 8.5 GHz map} \\
\multicolumn{12}{l}{s -- Spectral index estimated by dividing images made at
different frequencies} \\ 
\multicolumn{12}{l}{G -- Galactic source} \\
\multicolumn{12}{l}{EG -- Extragalactic source} \\
\multicolumn{12}{l}{U -- Unresolved source} \\
\multicolumn{12}{l}{SR -- Slightly resolved} \\
\multicolumn{12}{l}{D -- Double source} \\
\multicolumn{12}{l}{T -- Tripple} \\
\multicolumn{12}{l}{C+E -- Flat spectrum core with extended emission.} \\
\multicolumn{12}{l}{M -- Multiple sources in the field} \\
\multicolumn{12}{l}{$\dagger$ -- 1.4 GHz GPS flux density differs from NVSS by more than 20\%} \\
\multicolumn{12}{l}{(N) -- 1.4 GHz flux density taken from NVSS}  \\
\multicolumn{12}{l}{(L) -- Flux density taken from \citet{LAZIO1999}} \\
\multicolumn{12}{l}{** -- 1.4 GHz flux density unknown} \\
\multicolumn{12}{l}{(GC) -- P band flux density taken from \citep{LAROSA2000}} \\
\multicolumn{12}{l}{(GM) -- P band flux density taken from \citet{ROY2002}} \\
\multicolumn{12}{l}{(GM1) -- P band flux density taken from S. Bhatnagar
(private communication)} \\ 
\multicolumn{12}{l}{(i) -- Spectral index estimated from integrated flux
densities of components} \\

\end{longtable}
%
%

%
\setlongtables
\begin{longtable}{c c c c c c c c c c c c c}
\caption{Measured parameters of the polarised sources from VLA data} \\
\label{pol.src.prop.vla}\\
\hline
Source & Cmp & RA & DEC & S${maj}$ $\times$ S$_{min}$, PA & S$_p$ & S$_t$ & S$_{5}$ & \% & $\alpha_{8.5/4.8}$ & $\alpha_{4.8/1.4}$ & $\alpha_{1.4/0.3}$ & Notes\\
            &     & (J2000) & (J2000) &      &              & mJy  &  mJy        & Poln. &           &    &  &  \\
\hline
\endfirsthead

\hline
Source & Cmp & RA & DEC & S${maj}$ $\times$ S$_{min}$, PA & S$_p$ & S$_t$ & S$_{5}$ & \% & $\alpha_{8.5/4.8}$ & $\alpha_{4.8/1.4}$ & $\alpha_{1.4/0.3}$ & Notes \\
            &     & (J2000) & (J2000) &       &         & mJy    &  mJy   & Pol. &           &    &  &  \\
\hline
\endhead
353.462$-$0.691 & C & 17 31 55.40 & $-$34 49 59.9 & 0.4$\times$0.2, 165 & 98 & 103 &     & 0.6 & $-$0.9 & $-$0.9  & $-$0.7  & EG--U  \\
              &   &             &             &              &    &     &     &      &       &      &   &      \\
354.815+0.775 & E & 17 29 36.26 & $-$32 53 52.1 & 0.8$\times$0.5, 71  & 6.8 & 8.2 &     &     & $-$1.2 & $-$1.0 & $-$0.9 (i) & EG--D \\
              & W & 17 29 36.64 & $-$32 53 51.7 & 0.5$\times$0.4, 176 & 18.2 & 20.2 &   & 5   & $-$1.4 & $-$1.0 &            &  \\
              &   &             &             &              &      &     &    &     &    &      &              &     \\
355.424$-$0.809 & N & 17 37 32.54 & $-$33 14 36.7 & 2.7$\times$1.3, 57  & 6.5  & 22  &     & 20 & $-$1.0 s & $-$1.0 (N)& $-$0.9 & FR-II \\
              & C & 17 37 32.07 & $-$33 14 50.7 & 1.5$\times$1.0, 167 & 1.0 & 1.1  &     &    & $-$1.0 s &      &      &       \\
              & S & 17 37 31.40 & $-$33 15 10.4 & 1.5$\times$1.1, 176 & 11 & 24    &     & 12 & $-$0.8 s &      &      &       \\
              &   &             &             &              &    &       &     &    &      &      &     &     \\
356.905+0.082 & C & 17 37 43.79 & $-$31 31 13.7 & 3.6$\times$0.4, 179 & 40 & 43    & 32  & 3  & $-$0.1 &      & $-$0.4 (GM)(i) & FR-I \\
              &   &             &             &              &    &       &     &    &      &      &          &             \\
357.865$-$0.996 & C & 17 44 23.58 & $-$31 16 36.0 & 0.1$\times$0.0, 22  & 369 & 370 &    & 1.4 & 0.2  & 0.4\da & $-$0.5 (i) & C+E \\
(1741$-$312)    &   &             &             &              &    &    &      &     &      &      &          &     \\
              &   &             &             &              &        &      &     &      &      &              &      \\
358.149$-$1.675 & E & 17 47 45.32 & $-$31 23 37.4 & 2.1$\times$1.4, 124 & 8.9 & 57   &    & 25 & $-$1.0 s & $-$1.0 (N)& $-$0.77 & FR-II \\
              & C & 17 47 48.13 & $-$31 23 15.2 & 0.3$\times$0.0, 18  & 7.4 & 7.6  &     &    &  0.1 s &      &       &   \\
              & W & 17 47 50.13 & $-$31 23 03.9 & 0.7$\times$0.4, 60  & 13 & 98    &     & 12 & $-$0.8 s &      &     &     \\
              &   &             &             &              &      &     &     &    &      &      &            &  \\
358.156+0.028 & N & 17 41 02.91 & $-$30 29 22.1 & 0.5$\times$0.5, 113 & 23.7 & 30  & 31  & 13 & $-$0.9 & $-$0.6 & $-$1.0 (GC) & FR-II \\
              & C & 17 41 02.90 & $-$30 29 27.3 & 0.5$\times$0.0, 04  & 1.7  & 1.5 &     &    & $-$0.1 &      &               &  \\
              & S & 17 41 02.67 & $-$30 29 36.1 & 1.5$\times$0.7, 39  & 3.0 & 4.2  & 5.2 & 21 & $-$0.8 & $-$0.86 &            &    \\
              &   &             &             &              &      &     &     &    &       &      &            &   \\
358.930$-$1.197 & E & 17 47 46.36 & $-$30 28 17.8 & 1.5$\times$0.9, 27  & 8.9 & 22   &     & 26 & $-$0.6 s & $-$0.4 (i)& $-$1.4 (GC) & EG--D\\
              & W & 17 47 47.54 & $-$30 28 11.7 & 2.6$\times$1.3, 22  & 4.5 & 14.8 &     &    & $-$0.7 s &     &       & \\
              &   &             &             &              &      &     &     &    &       &      &          &  \\
359.392+1.272 & E & 17 39 12.44 & $-$28 47 02.7 & 2.2$\times$1.2, 33  & 6.9 & 24.5 &     & 16 & $-$0.6 & $-$0.9 (N)& $-$0.76 (GC) & FR-II\\
              & W & 17 39 13.46 & $-$28 46 53.7 & 2.8$\times$1.5, 40  & 5.1 & 20.4 &     & 20 & $-$0.7  &      &   &     \\
              &   &             &             &              &     &      &     &    &       &      &          &      \\
359.2$-$0.8     & C & 17 47 15.78 & $-$29 58 01.0 & 3.2$\times$2.2, 20  & 11.3 & 20.2&     & 13 & 0.3   &      & $-$0.3 (GC) (i) & G \\
(Mouse)       &   &             &             &              &      &     &     &    &       &      &          &  \\
359.604+0.306 & E & 17 43 28.39 & $-$29 06 47.6 & 2.8$\times$1.0, 77  & 2.6 & 7.7  &  $-$  &  & $-$1.0  &  **  & $-$0.8 (GC) (i)& FR-II \\
              & C & 17 43 28.77 & $-$29 06 47.0 & 0.7$\times$0.1, 81  & 2.5 & 2.7  & 3.3 &    & $-$0.1  &      & (0.3/4.8) & \\
              & W & 17 43 29.26 & $-$29 06 47.1 & 1.4$\times$0.9, 14  & 6.7 & 10.8 & 7.6 & 12 & $-$1.0  &      &           & \\
              &   &             &             &              &      &     &     &    &       &      &          &  \\
359.710$-$0.904 & E & 17 48 27.73 & $-$29 39 10.3 & 4.0$\times$1.9, 67  & 2.4 & 23   &     & 23 & $-$1.4 &  ** & $-$0.75 (GC) & FR-II?\\
              & C & 17 48 28.58 & $-$29 39 10.9 & 0.4$\times$0.0, 52  & 0.7 & 0.7  &     &    & $-$1.5  &      & (0.3/4.8) & \\
              & W & 17 48 29.39 & $-$29 39 11.8 & 5.5$\times$2.5, 90  & 2.3 & 23.6 &     & 15 & $-$1.7  &      &            & \\
              &   &             &             &              &      &     &     &    &       &      &          &     \\
359.871+0.179 & E & 17 44 36.22 & $-$28 57 14.9 & 1.7$\times$0.0, 78  & 1.5 & 2.2  &   &  5 & $-$0.8  & $-$0.5 (L)& $-$0.26 (GC) & FR-II\\
              & C & 17 44 36.84 & $-$28 57 11.5 & 0.2$\times$0.0, 11  & 7.9 & 8.1  &     &    & $-$0.8  & $-$0.4 (L)&               &  \\
              & W & 17 44 37.08 & $-$28 57 09.7 &              & 41  & 46   & 34  &    & $-$1.3  & $-$1.0 (L)&      & \\
              &   &             &             &              &      &    &     &    &        &      &     &   \\
0.404+1.061 & E & 17 42 27.93 & $-$28 02 08.5 & 0.4$\times$0.1, 130 & 13.5 & 14.8 &     & 11 & $-$1.3   & $-$1.0\da(i)& $-$0.9 (GC) & EG--D \\
            & W & 17 42 28.25 & $-$28 02 10.9 & 0.7$\times$0.1, 97  & 07.5 & 10.0 &     &    & $-$1.3   &       &     &      \\
            &   &             &             &              &      &      &     &    &        &       &  &          \\
1.826+1.070 & N & 17 45 47.67 & $-$26 49 05.6 & 1.8$\times$1.3, 176 & 7.5 & 20.2  &     &  8 & $-$1.3   & $-$0.86 & $-$0.9 (i) & FR-II \\
            & S & 17 45 47.17 & $-$26 49 21.5 & 0.9$\times$0.0, 08  & 9.8 & 16.4  &     & 23 & $-$1.0   & $-$0.6  &            &     \\
            &   &             &             &              &      &      &     &    &        &       &  &         \\
2.922+1.028 & N & 17 48 29.18 & $-$25 54 15.8 & 1.1$\times$0.9, 37  & 13.8 & 16.7 &     &  6 & $-$1.1 s & $-$0.8 (i)& $-$0.94  &  EG--D  \\
            & S & 17 48 29.09 & $-$25 54 18.0 & 0.6$\times$0.5, 11  & 21.3 & 25.4 &     &  8 & $-$1.1 s &       &   &     \\
            &   &             &             &              &      &      &     &    &        &       &          &   \\
3.347$-$0.327 & C & 17 54 38.31 & $-$26 13 50.9 & 1.7$\times$0.9, 168 & 4.8 & 10.3  & 6  & 20 & $-$1.2   & $-$1.1\da & $-$1.2 (GM) & EG--SR \\
            &   &             &             &              &      &      &     &    &        &       &   &     \\
3.745+0.635 & C & 17 51 51.26 & $-$25 24 0.0 & 0.0$\times$0.0, 0 & 446 & 446 &   & 2.6 & $-$1.0 & $-$1.0 & 0.1 (GM1) & EG--U  \\
(1748$-$253)  &   &             &             &              &    &    &     &     &      &         &    &        \\
            &   &             &             &              &    &    &     &     &      &      &    &    \\
3.748$-$1.221 & E & 17 58 58.66 & $-$26 20 02.1 & 0.4$\times$0.0, 27  & 11.3 & 12.9 &     & 10 & $-$1.3   & $-$1.0\da(N) (i) & $-$0.77 (GM) & EG--D \\
            & W & 17 58 59.75 & $-$26 19 56.8 & 0.0,0.0, 00  & 2.8 & 3.1   &     &    &  0.1   &       &  &      \\
            &   &             &             &              &      &      &     &    &        &       &    &  \\
3.928+0.253 & C & 17 53 43.37 & $-$25 26 10.3 & 0.2$\times$0.0, 178 & 120 & 122   & 67.9&  2 &  0.2   &  **   & 0.0 (GM1) &  EG--U \\
            &   &             &             &              &     &       &     &    &        &       & (0.3/4.8)          &    \\
4.752+0.255 & C & 17 55 33.41 & $-$24 43 30.6 & 1.2$\times$1.1, 143 & 26  & 46    & 33.2&  7 & $-$1.3   & $-$1.1\da & $-$0.8 &  EG--SR \\
            &   &             &             &              &      &      &     &    &        &       &  &    \\
5.791+0.794 & E & 17 55 48.31 & $-$23 33 21.9 & 3.1$\times$1.3, 77  & 17.7 & 64   &     & 17 & $-$1.4 s &       &   & EG--T   \\
            & C & 17 55 48.60 & $-$23 33 21.1 & 1.9$\times$0.0, 77  & 10.8 & 13.9 &     & 17 & $-$1.0 s & $-$1.0\da(N) & $-$0.8 & \\
            & W & 17 55 48.82 & $-$23 33 20.5 & 1.6$\times$1.5, 68  & 12.7 & 49   &     & 30 & $-$1.3 s &       &      &  \\
            &   &             &             &              &      &      &     &    &        &       &          &       \\
5.852+1.041 & N & 17 55 00.75 & $-$23 22 32.4 & 1.5$\times$0.4, 01  & 9.2 & 21.8  &     & 12 & $-$1.0   & $-$0.9  & $-$1.1 (i) & EG--D \\
            & S & 17 55 00.80 & $-$23 22 52.0 & 3.4$\times$1.2, 119 & 2.8 & 12.8  &     & 24 & $-$1.5   & $-$0.75 &     &   \\
\hline
\end{longtable}


\begin{longtable}{c c c c c c c c c c c c}
\label{unpol.src.prop.at} \\
\caption{Measured parameters of the unpolarised sources from the ATCA data} \\
\hline

Source & Cmp & RA & DEC & S${maj}$ $\times$ S$_{min}$, PA & S$_p$ & S$_t$ & S$_{5}$ & $\alpha_{8.5/4.8}$ & $\alpha_{4.8/1.4}$ & $\alpha_{1.4/0.3}$ & Notes \\
            &     & (J2000) & (J2000) &       &       &       &             & &    &   &  \\
\hline
\endfirsthead

\hline
Source & Cmp & RA & DEC & S${maj}$ $\times$ S$_{min}$, PA & S$_p$ & S$_t$ & S$_{5}$ & $\alpha_{8.5/4.8}$ & $\alpha_{4.8/1.4}$ & $\alpha_{1.4/0.3}$ & Notes \\
            &     & (J2000) & (J2000) &       &       &       &             & &    &      &  \\
\hline
\endhead

354.719-1.117 & N & 17 36 57.84 & $-$34 00 30.0 & 0.4$\times$0.0, 01  & 92 & 91 &      &  $-$0.8 & $-$0.8 & $-$0.6 (i) & EG--D \\
              & S & 17 36 57.19 & $-$34 00 37.6 & 1.2$\times$0.0, 01  & 06 & 07 &      &  $-$1.2 & $-$1.1 &               &  \\
              &   &             &             &              &    &    &      &       &      &            &   \\
357.435-0.519 & C & 17 41 26.32 & $-$31 23 30.3 & 6.0$\times$2.0, 167 & 49 & 73 &      &  --   & $-$0.8 & $-$0.6 (GM)  & EG--SR\\
                 &   &             &             &              &    &    &      &       &      &             &    \\
358.982+0.580 & C & 17 40 54.52 & $-$29 29 50.6 & 0.3$\times$0.0, 7   & 44 & 44 &      &  --   & $-$0.6\da & $-$0.5 (GC) & EG--U \\
              &   &             &             &              &    &    &      &       &        &            &  \\
359.568+1.146 & C & 17 40 07.79 & $-$28 42 03.0 & 0.4$\times$0.4, 38  & 113&114 &      & --   & $-$0.5 & $-$0.4 (GC)   & EG--U \\
              &   &             &             &               &    &    &      & &      &         &   \\
358.591+0.046 & C & 17 42 02.71 & $-$30 06 41.7 & 1.0$\times$0.0, 26  & 25 & 25 &  17  & $-$0.8 & $-$1.0\da & $-$1.1 (GC) & EG--U\\
              &   &             &             &              &    &    &      &       &       &    &        \\
359.546+0.988 & C & 17 40 41.56 & $-$28 48 10.7 & 0.0$\times$0.0, 0   & 35 & 35 &    &  $-$0.8 & $-$0.8\da & $-$0.5 (GC) & EG--U \\
              &   &             &             &              &    &    &      &       &      &    &        \\
0.846+1.173   & C & 17 43 05.43 & $-$27 36 05.2 & 1.3$\times$0.0, 7   & 43 & 46 &      &  $-$0.9 & $-$1.0 & $-$0.8 (GC)& EG--U \\
              &   &             &             &            &    &     &     &       &      &        &  \\
1.954-1.702   & C & 17 56 49.65 & $-$28 07 37.4 & 0.7$\times$0.0, 171 & 60 & 60 &    &  $-$1.6 & $-$1.1\da & $-$0.4 &  EG--U \\
              &   &             &             &              &    &     &    &      &      &      &  \\
2.423-1.660   & C & 17 57 43.61 & $-$27 42 02.0 & 0.6$\times$0.3, 11  & 46 & 46  &    & $-$1.1 & $-$0.8 & $-$1.0 & EG--U \\
              &   &             &             &              &    &     &           &      &      &    \\
4.898+1.292 & C & 17 51 57.26 & $-$24 04 24.9 & 0.9$\times$0.0, 26   & 51 & 52 &     &  $-$1.3 & $-$1.1 & $-$0.9 & EG--U  \\
\hline
\end{longtable}
\begin{longtable}{c c c c c c c c c c c c}
\label{unpol.src.prop.vla} \\
\caption{Estimated parameters of the unpolarised sources from the VLA data} \\
\hline

Source & Cmp & RA & DEC & S${maj}$ $\times$ S$_{min}$, PA & S$_p$ & S$_t$ & S$_{5}$ & $\alpha_{8.5/4.8}$ & $\alpha_{4.8/1.4}$ & $\alpha_{1.4/0.3}$ & Notes \\
            &     & (J2000) & (J2000) &       &       &       &             & &    &   &  \\
\hline
\endfirsthead

\hline
Source & Cmp & RA & DEC & S${maj}$ $\times$ S$_{min}$, PA & S$_p$ & S$_t$ & S$_{5}$ & $\alpha_{8.5/4.8}$ & $\alpha_{4.8/1.4}$ & $\alpha_{1.4/0.3}$ & Notes \\
            &     & (J2000) & (J2000) &       &       &       &             & &    &      &  \\
\hline
\endhead

353.410-0.360 & C & 17 30 26.14 & $-$34 41 44.9 & 2.6$\times$2.1, 51 & 149 & 469 & 155 &  0.0  & 0.1 &  & G--HII \\
            &   &               &               &              &    &     &    &       &      &         &  \\
355.739+0.131 & EX & 17 34 34.24 & $-$32 28 34.3 & $\sim$12$\times$10, -  & 28 & 256 & 155 &  $-$0.1  & 0.1 &   & G--HII \\
            &   &               &               &              &    &     &    &       &      &       &    \\
358.643-0.034 & EX & 17 42 29.08 & $-$30 06 34.6 & $\sim$20$\times$20, -  & 07 & 169 & 48  & $-$1.2   & 0.4 &   & G--HII \\
            &   &             &             &              &    &    &     &      &         &         &   \\
358.605+1.440 & C & 17 36 38.76 & $-$29 21 26.4 & 0.4$\times$0.0, 17 & 39 & 40 &    & $-$0.8 &  $-$0.7 & $-$0.7 & EG--U  \\
            &   &             &             &              &    &    &     &      &         &          &  \\
359.717-0.036 & R & 17 45 05.08 & $-$29 11 45.0 & $\sim$22$\times$20, --  & 12 & 198\footnote{flux density of the component 
near the bottom of its map (Fig.~\ref{5ghz.vla.unpol.src.maps})} & 28  & $-$1.3 &  -- &  &  G--HII \\
            &   &             &             &              &    &    &     &      &         &       &   \\
4.005+0     & C & 17 54 38.87  & $-$25 27 30.7 & 1.0$\times$0.0, 165 & 08 & 09\footnote{flux density of the component 
near the top of its map (Fig.~\ref{5ghz.vla.unpol.src.maps})} &     & $-$0.8 & $-$0.8    & -- & EG--M \\
            &   &             &             &              &    &    &     &      &         &             &  \\
4.619+0.288 & C & 17 55 07.96 & $-$24 49 21.7 & 0.8$\times$0.3, 10  & 22 & 25 & 21  & $-$1.1 & $-$0.9\da   & $-$0.45 & EG--SR \\
\hline
\end{longtable}
\twocolumn
\end{landscape}

\normalsize
\begin{figure*}
\centering
\vbox{
\hbox{
\includegraphics[width=4.7cm, angle=0,clip=]{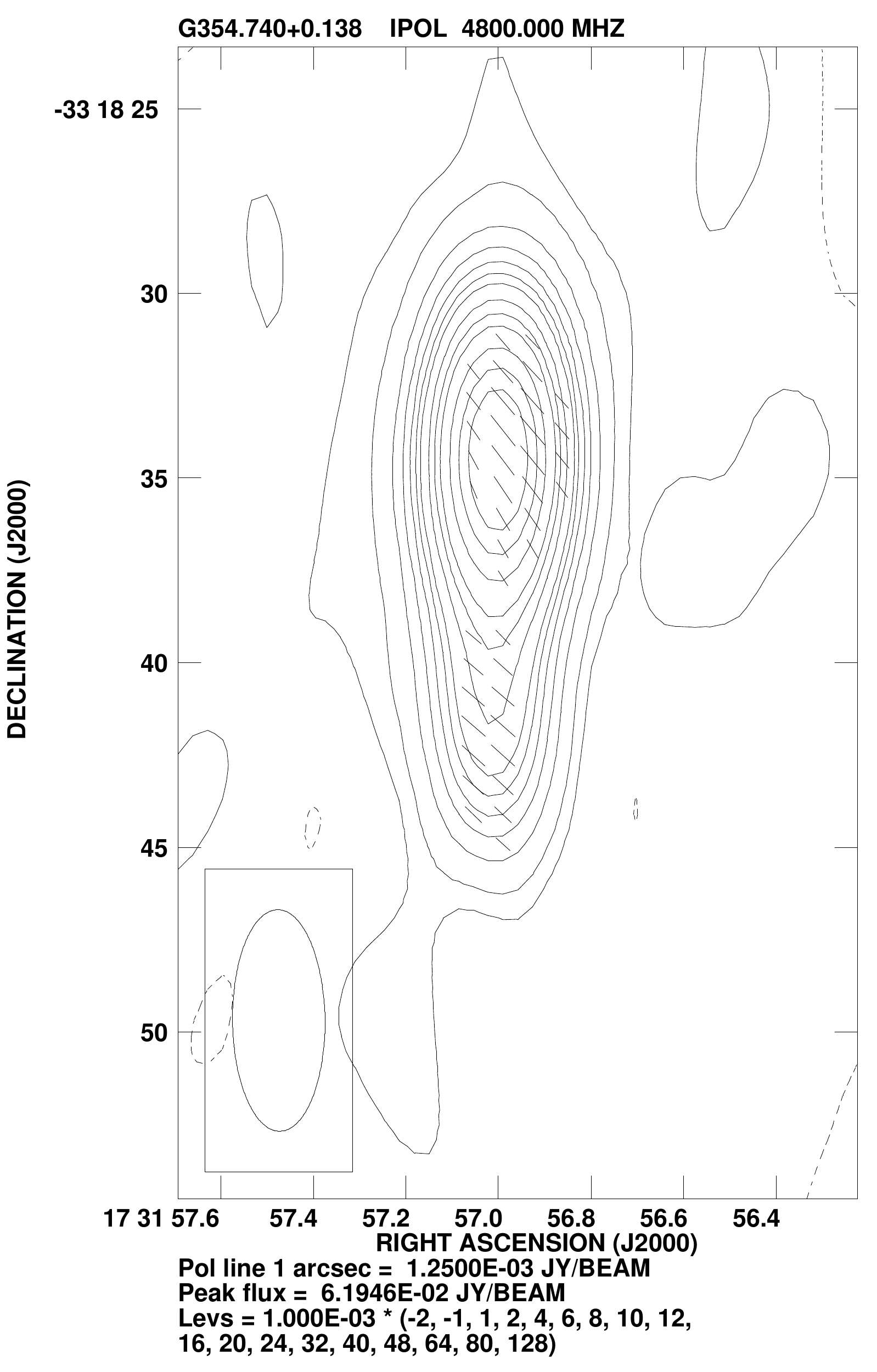}
\includegraphics[width=6.0cm, angle=0,clip=]{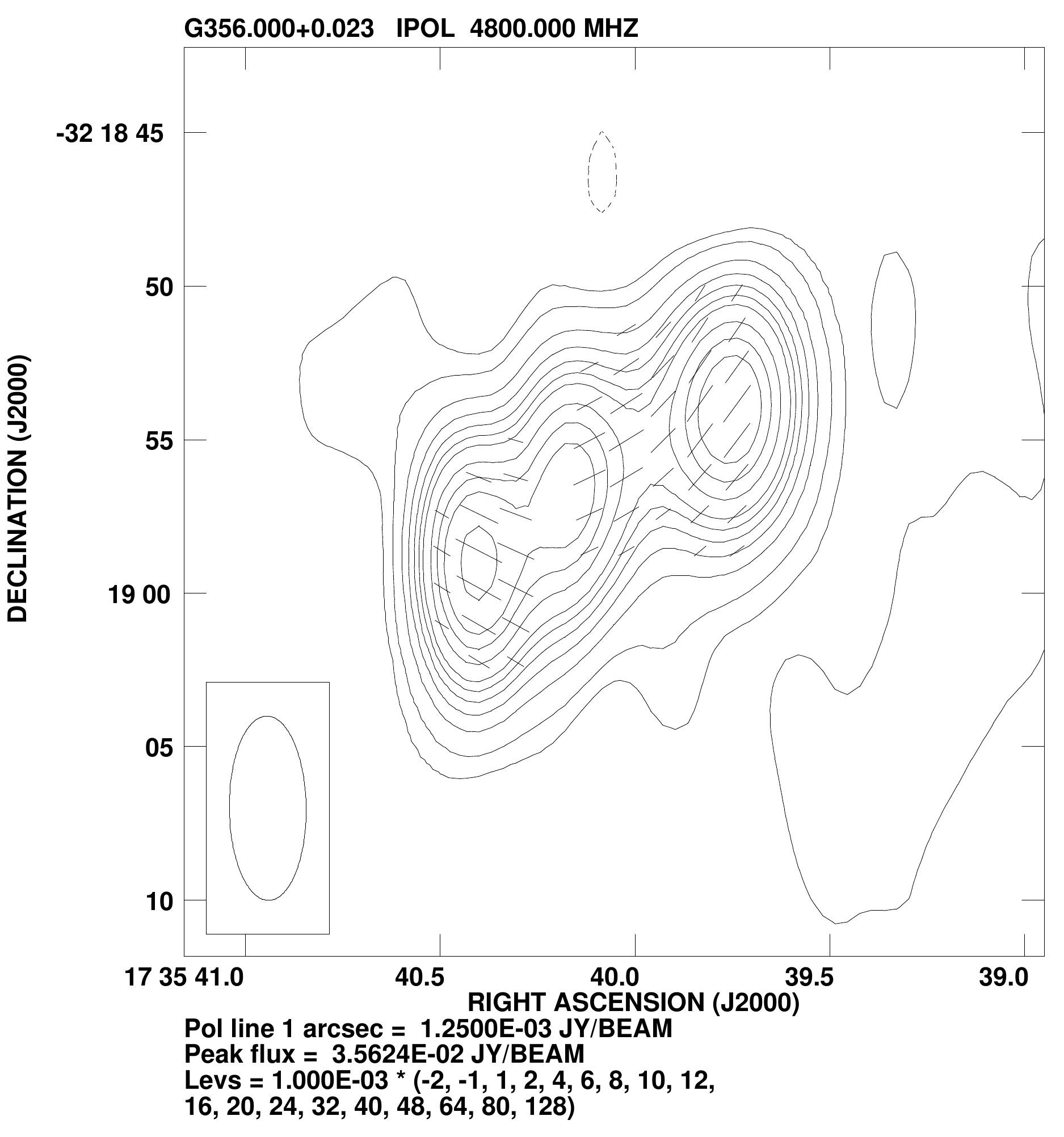}
\includegraphics[width=5.3cm, angle=0,clip=]{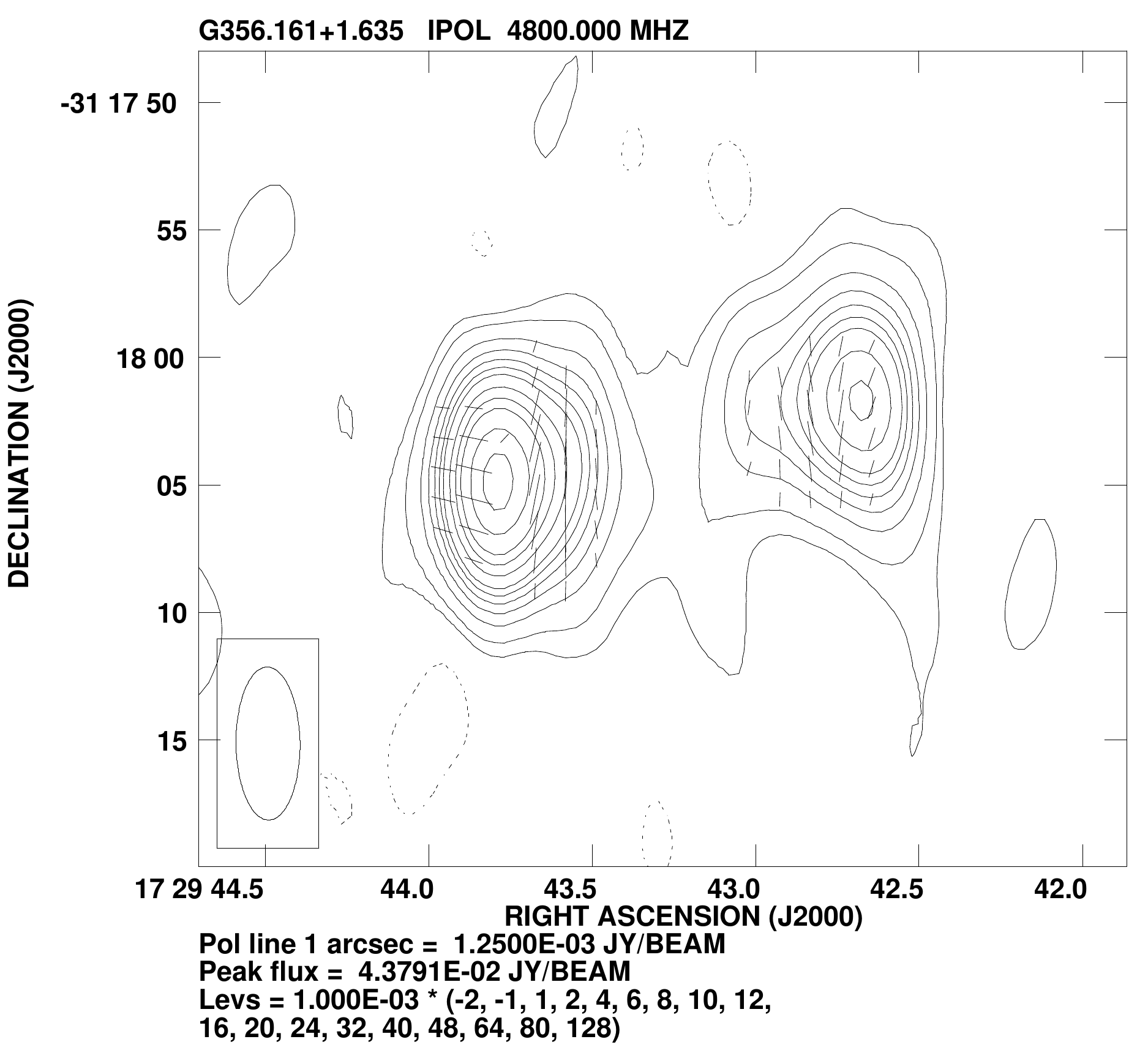}
}
\hbox{
\includegraphics[width=4.7cm, angle=0,clip=]{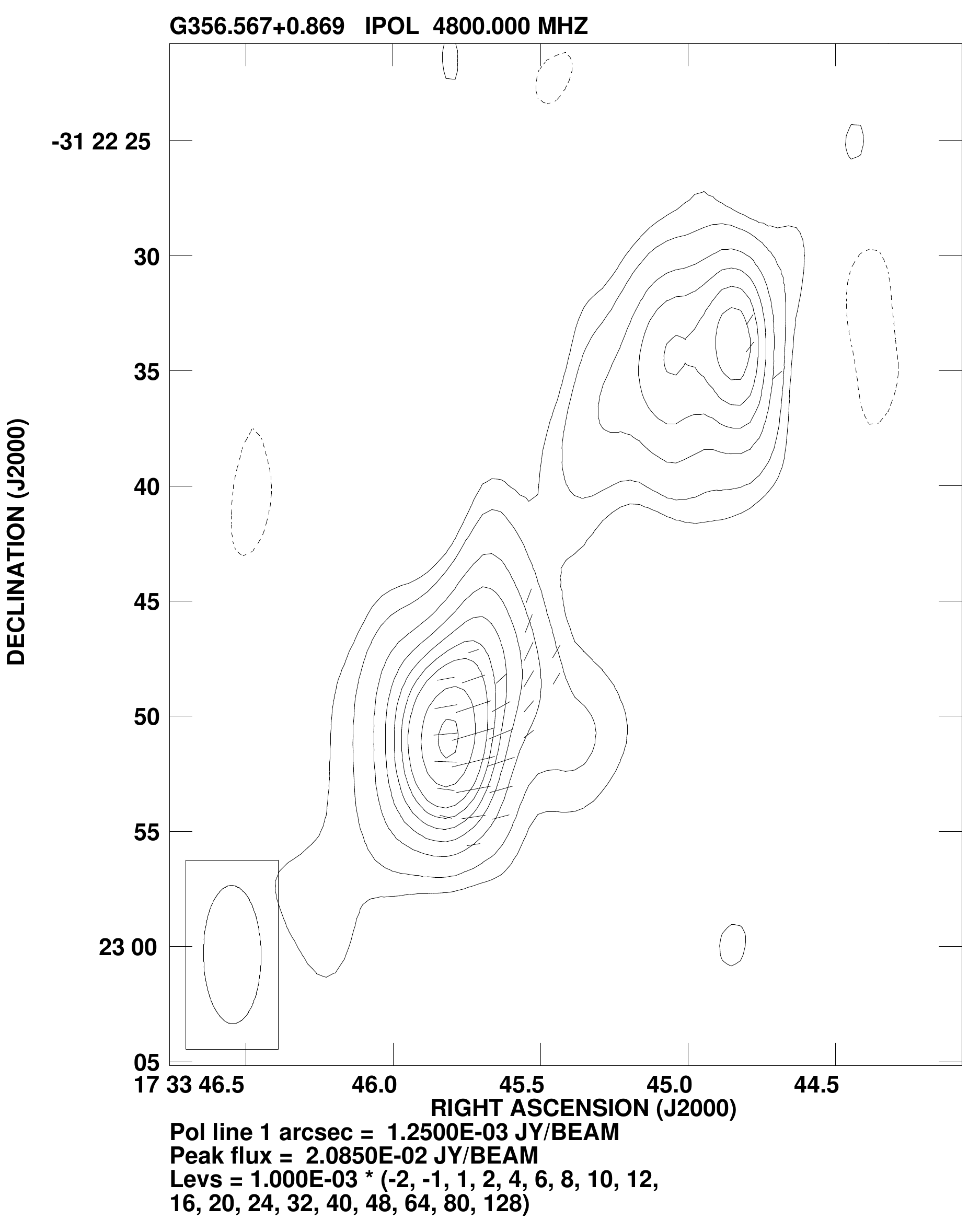}
\includegraphics[width=5.3cm, angle=0,clip=]{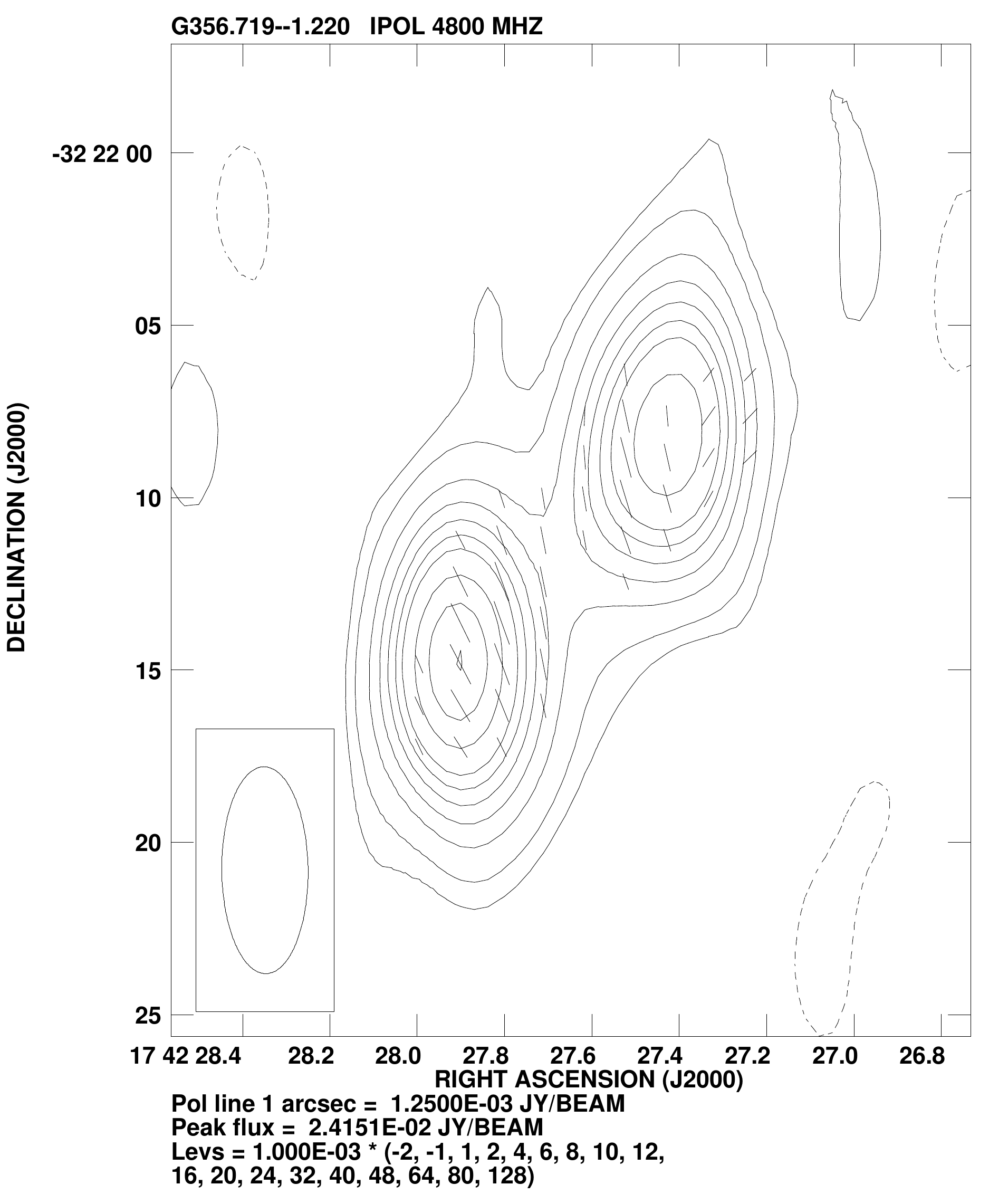}
\includegraphics[width=5.3cm, angle=0,clip=]{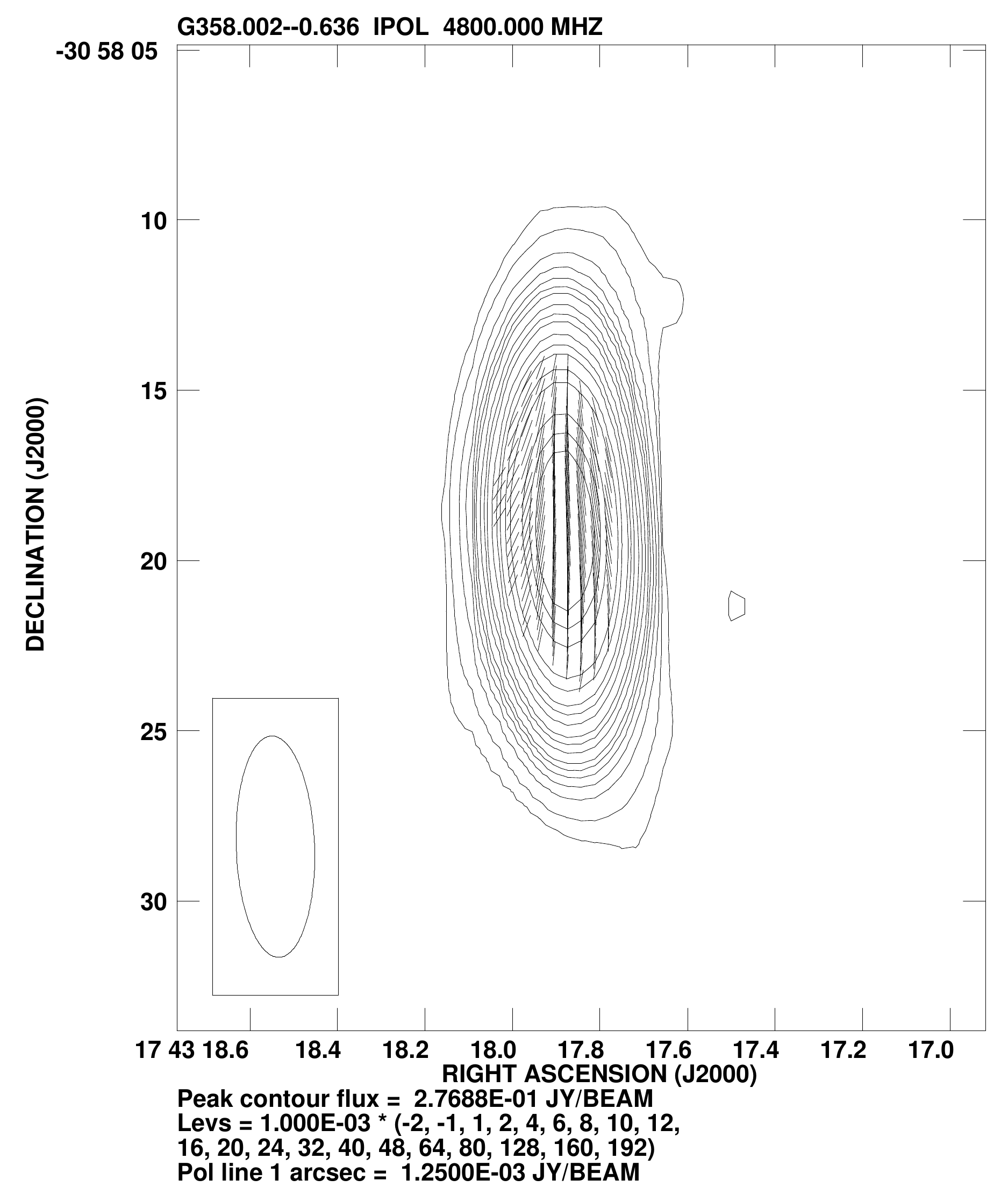}
}
\hbox{
\includegraphics[width=5cm, angle=0,clip=]{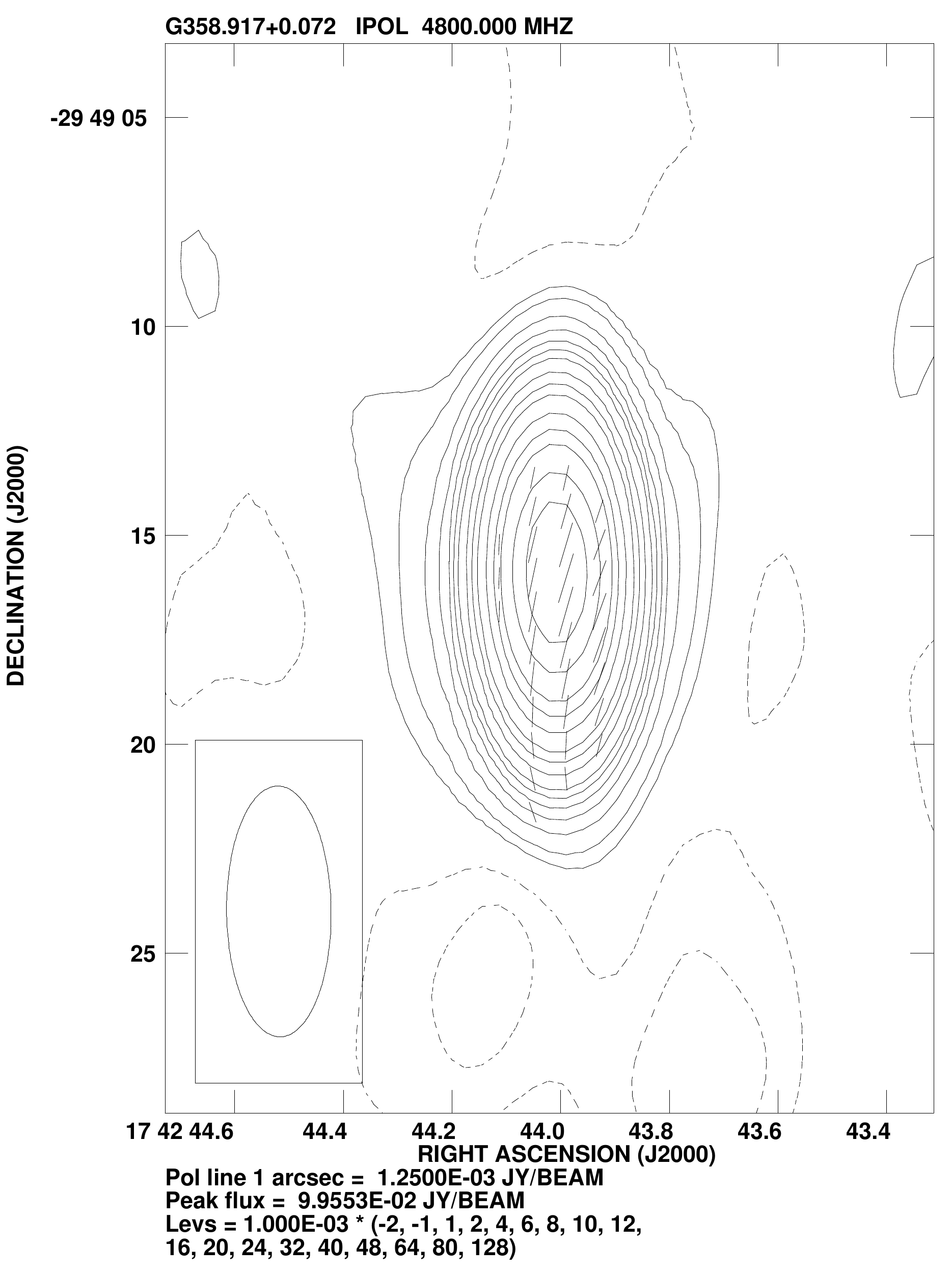}
\includegraphics[width=5.0cm, angle=0,clip=]{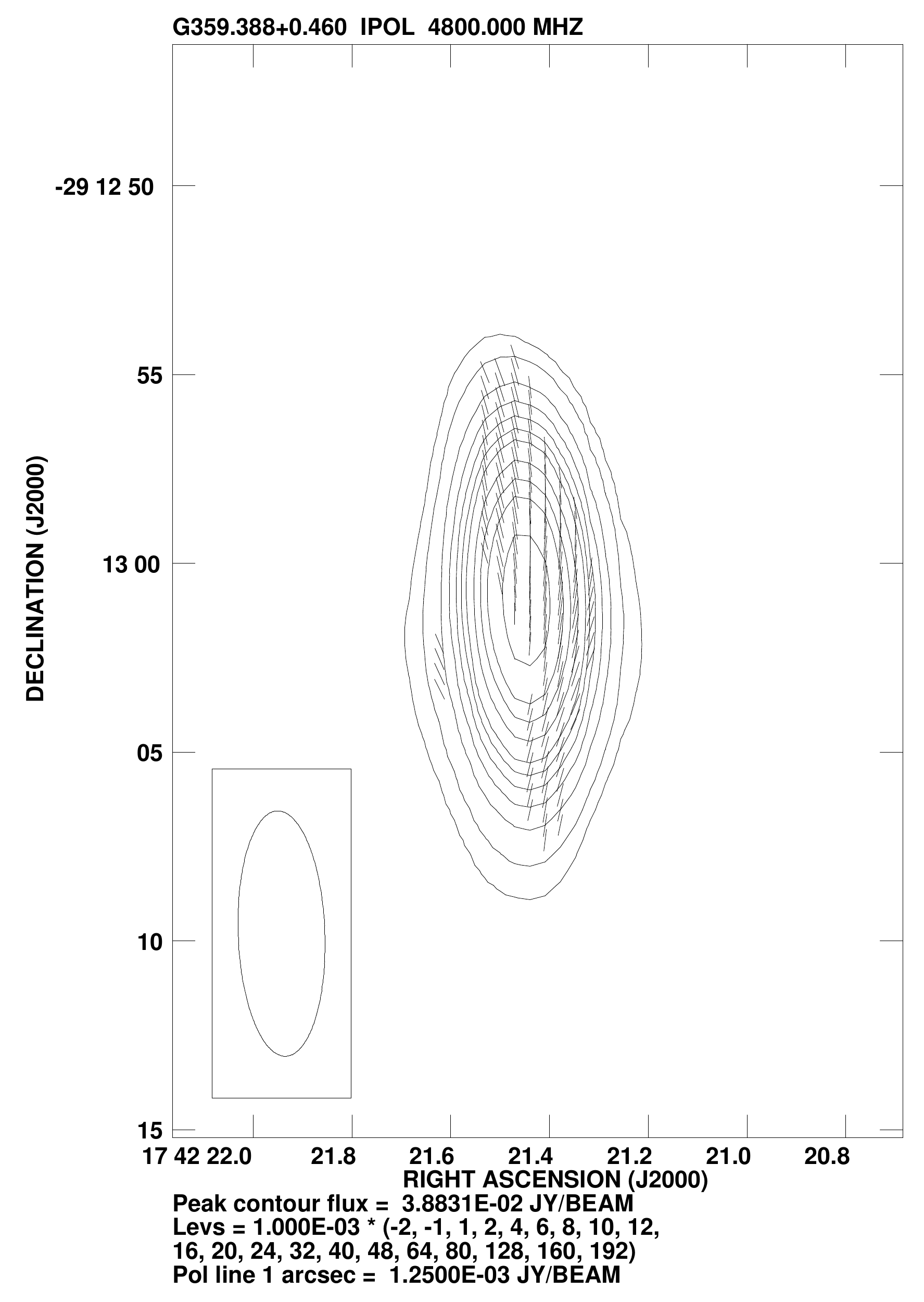}
\includegraphics[width=5cm, angle=0,clip=]{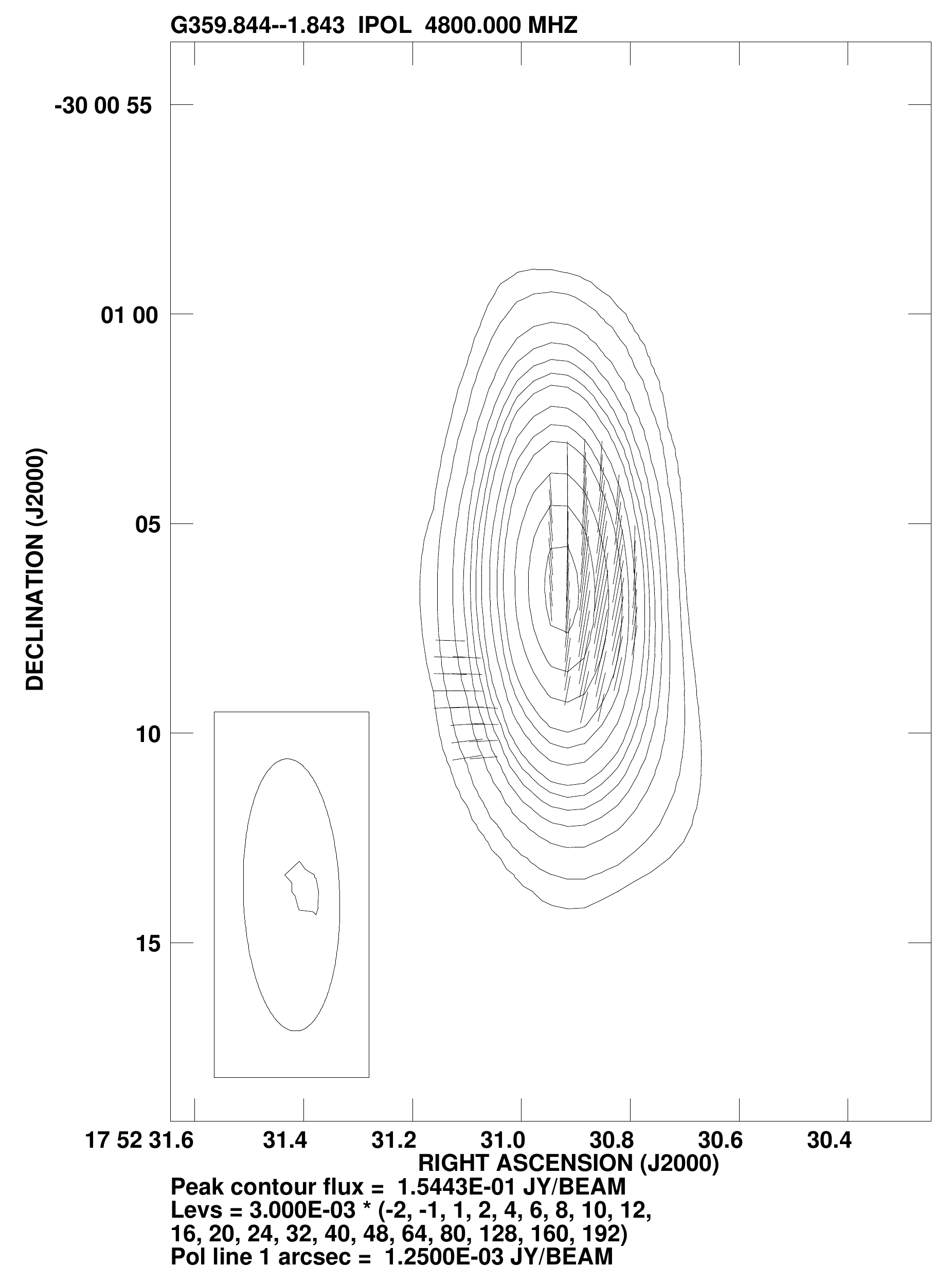}
}
}
\caption{4.8 GHz ATCA continuum images of the polarised sources with the
polarisation vectors representing the projected electric field superposed on
them.
Typical rms noise in Stokes I is about 0.23 mJy/beam and about 0.15 mJy/beam
in Stokes Q and U. Typical beamsize of these images are $\approx$6$^{''} \times
2^{''}$.}
\label{5ghz.atca.larger.maps}
\end{figure*}
\vspace{-0.4cm} 

\addtocounter{figure}{-1}
\begin{figure*}
\centering
\vbox{
\hbox{
\includegraphics[width=5cm, angle=0,clip=]{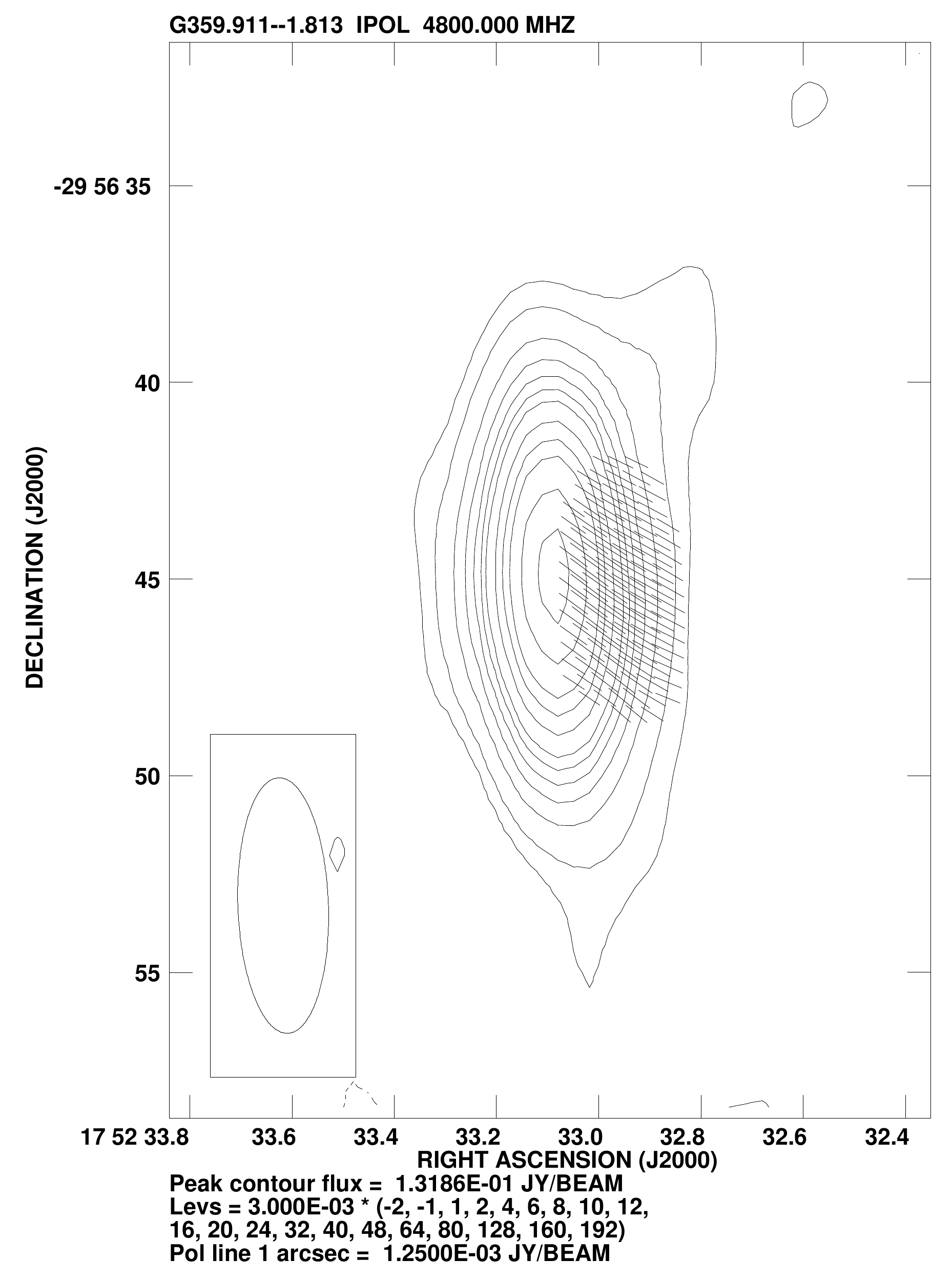}
\includegraphics[width=5cm, angle=0,clip=]{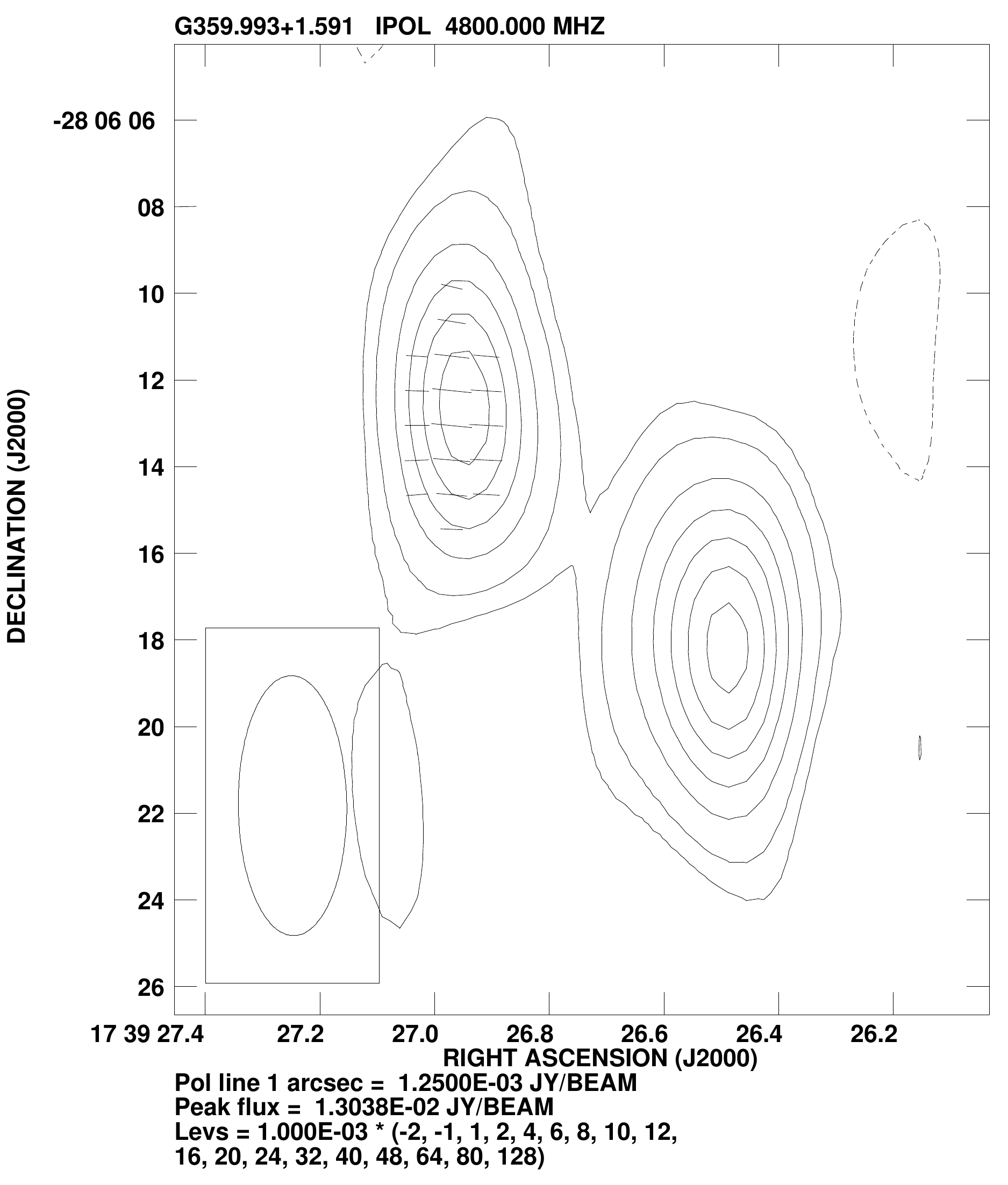}
\includegraphics[width=5cm, angle=0,clip=]{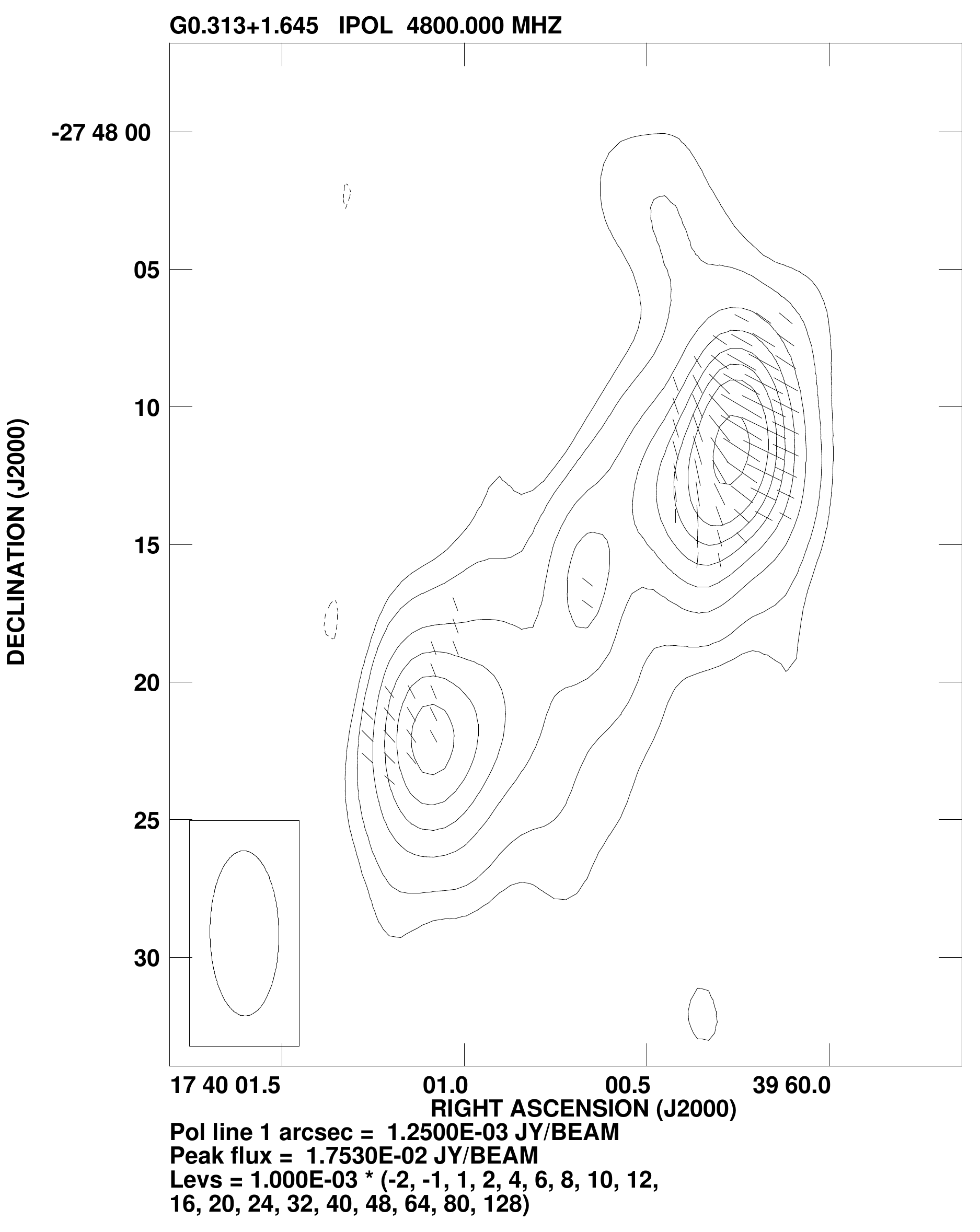}
}
\hbox{
\includegraphics[width=5.3cm, angle=0,clip=]{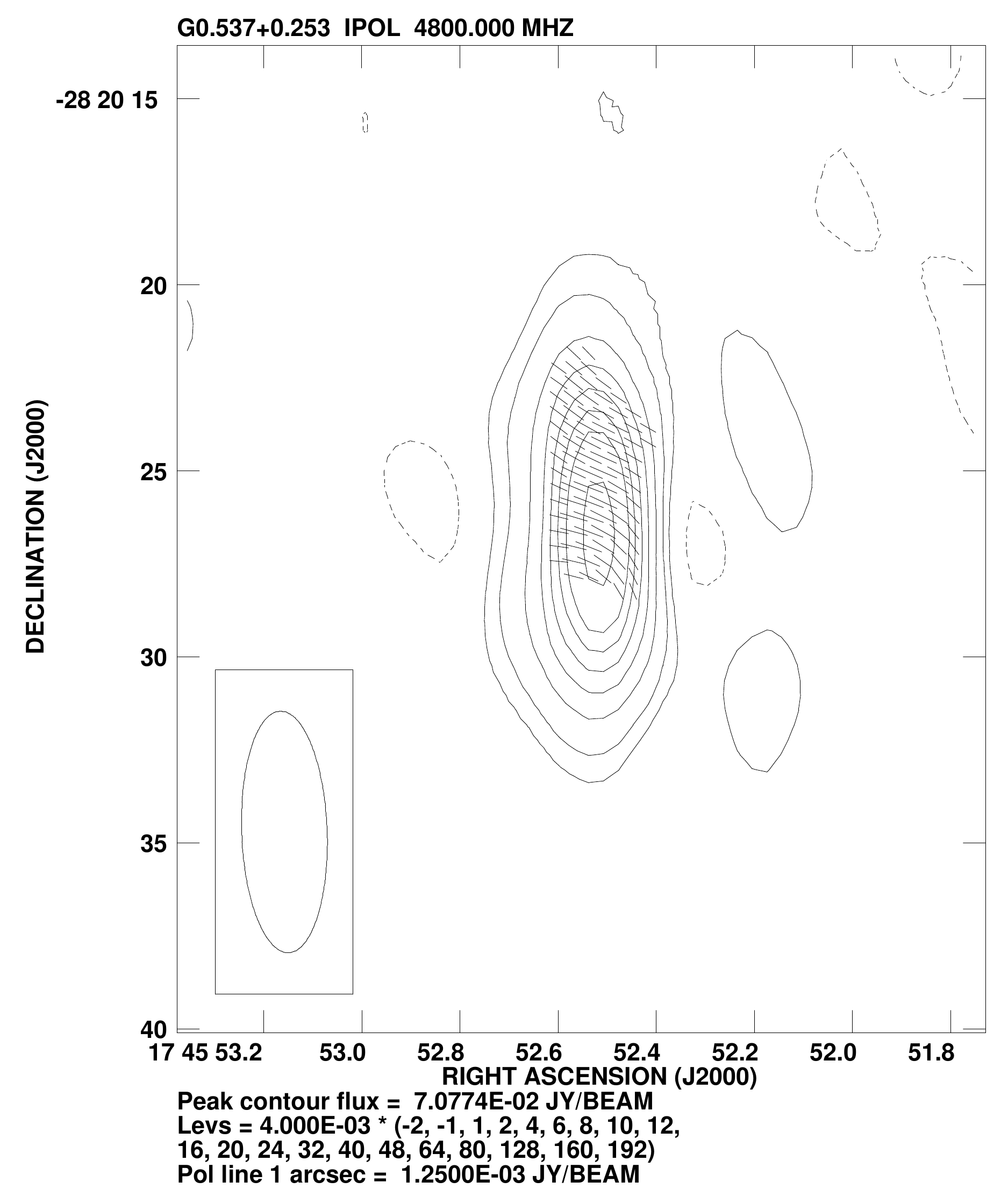}
\hspace{1.0cm}
\includegraphics[width=5cm, angle=0,clip=]{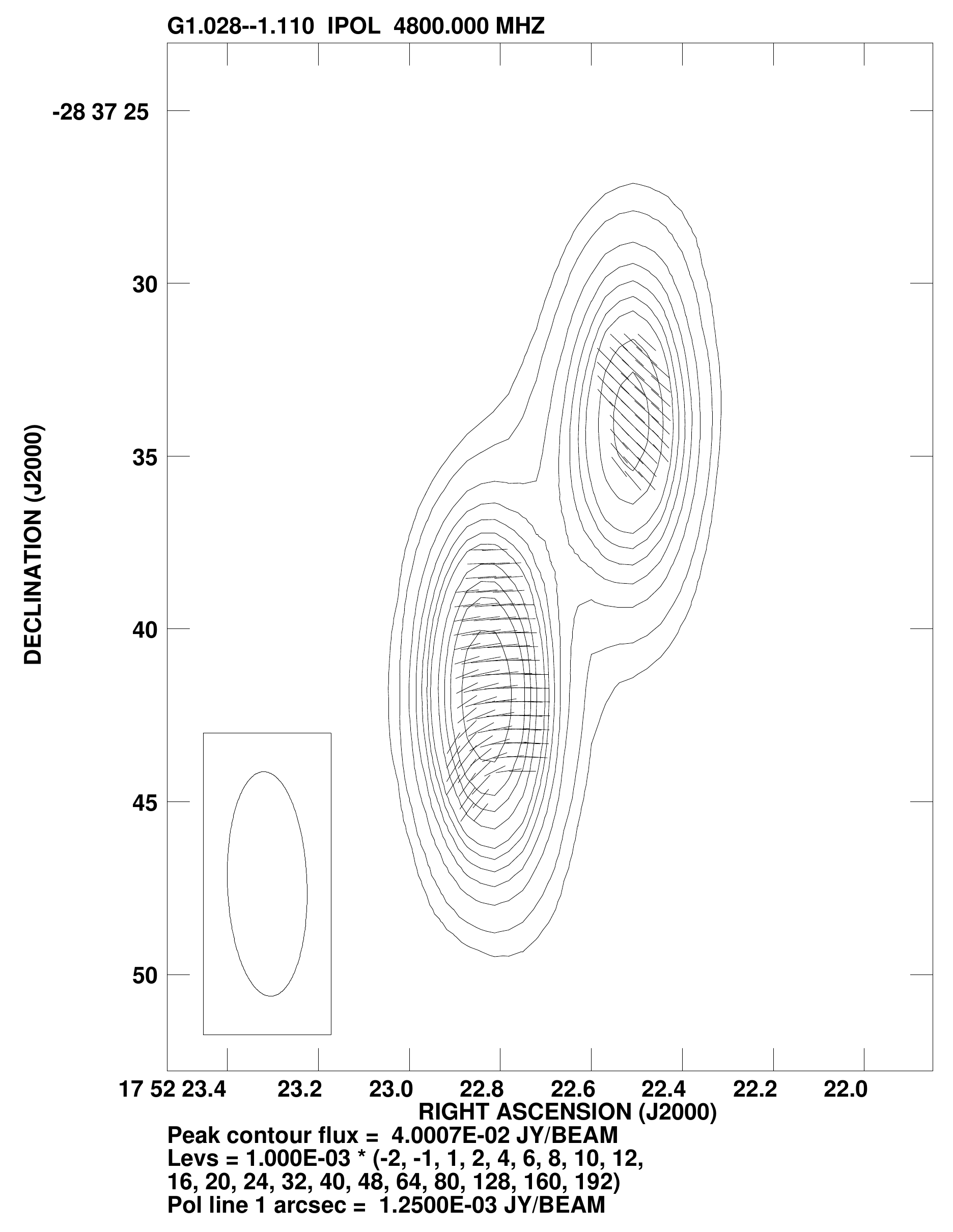}
\includegraphics[width=3.3cm, angle=0,clip=]{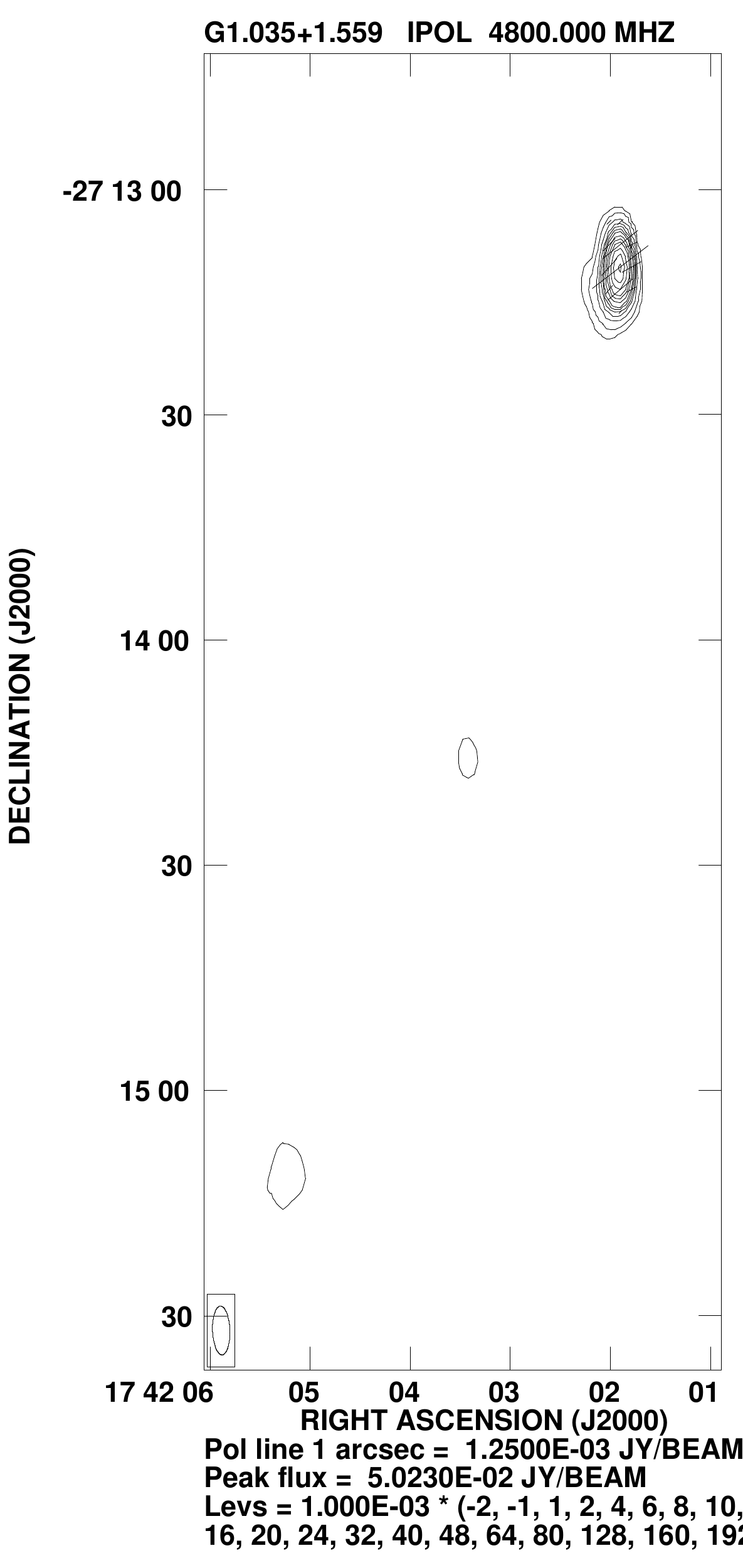}
}
\hbox{
\includegraphics[width=5cm, angle=0,clip=]{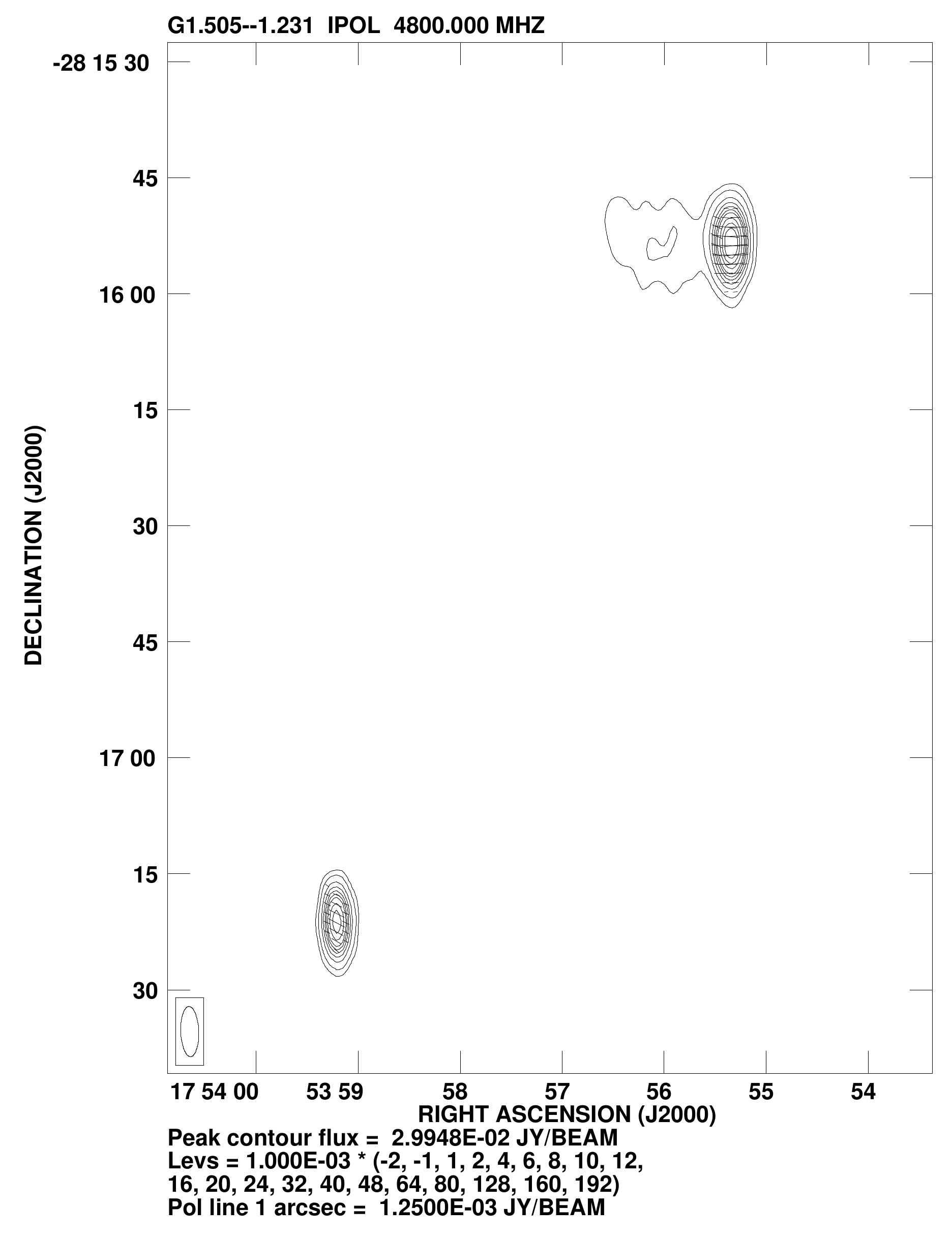}
\includegraphics[width=4.4cm, angle=0,clip=]{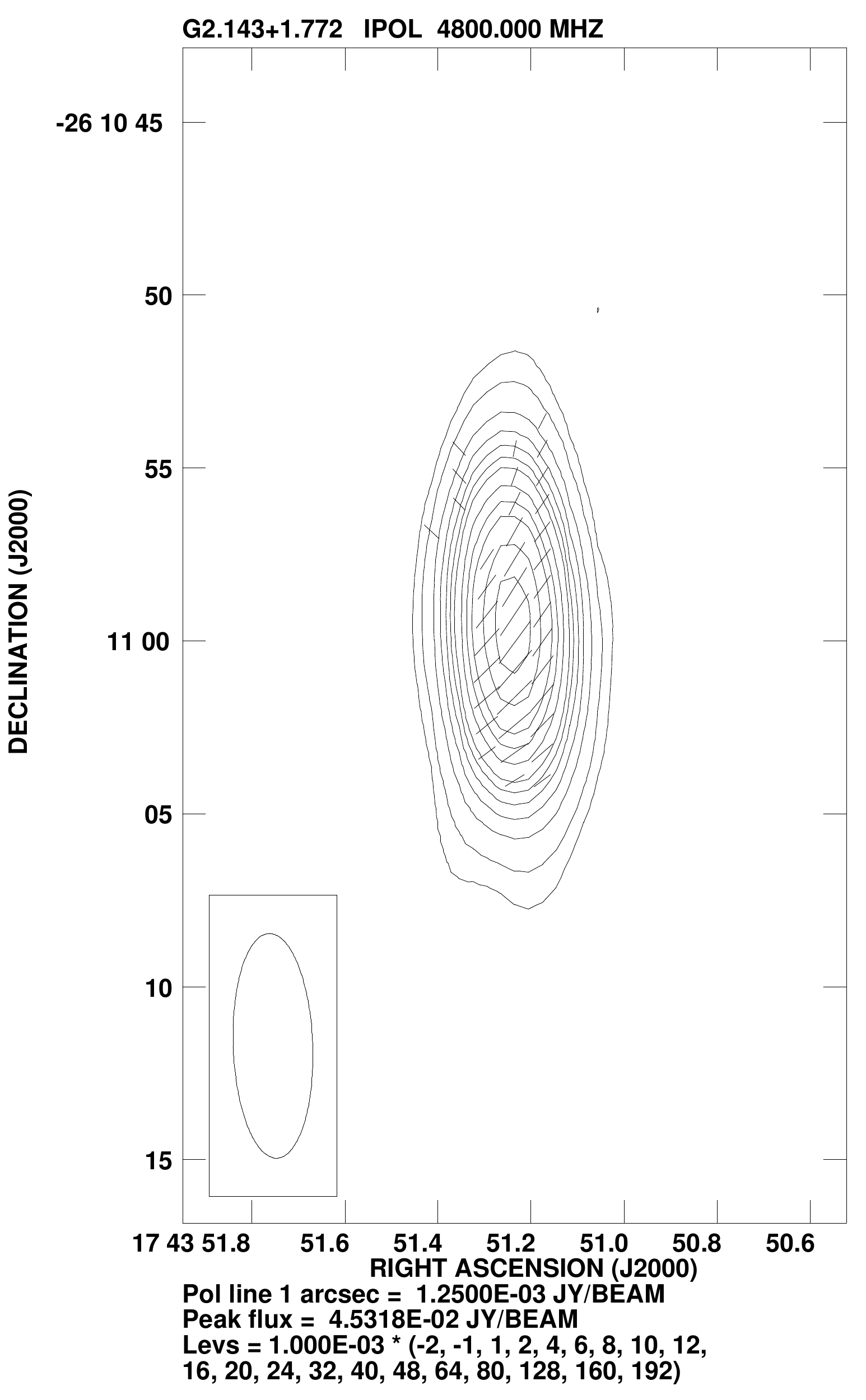}
\includegraphics[width=5.3cm, angle=0,clip=]{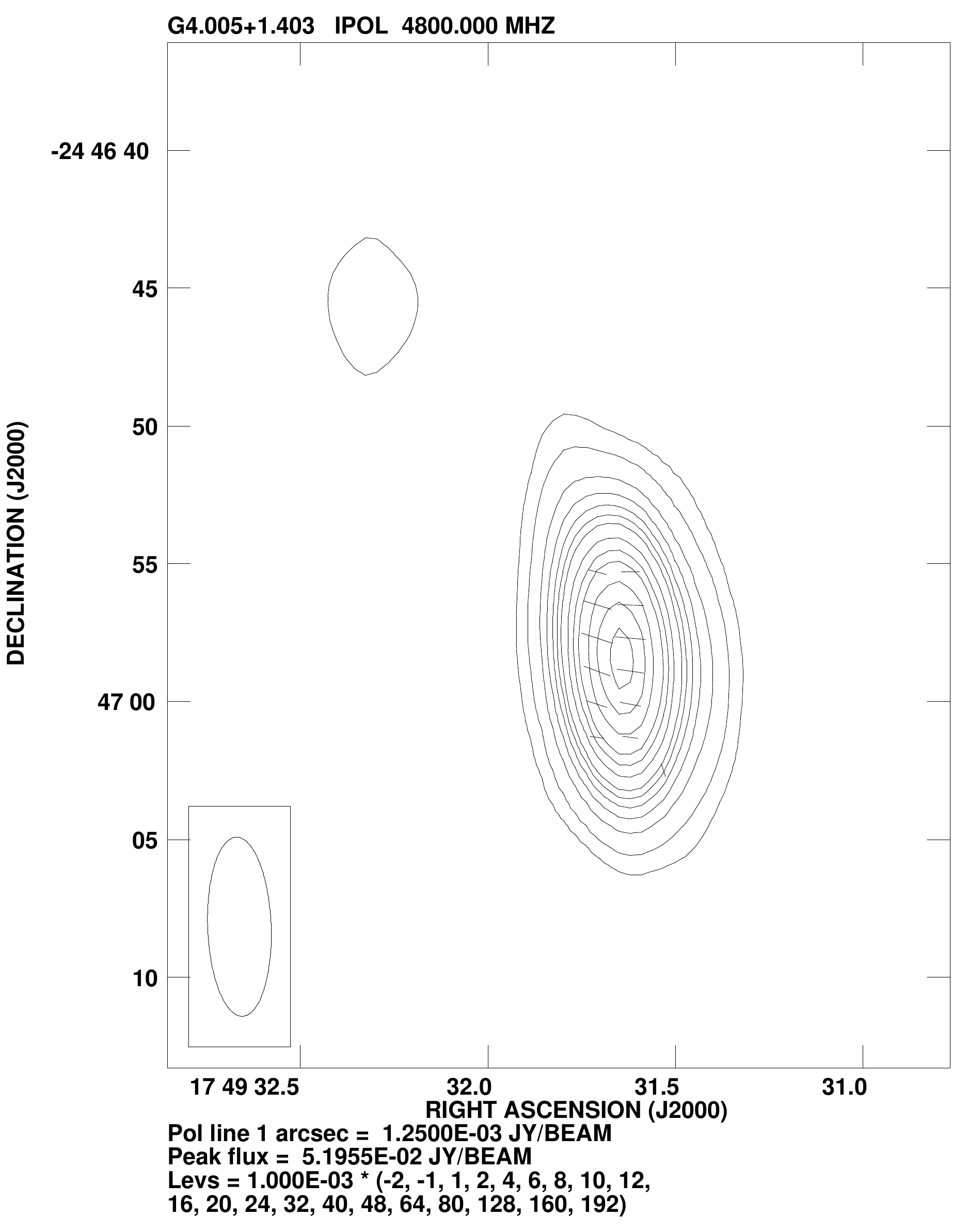}
}
}

\caption{Continued}
\end{figure*}
\vspace{-0.4cm} 

\addtocounter{figure}{-1}
\begin{figure*}
\centering
\vbox{
\hbox{
\includegraphics[width=4.7cm, angle=0,clip=]{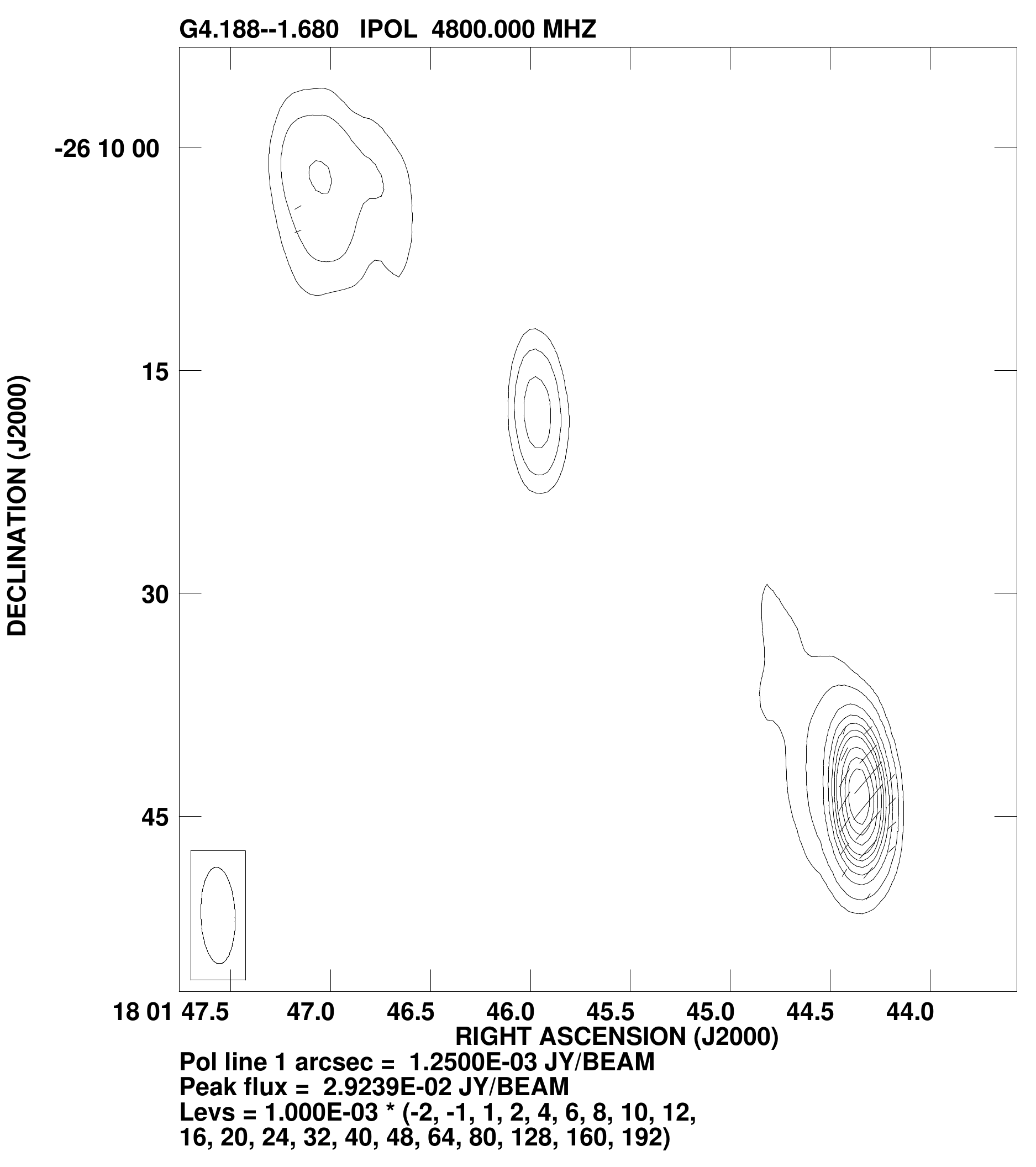}
\includegraphics[width=5.3cm, angle=0,clip=]{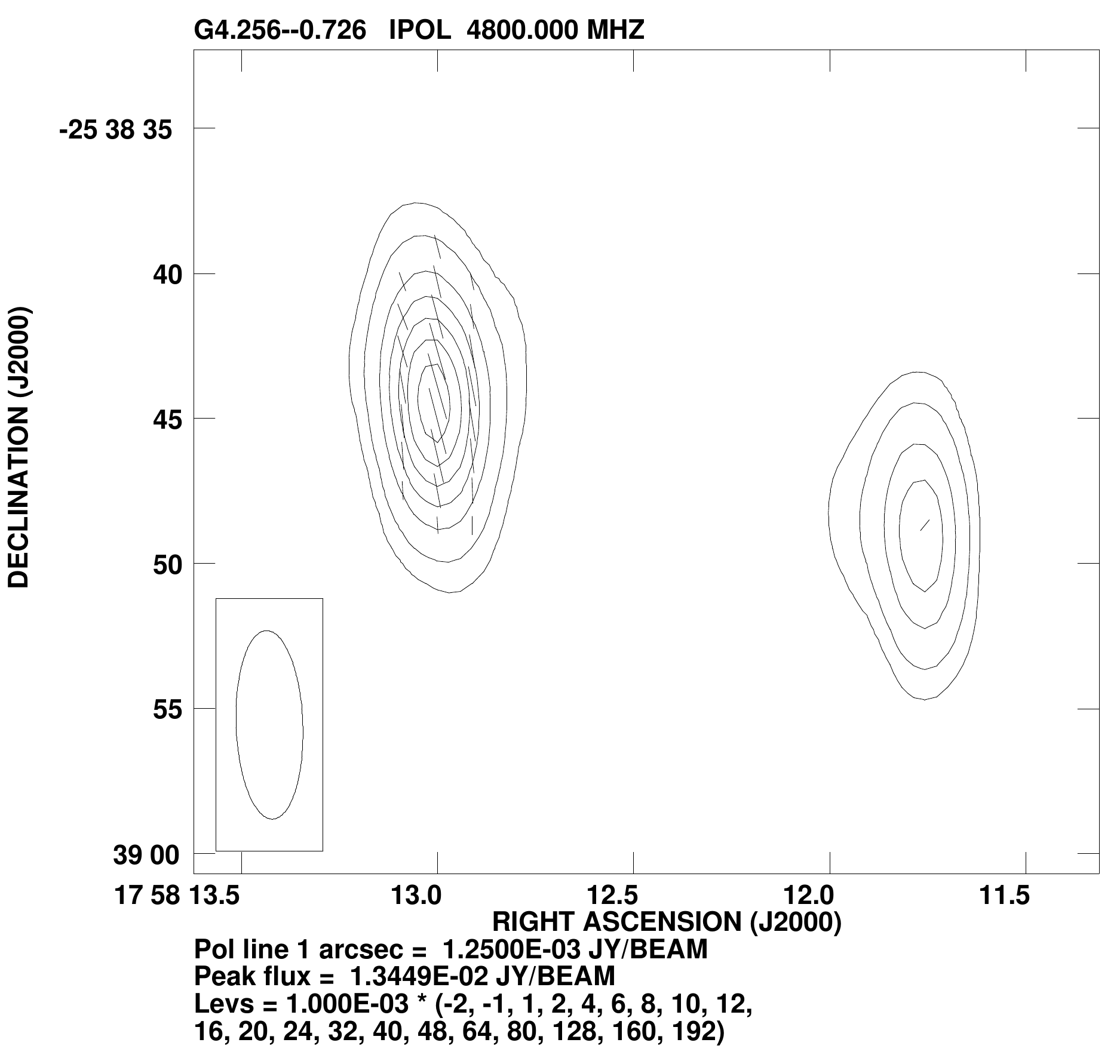}
\includegraphics[width=4.7cm, angle=0,clip=]{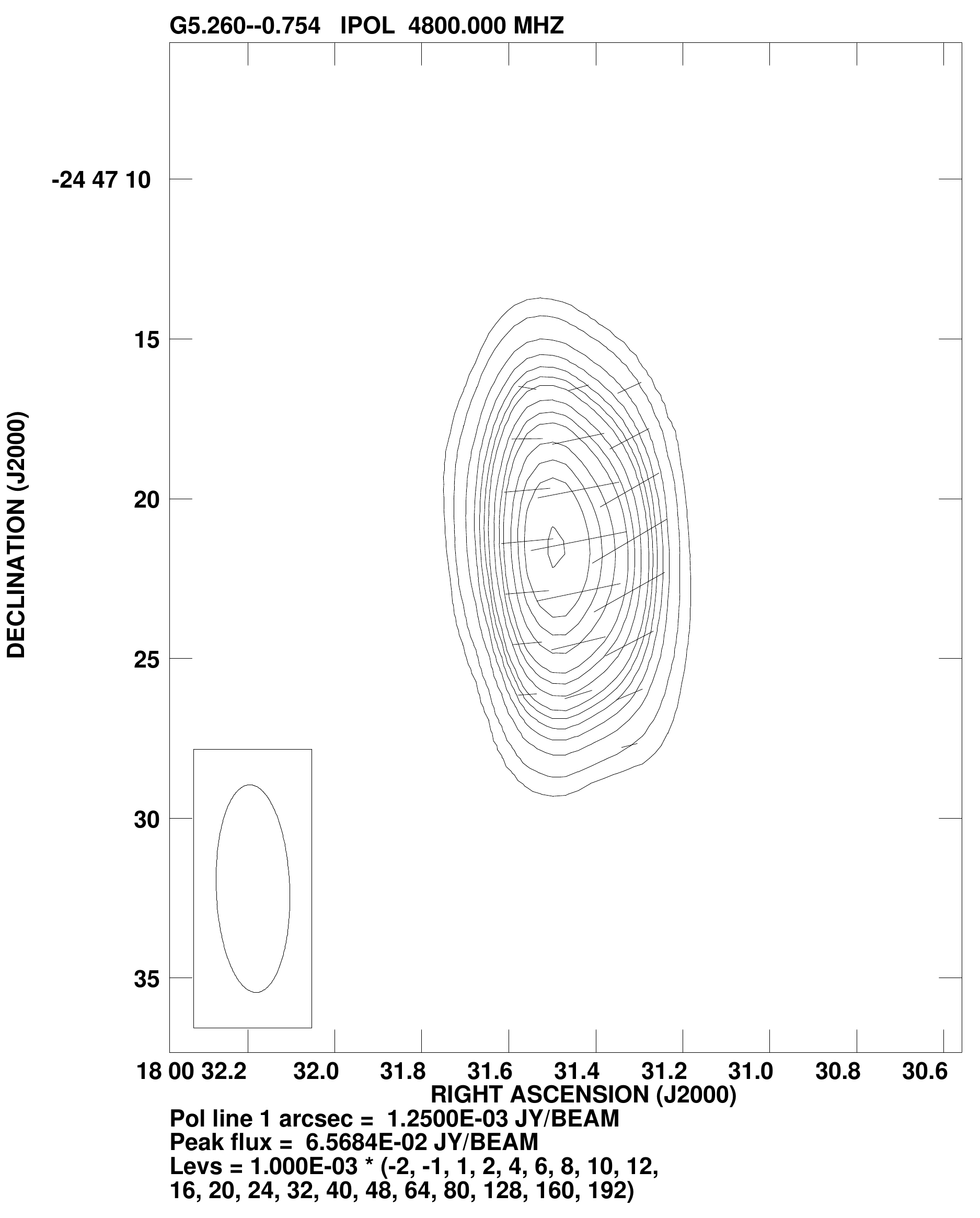}
}
\hbox{
\includegraphics[width=5.3cm, angle=0,clip=]{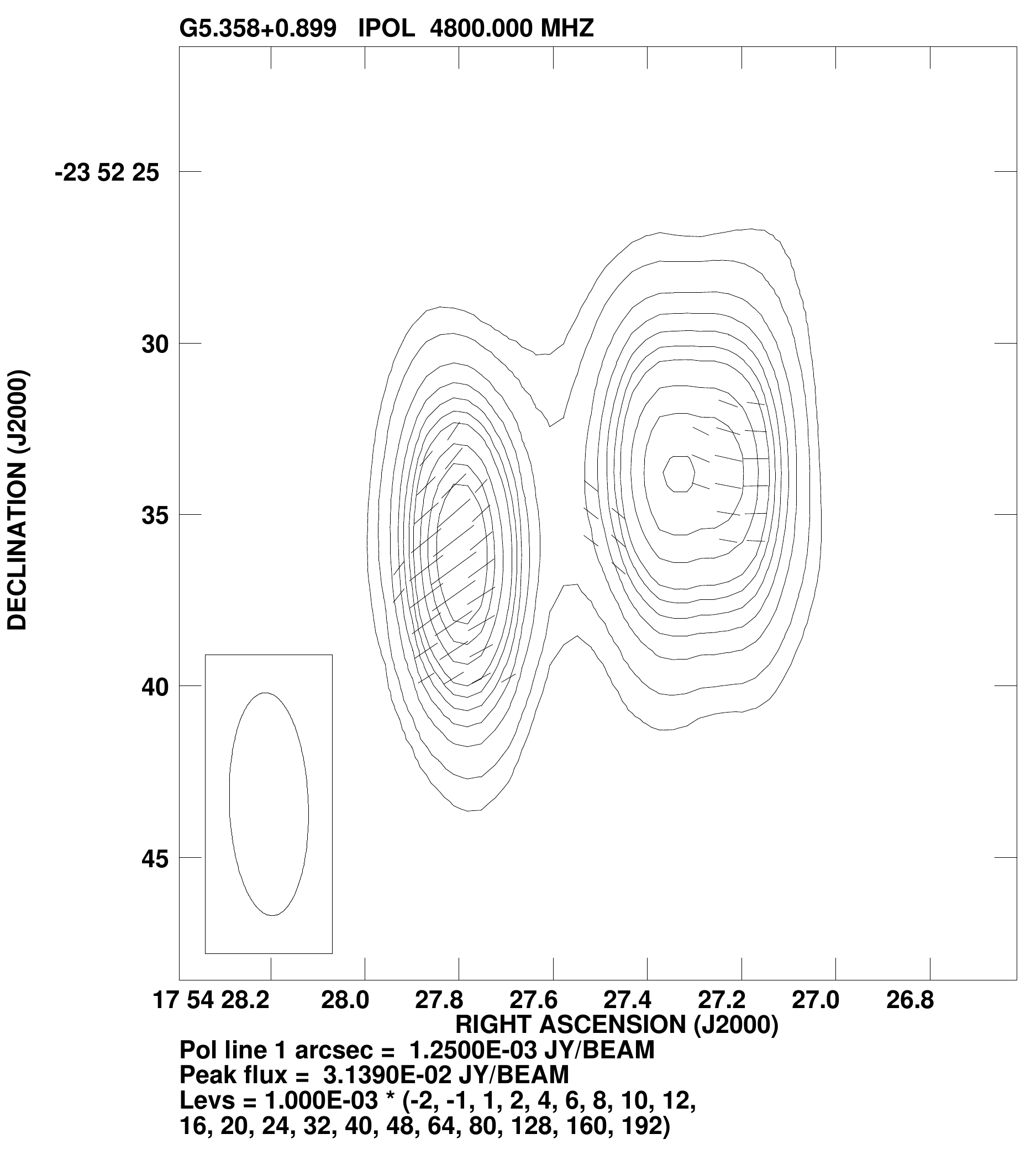}
\includegraphics[width=5cm, angle=0,clip=]{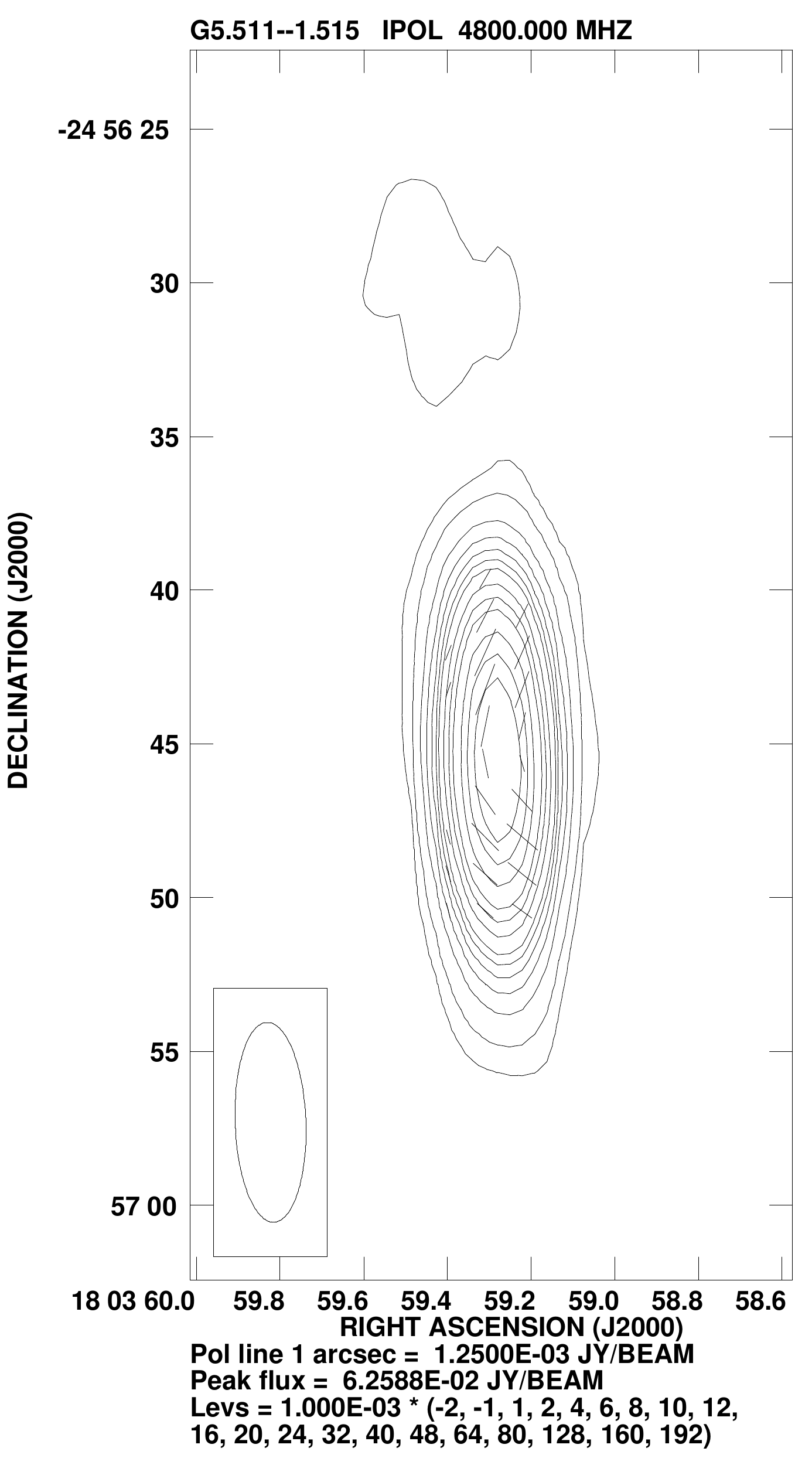}
\includegraphics[width=5cm, angle=0,clip=]{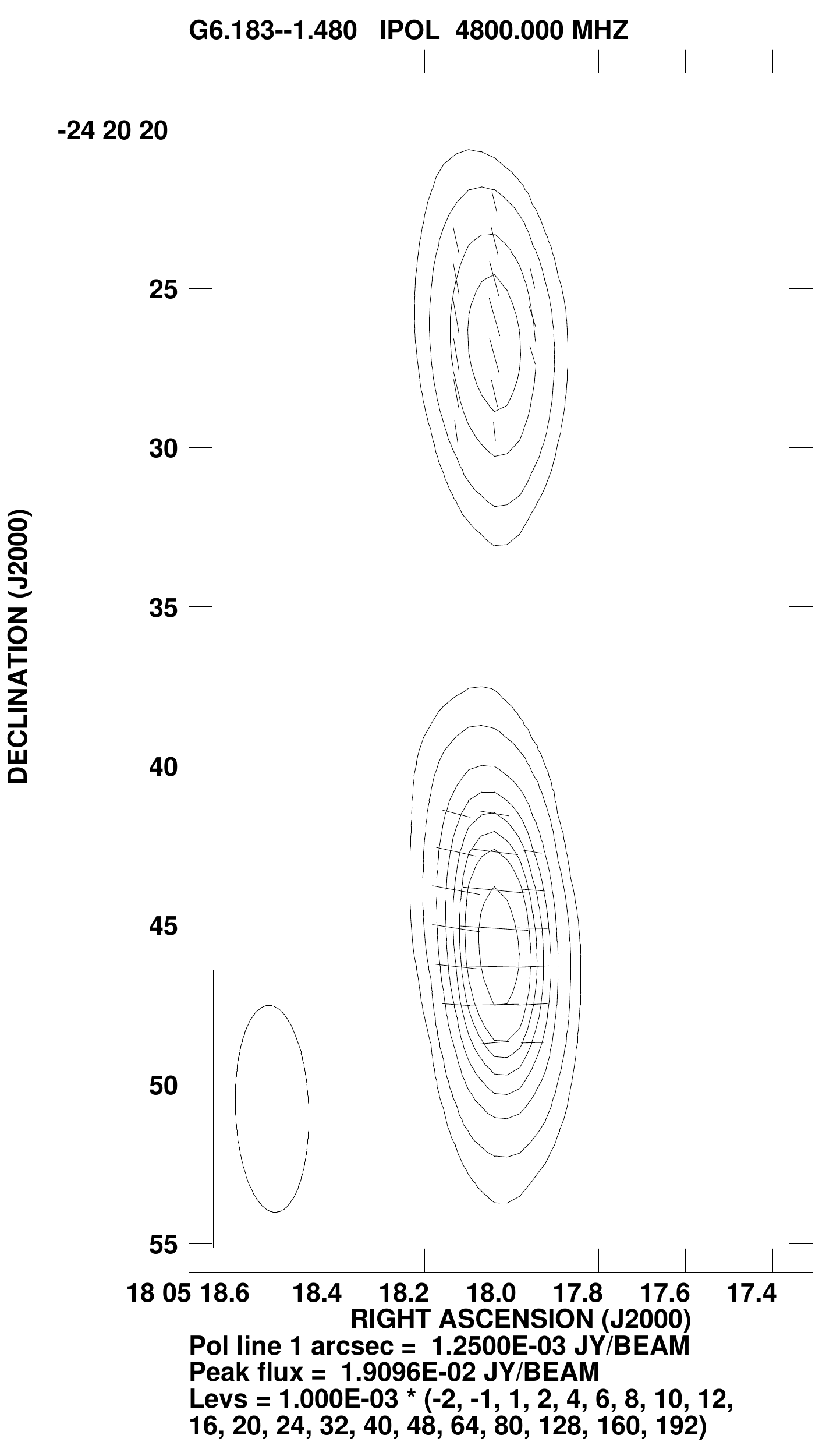}
}
}
\caption{Continued}
\end{figure*}
\vspace{-0.4cm} 

\begin{figure*}
\centering
\vbox{
\hbox{
\includegraphics[width=4.7cm, angle=0,clip=]{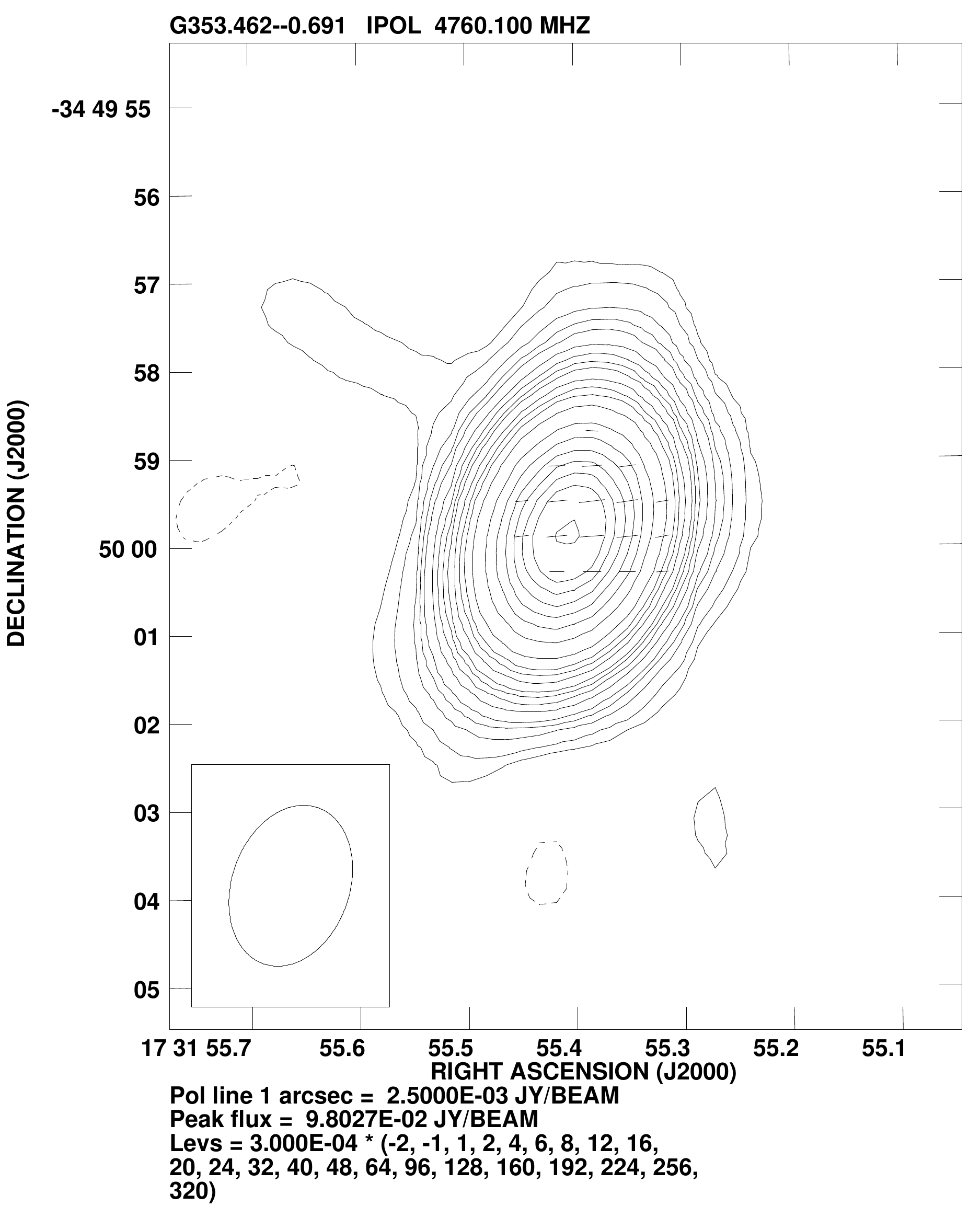}
\includegraphics[width=6.2cm, angle=0,clip=]{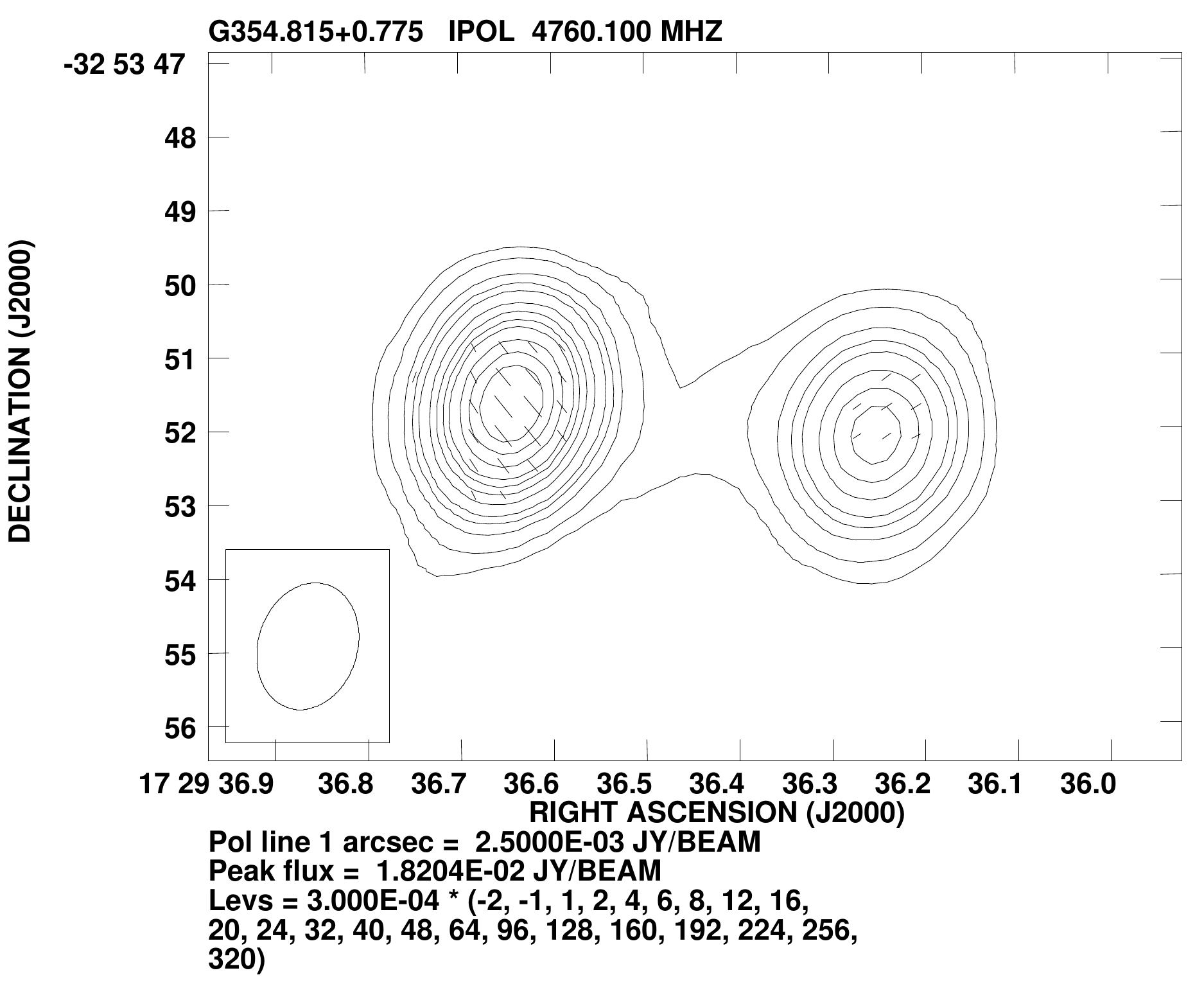}
\includegraphics[width=5.3cm, angle=0,clip=]{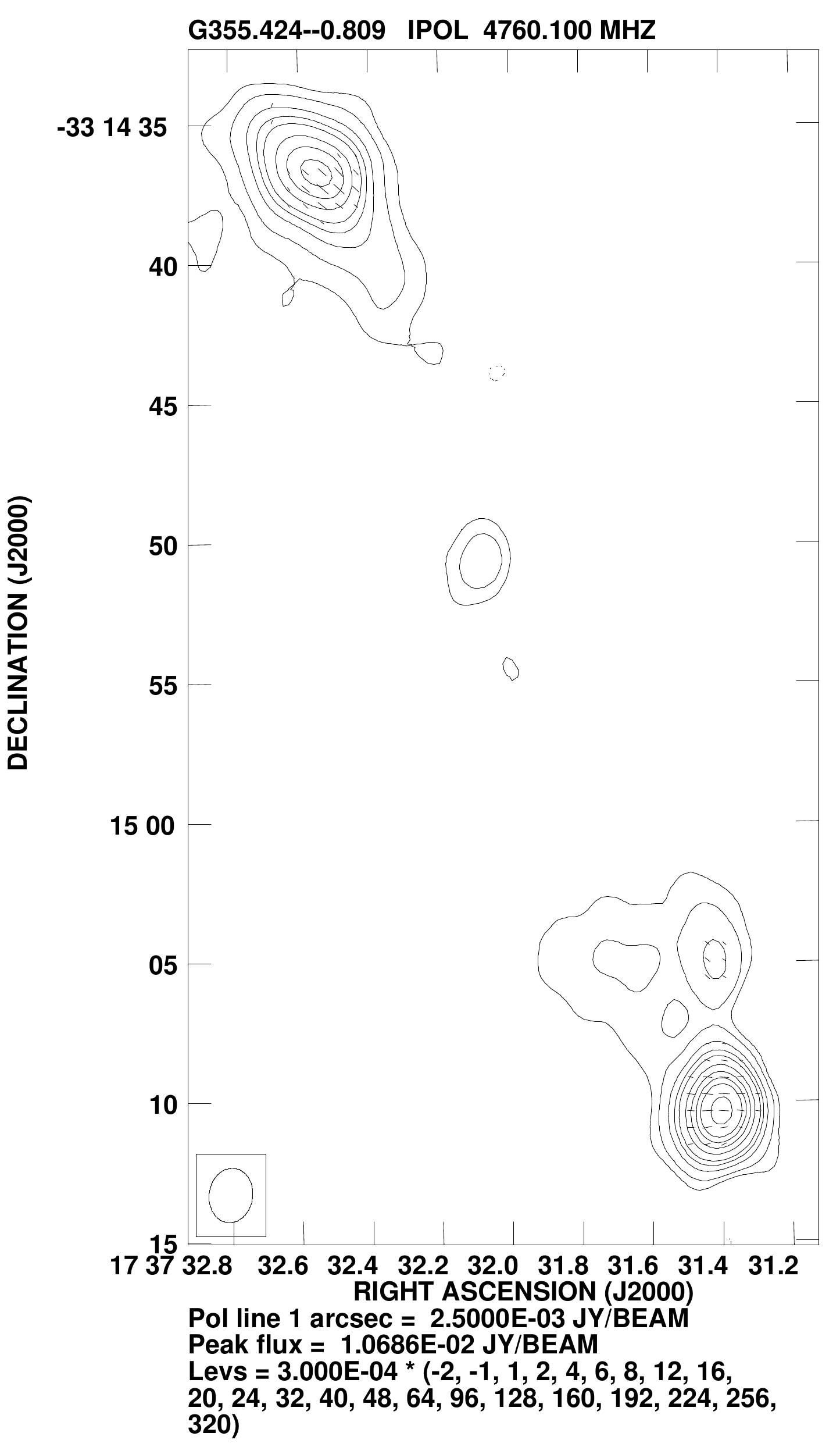}
}
\hbox{
\includegraphics[width=5.8cm, angle=0,clip=]{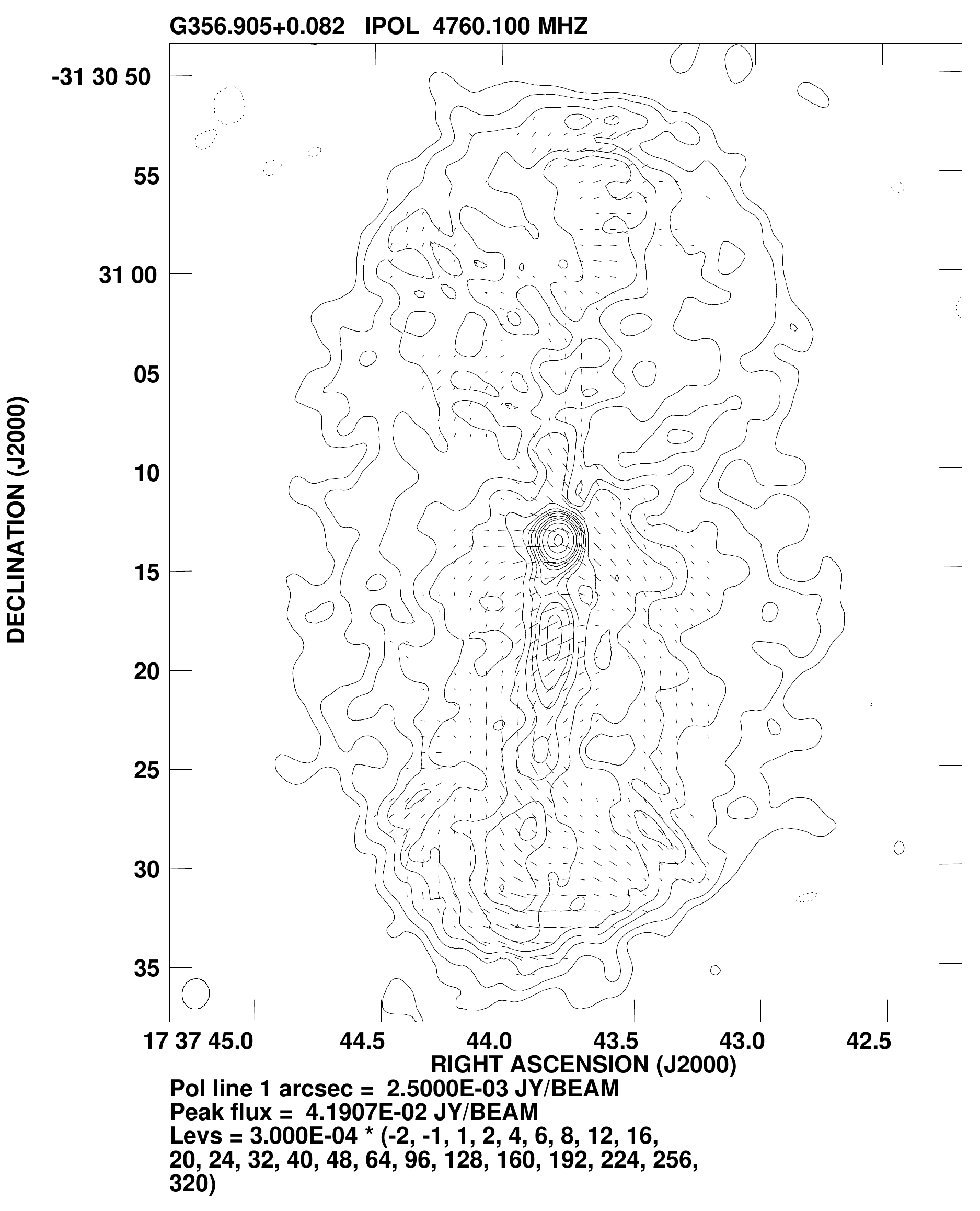}
\includegraphics[width=6.2cm, angle=0,clip=]{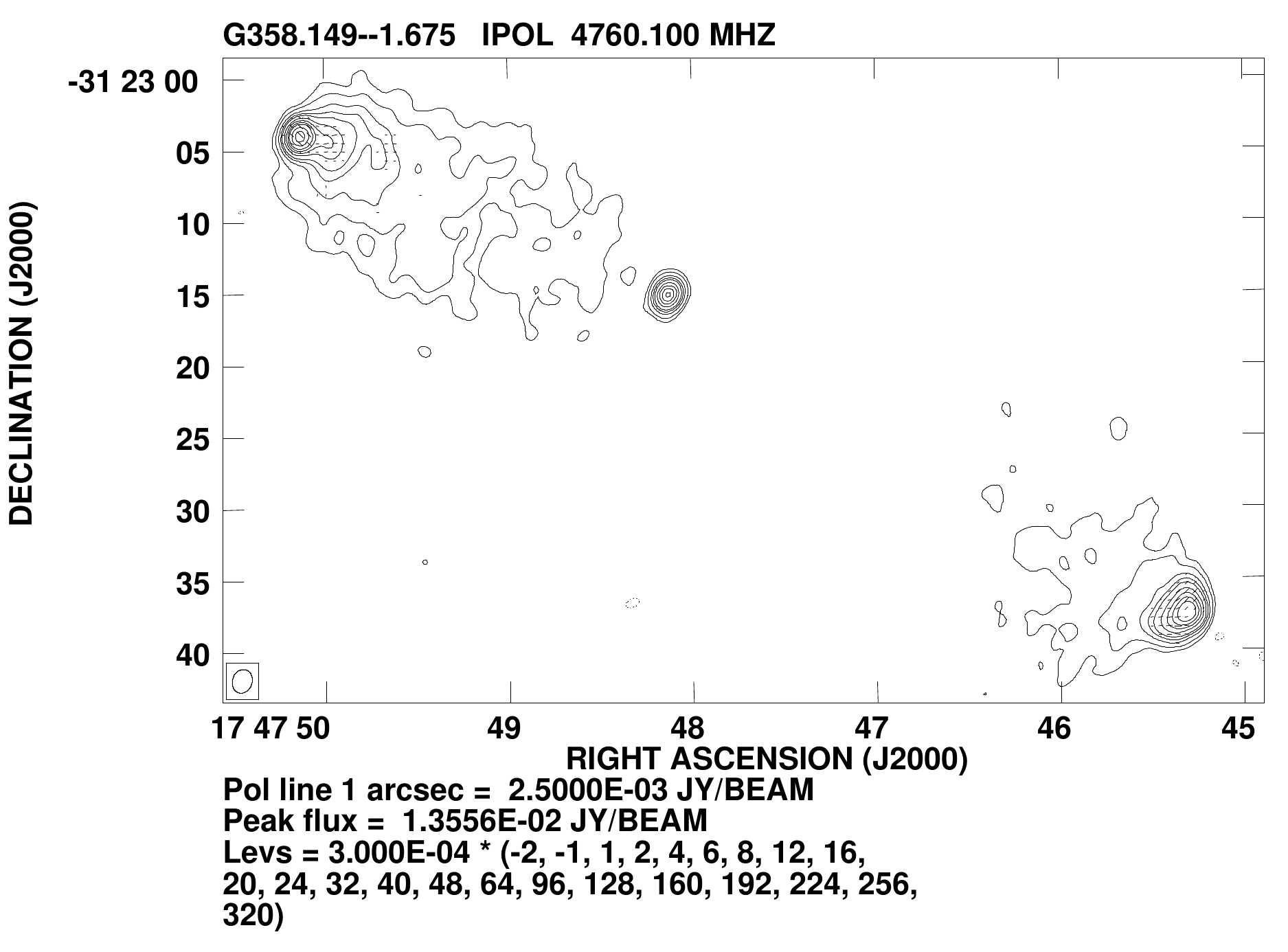}
\includegraphics[width=4.7cm, angle=0,clip=]{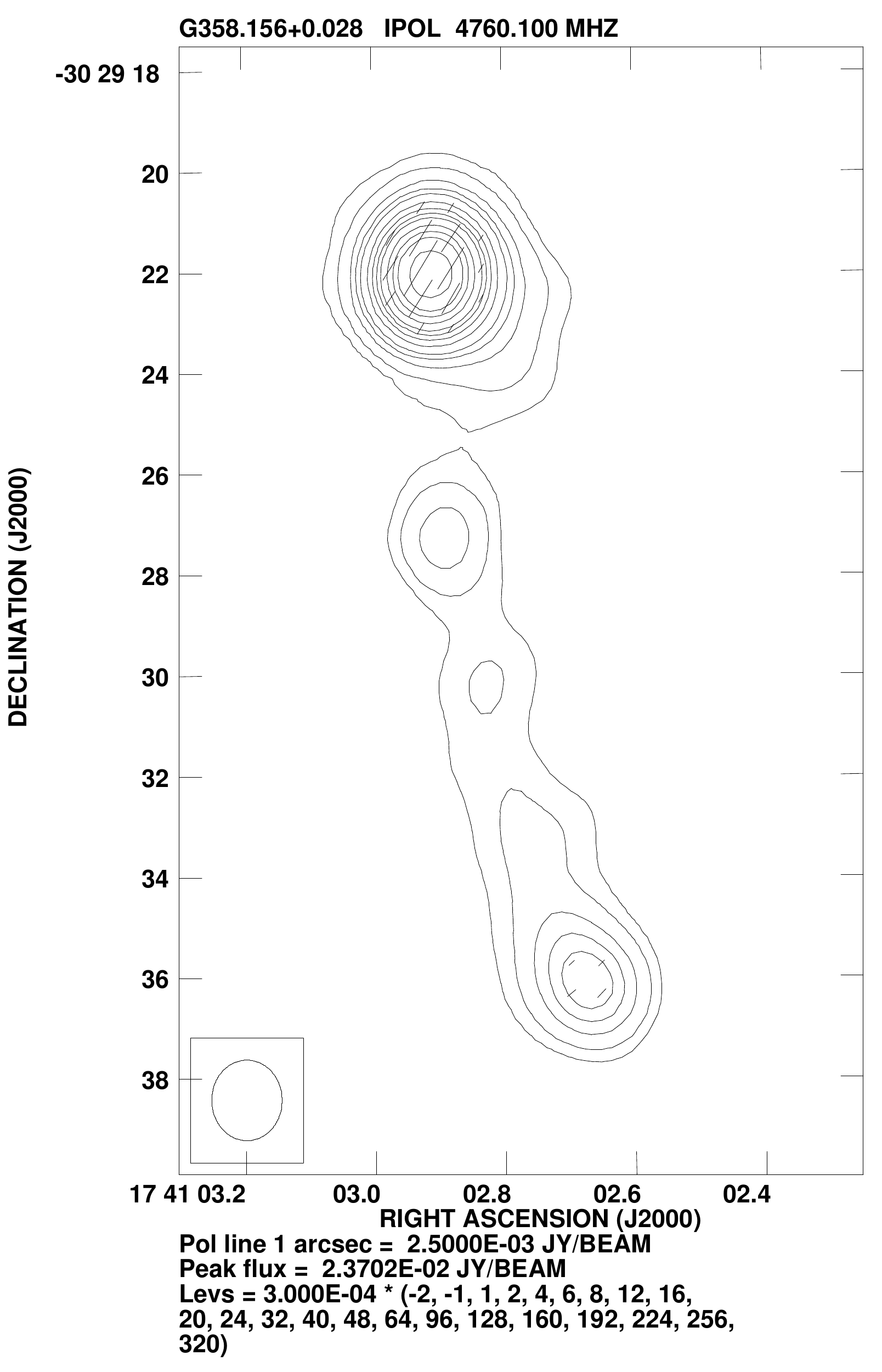}
}
\hbox{
\includegraphics[width=5.5cm, angle=0,clip=]{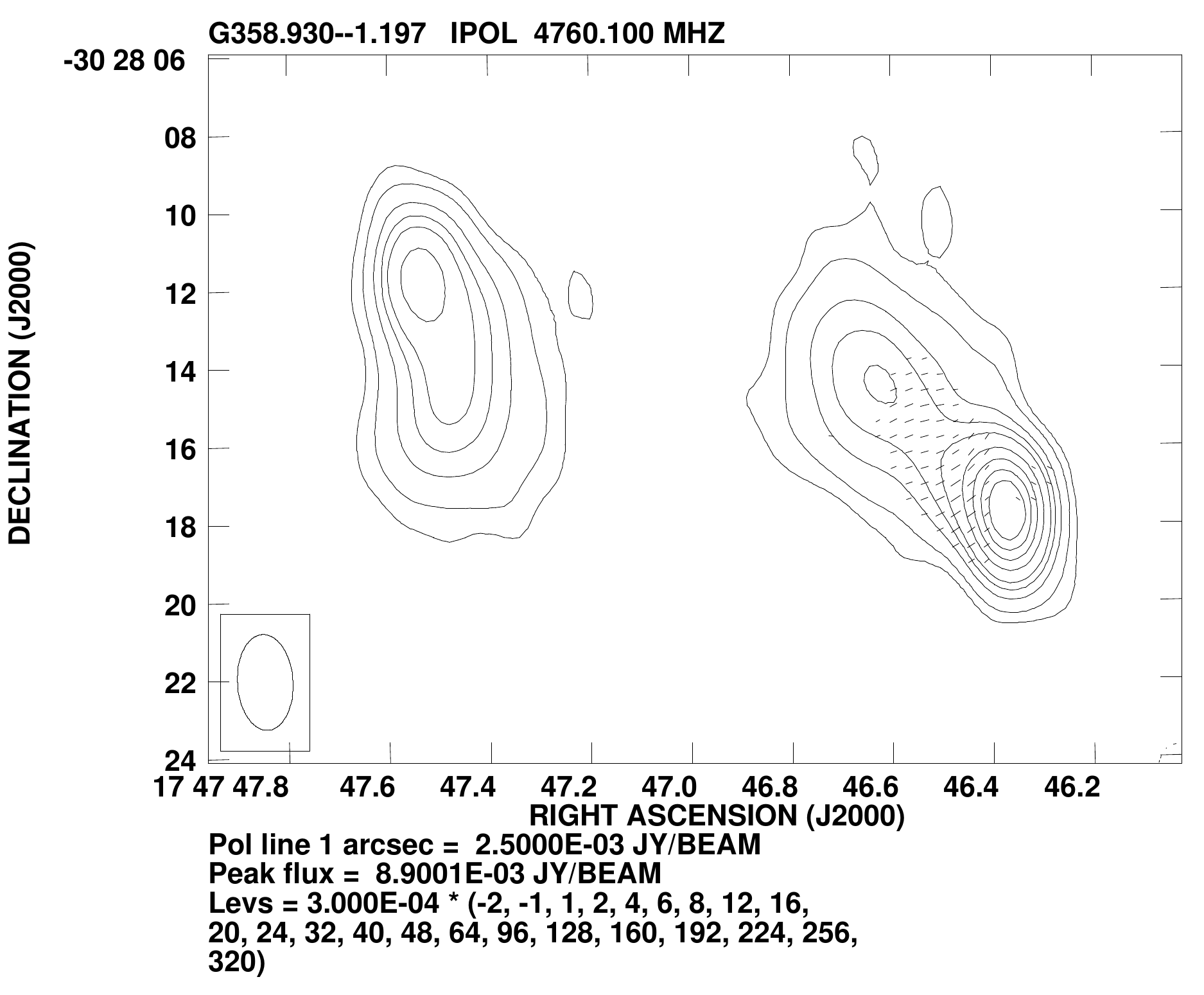}
\includegraphics[width=5.5cm, angle=0,clip=]{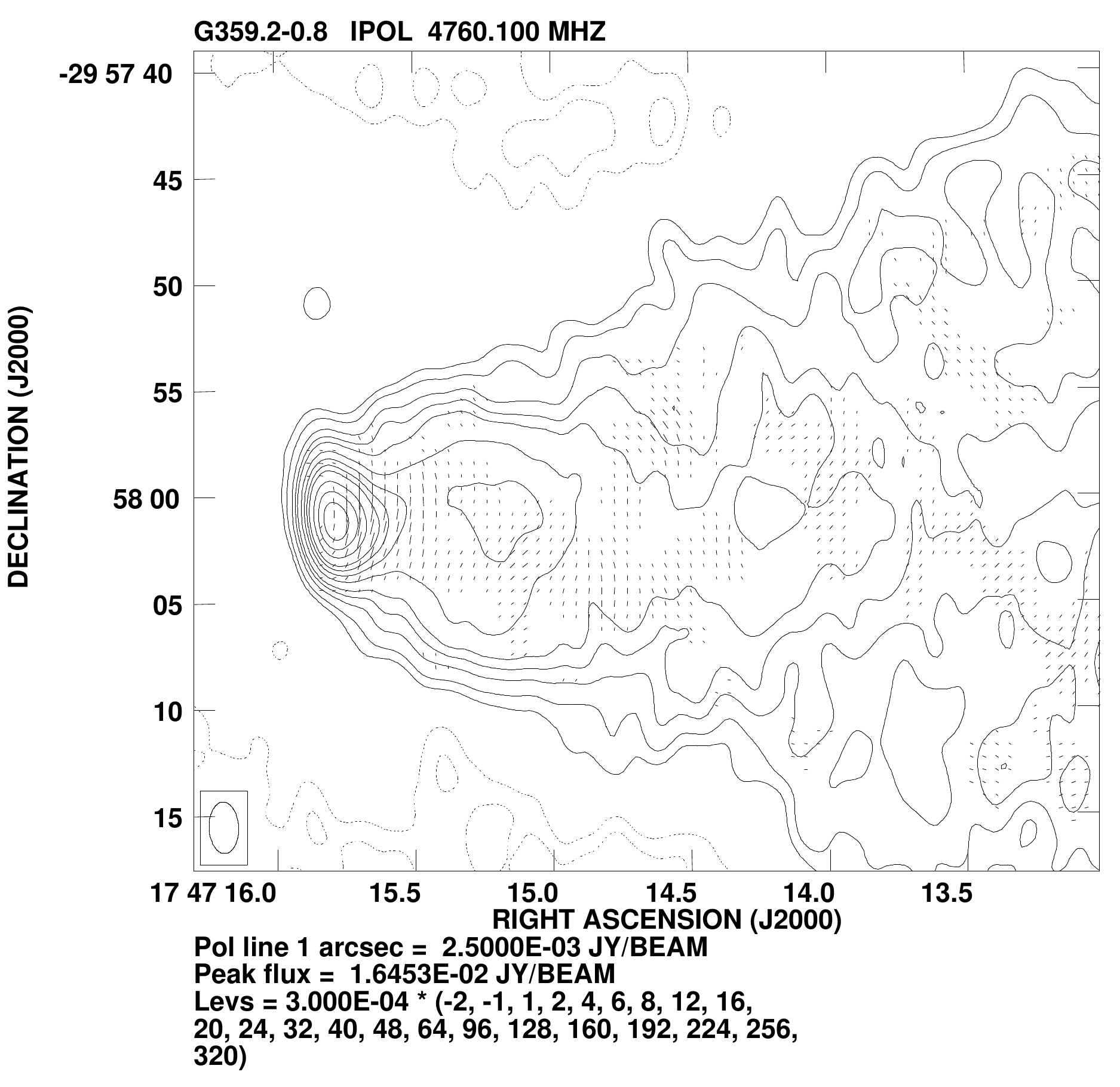}
\includegraphics[width=5.5cm, angle=0,clip=]{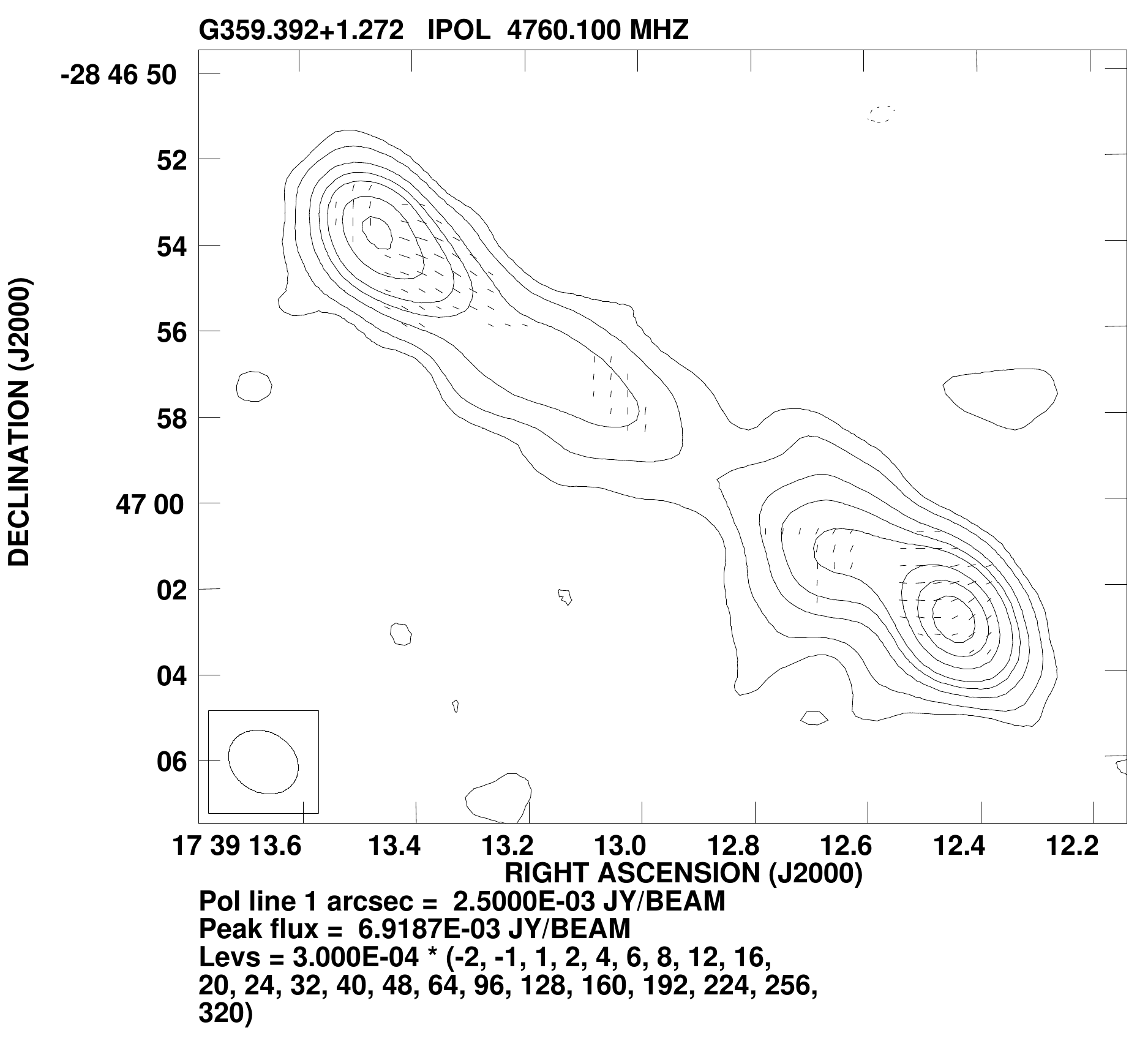}
}
}
\caption{4.8 GHz VLA continuum maps of the polarised sources with the
polarisation vectors superposed on them. Typical rms noise in Stokes I is about
90 $\mu$Jy/beam and about 75 $\mu$Jy/beam in Stokes Q and U. Typical beamsize
of these images are $\approx$2$^{''} \times 1.5^{''}$.}
\label{5ghz.vla}
\end{figure*}
\vspace{-0.4cm}

\addtocounter{figure}{-1}
\begin{figure*}
\centering
\vbox{
\hbox{
\includegraphics[width=5.5cm, angle=0,clip=]{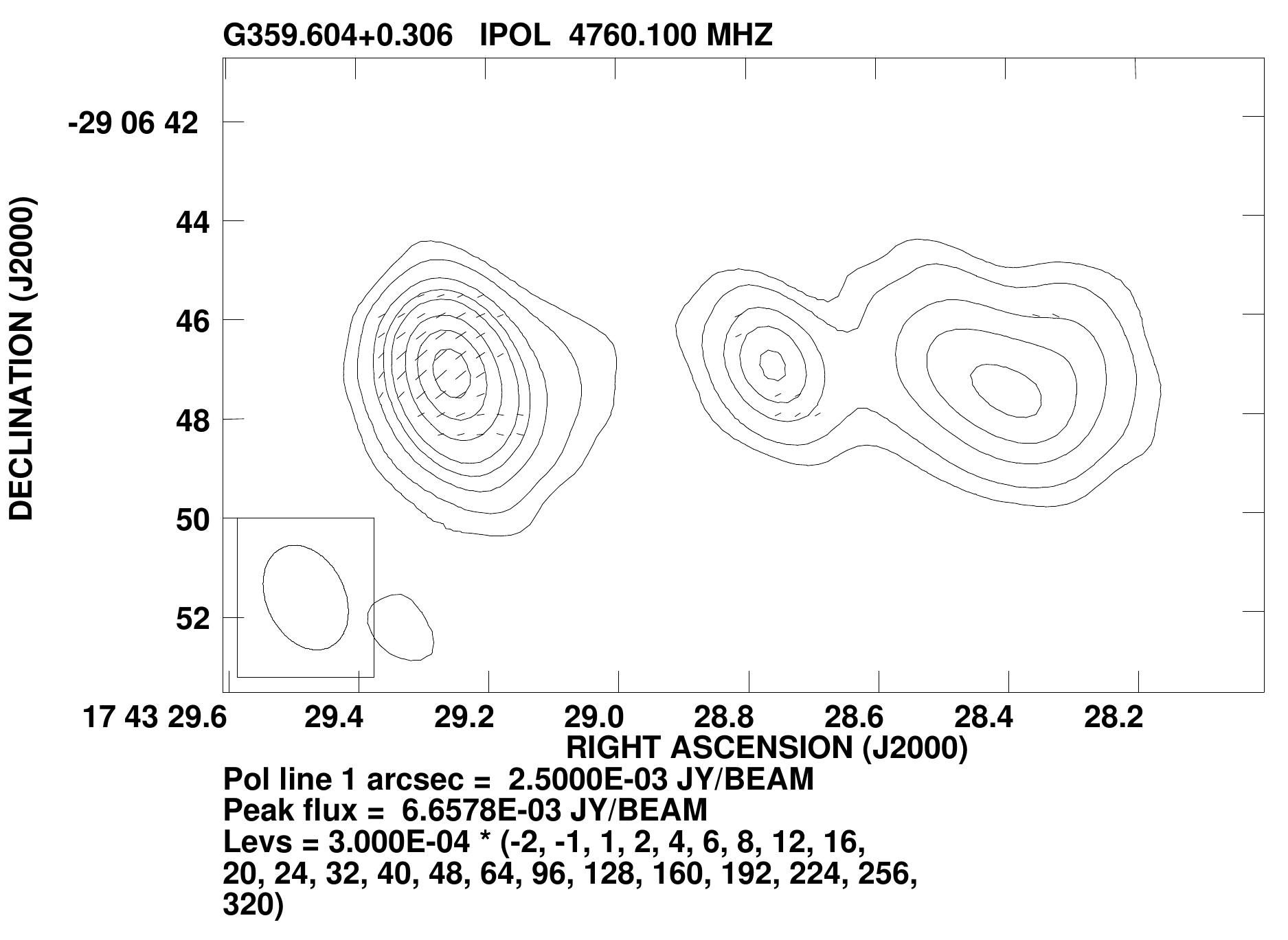}
\includegraphics[width=5.5cm, angle=0,clip=]{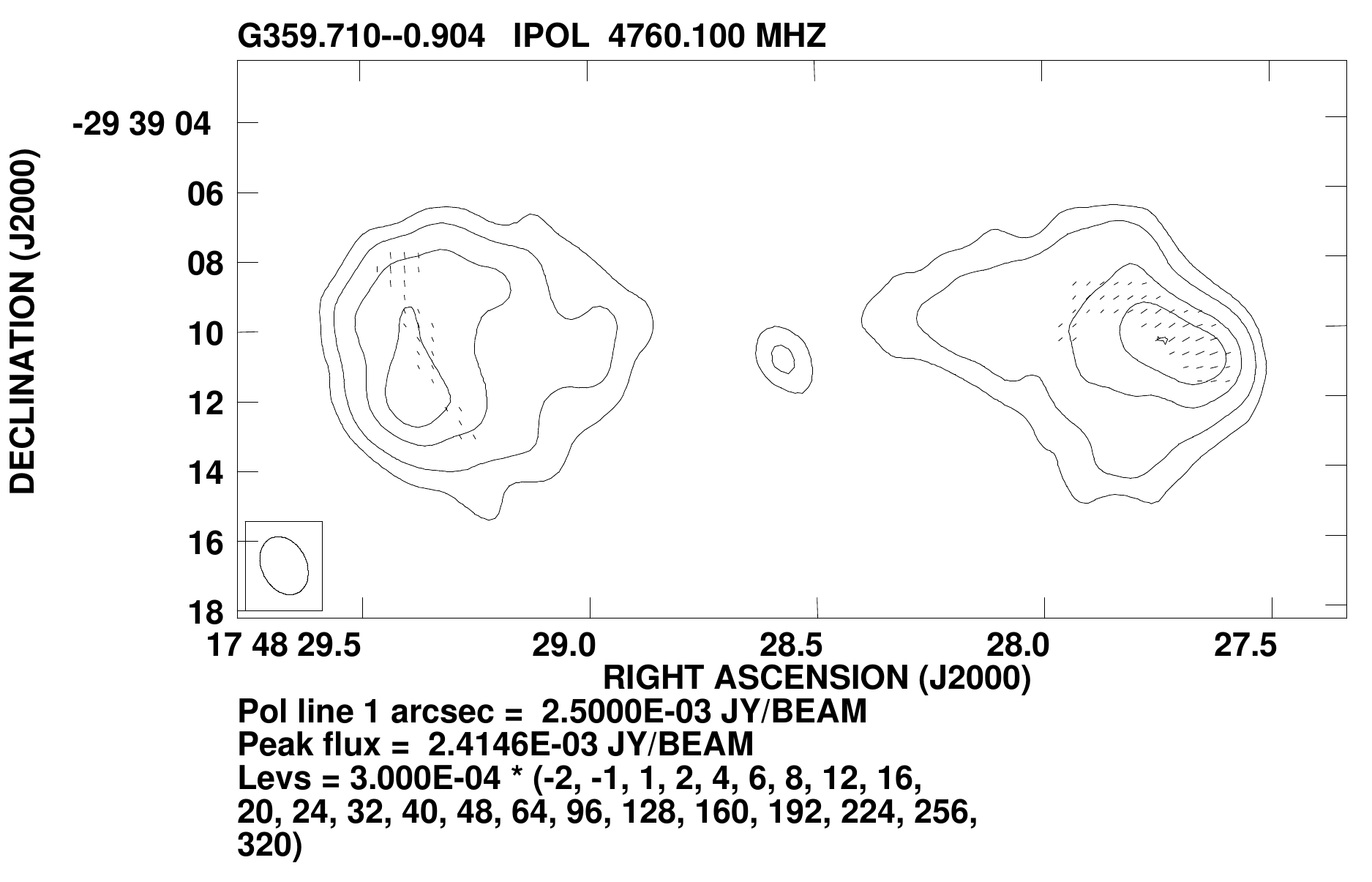}
\includegraphics[width=5.5cm, angle=0,clip=]{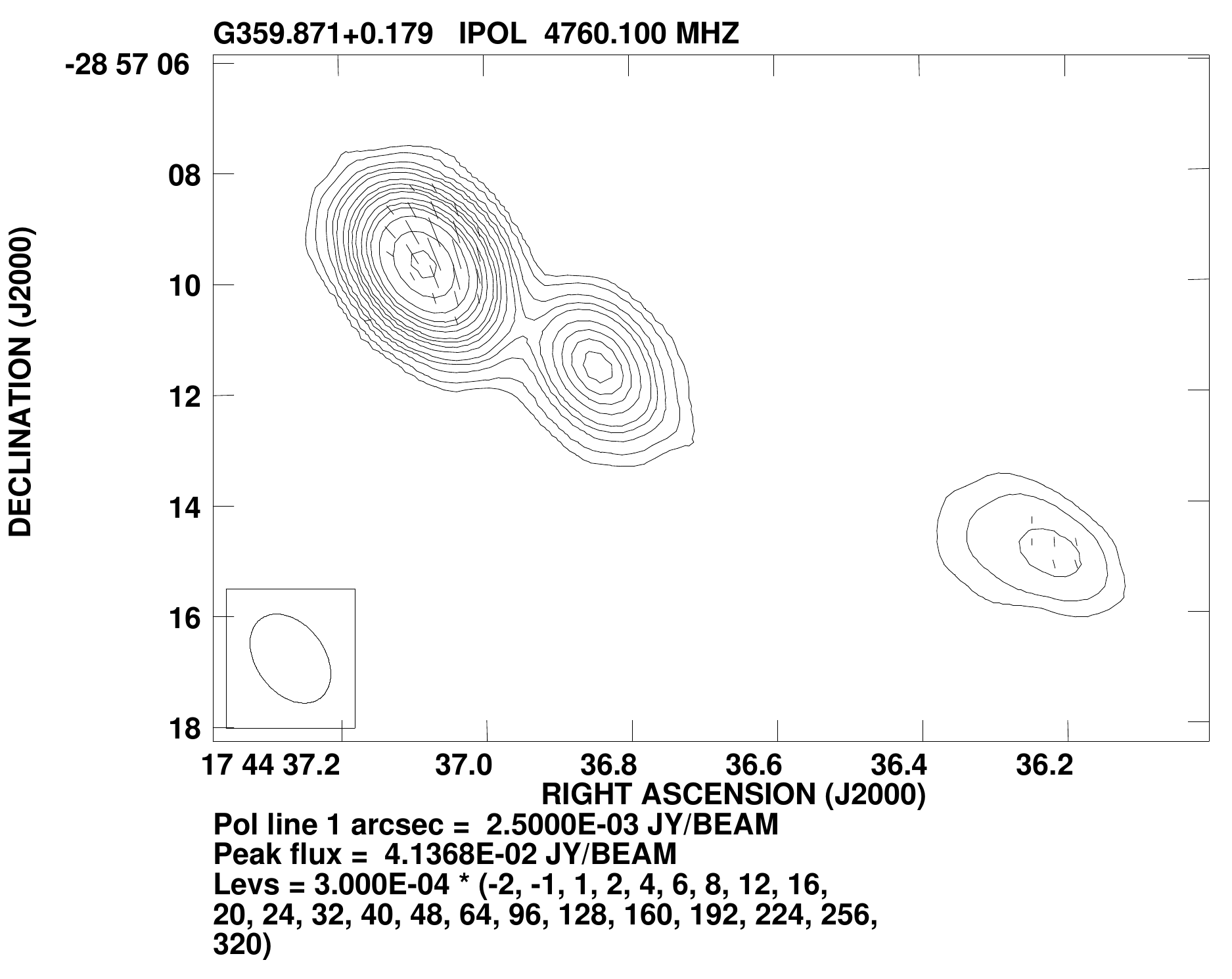}
}
\hbox{
\includegraphics[width=5.8cm, angle=0,clip=]{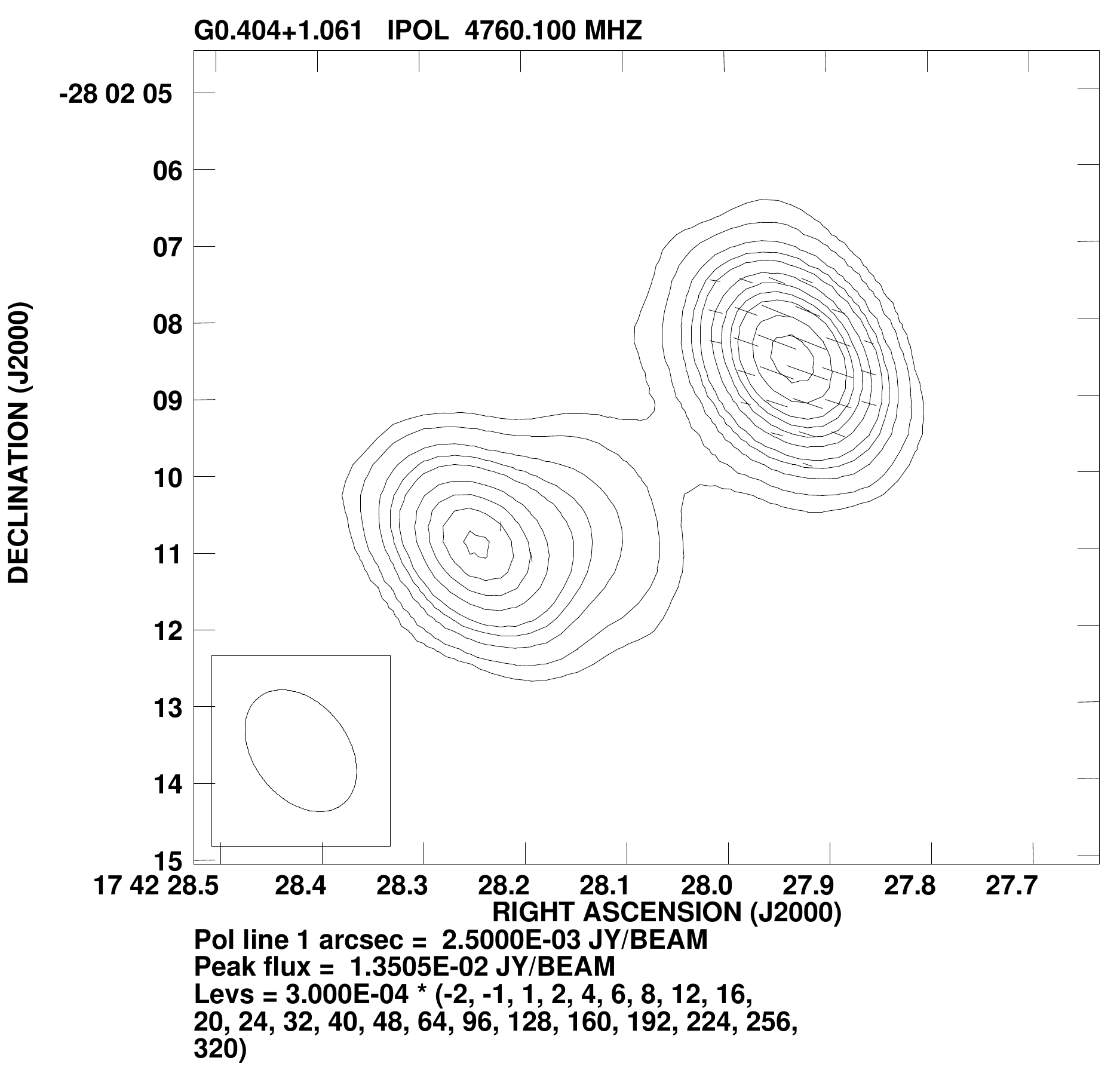}
\includegraphics[width=5.2cm, angle=0,clip=]{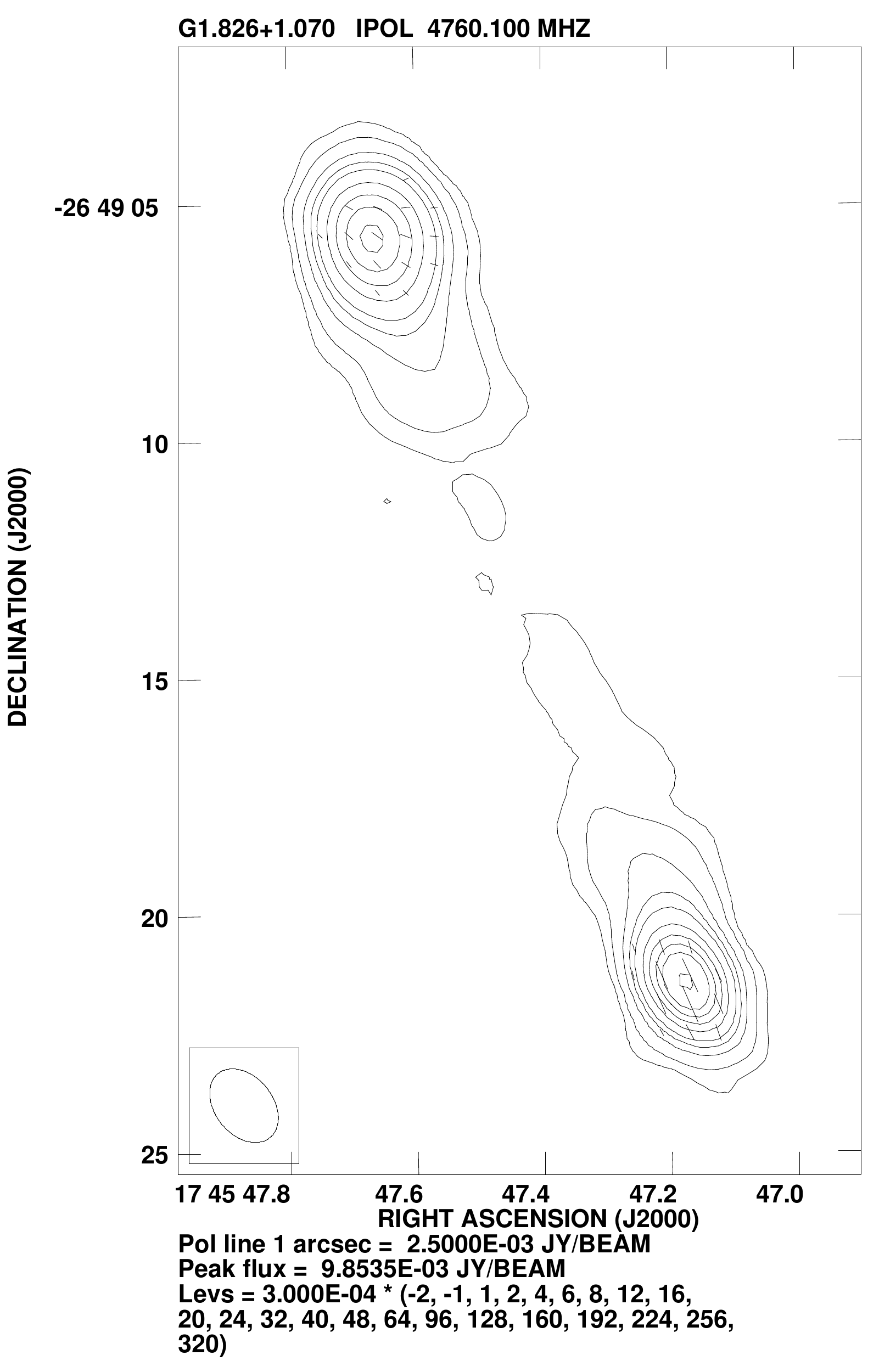}
\includegraphics[width=5.5cm, angle=0,clip=]{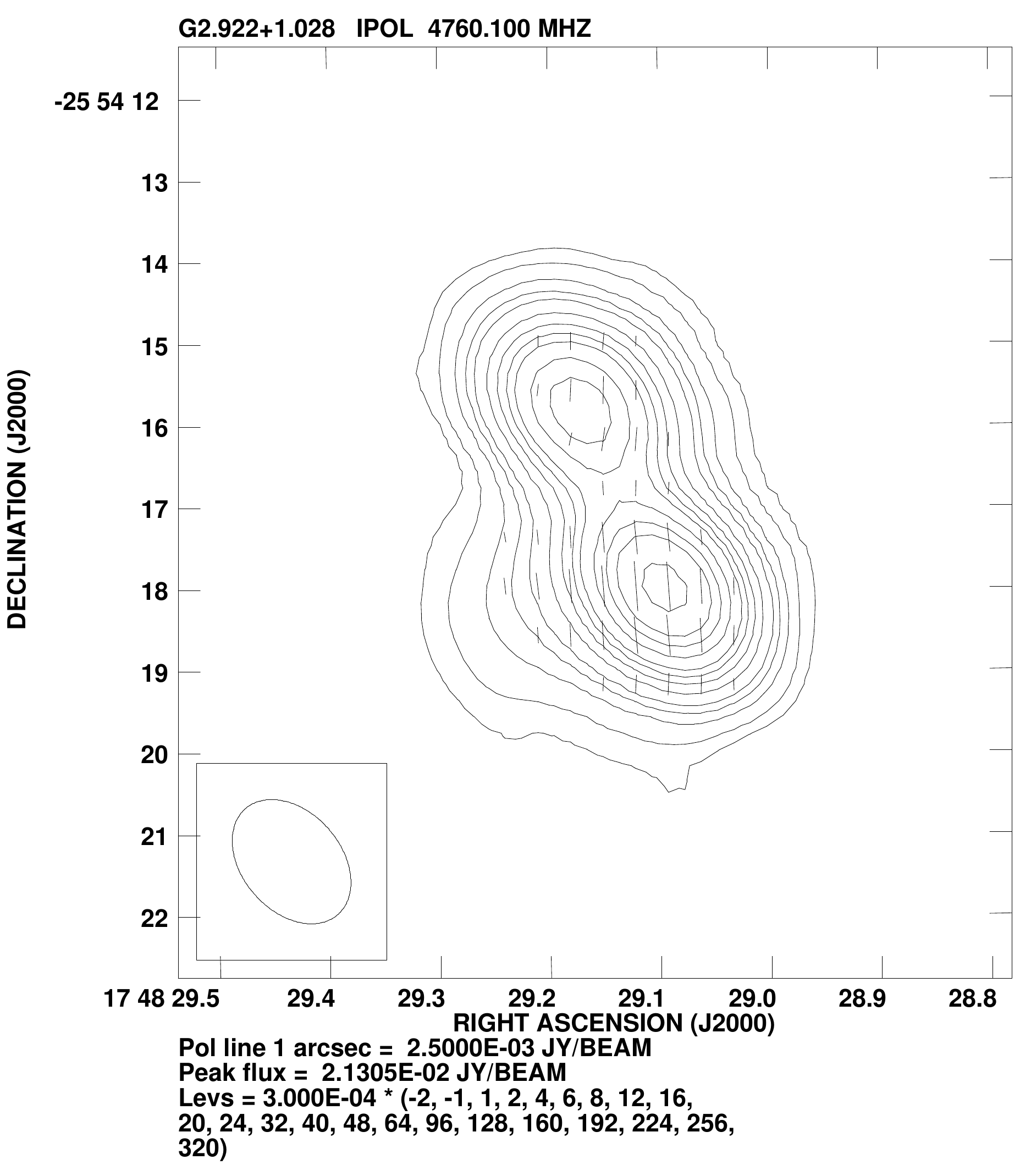}
}
\hbox{
\includegraphics[width=5.5cm, angle=0,clip=]{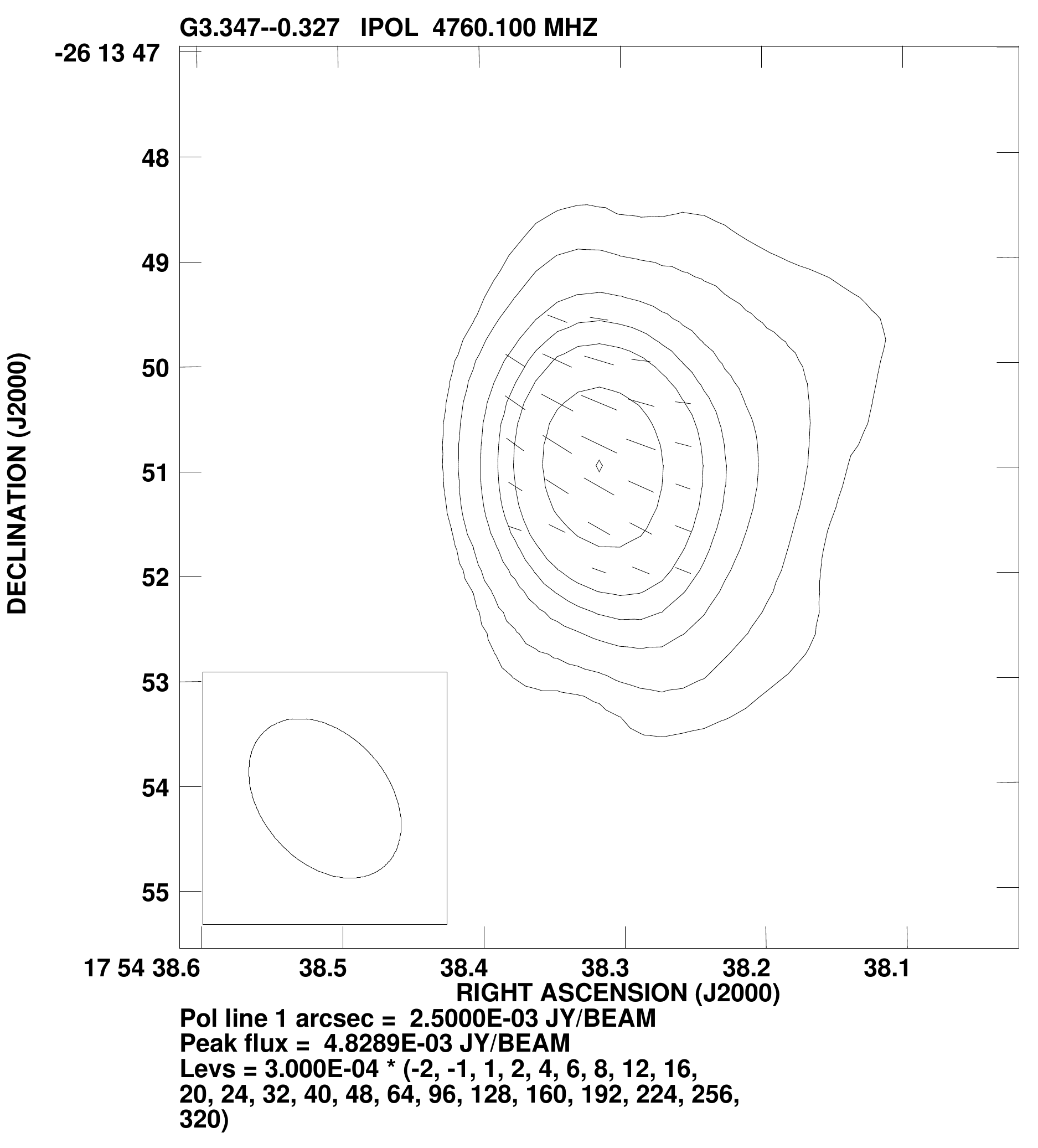}
\includegraphics[width=5.8cm, angle=0,clip=]{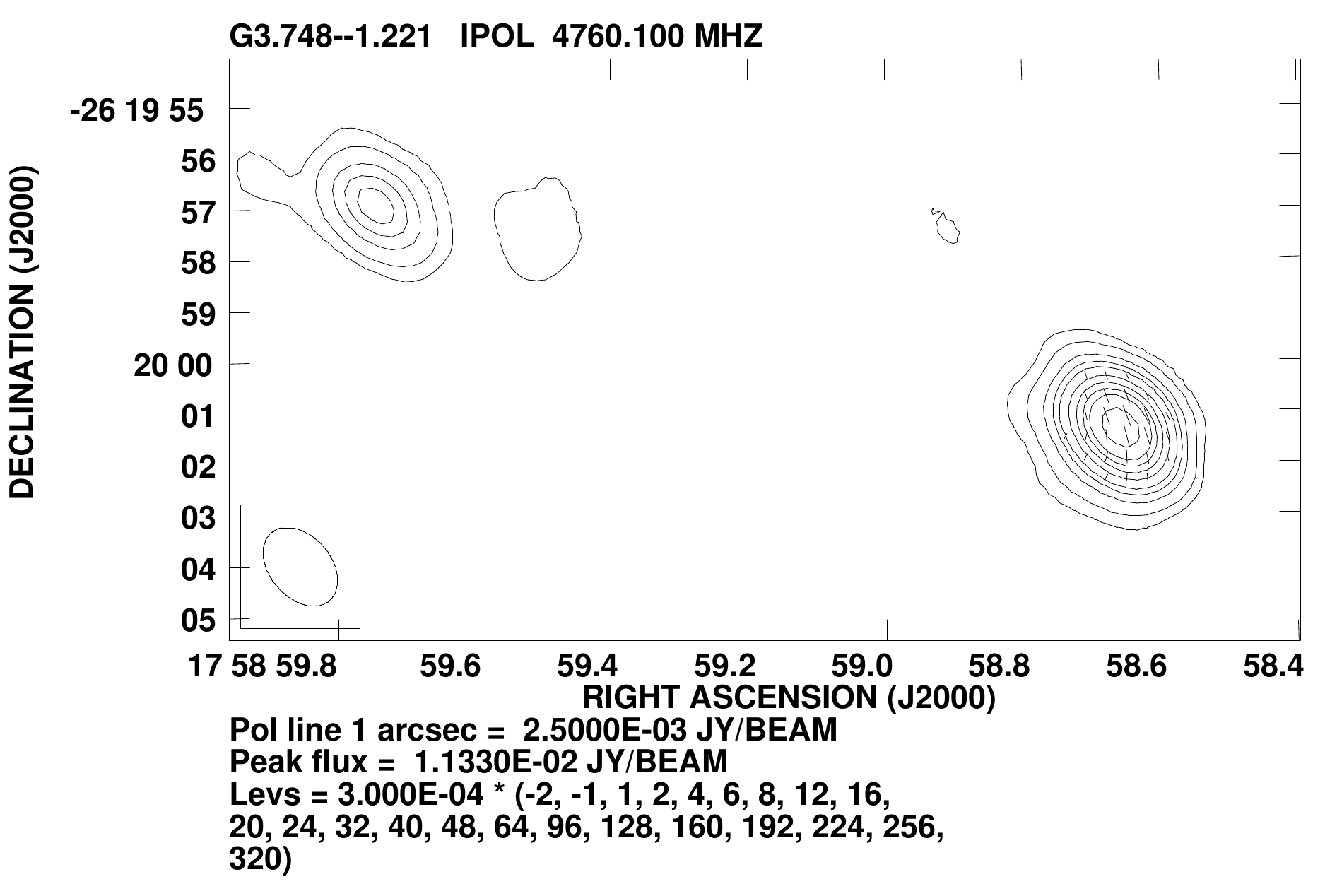}
\includegraphics[width=5.2cm, angle=0,clip=]{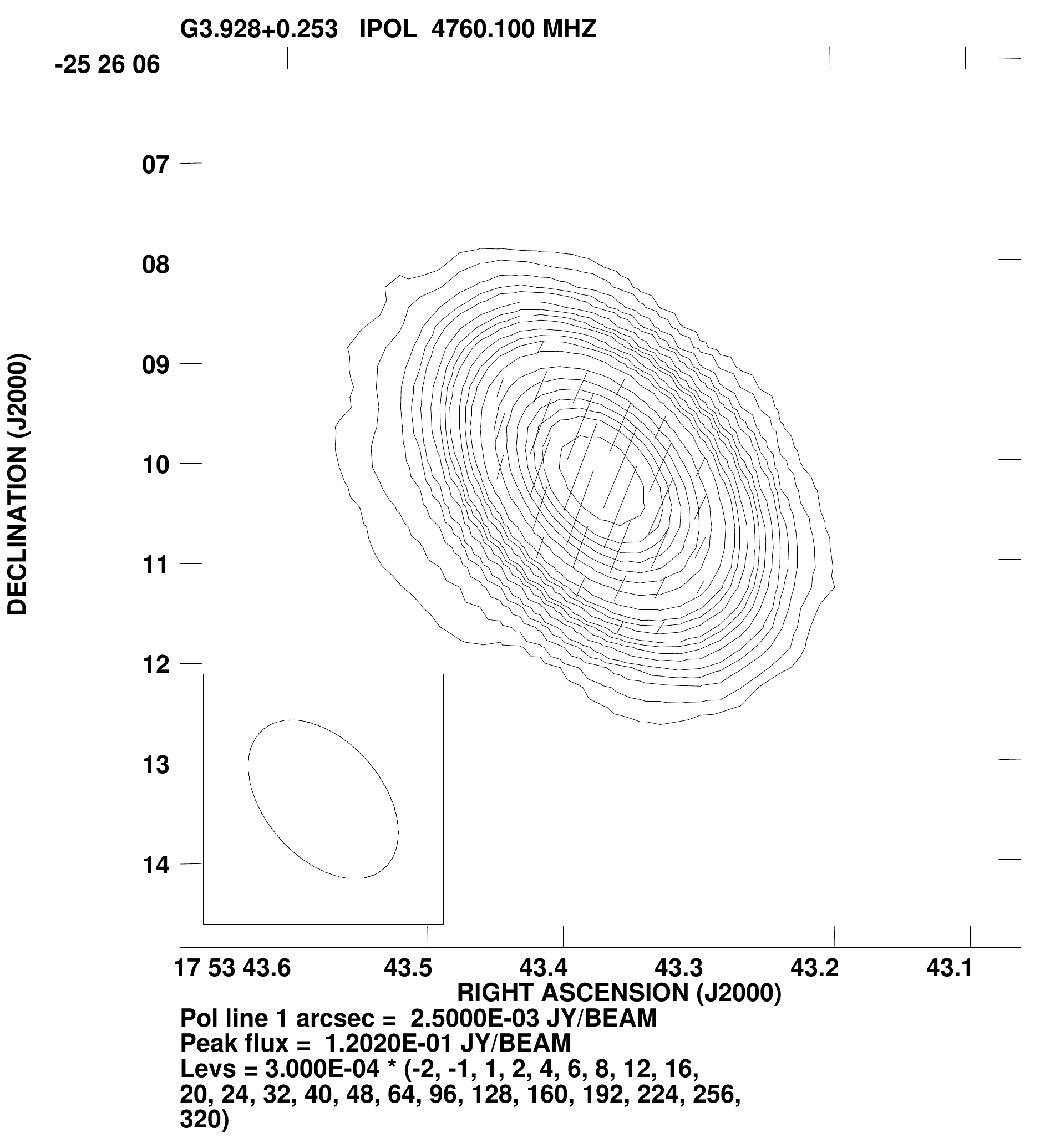}
}
}
\caption{Continued}
\end{figure*}
\vspace{-0.4cm}

\addtocounter{figure}{-1}
\begin{figure*}
\centering
\vbox{
\hbox{
\includegraphics[width=5.2cm, angle=0,clip=]{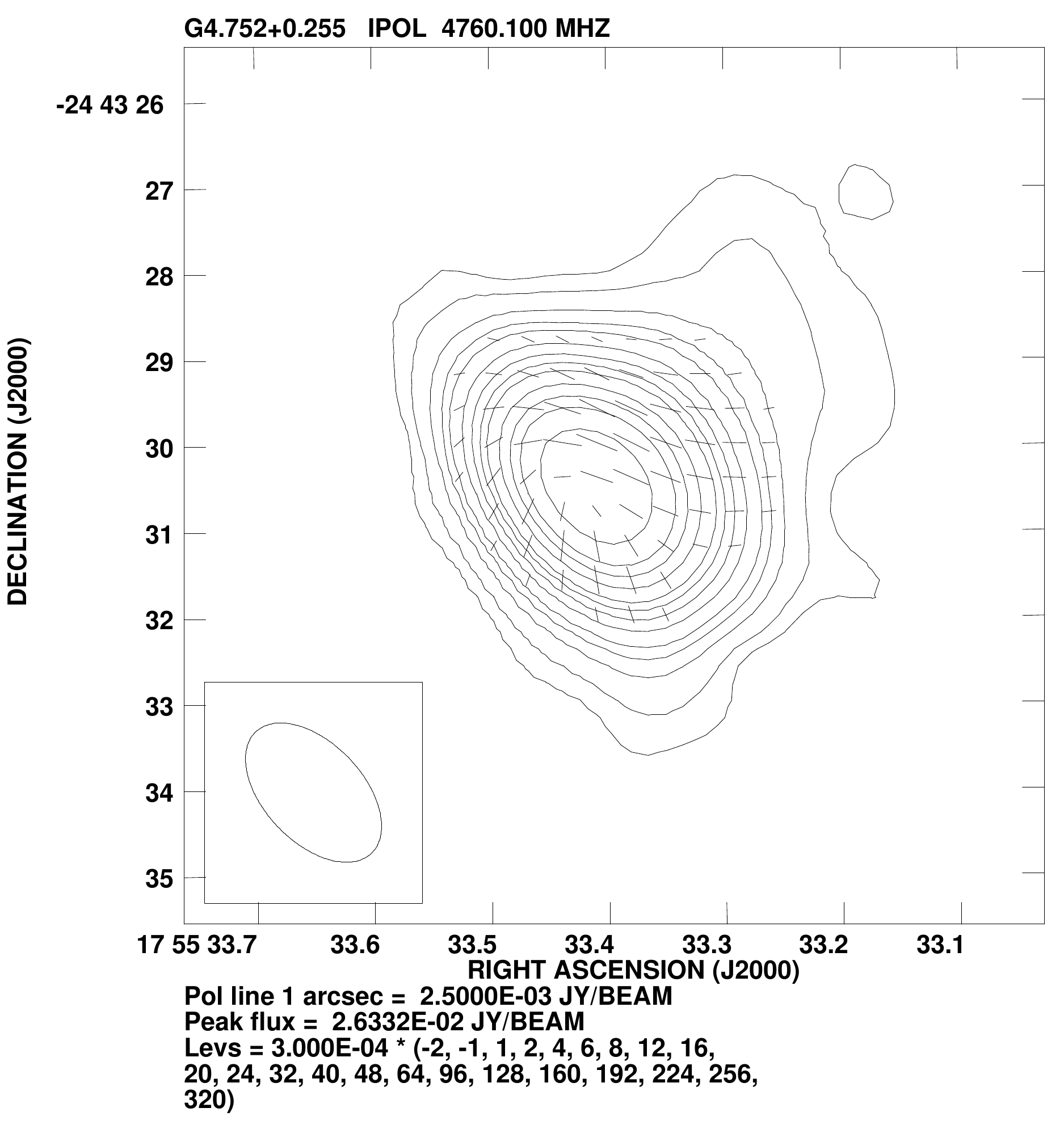}
\includegraphics[width=5.9cm, angle=0,clip=]{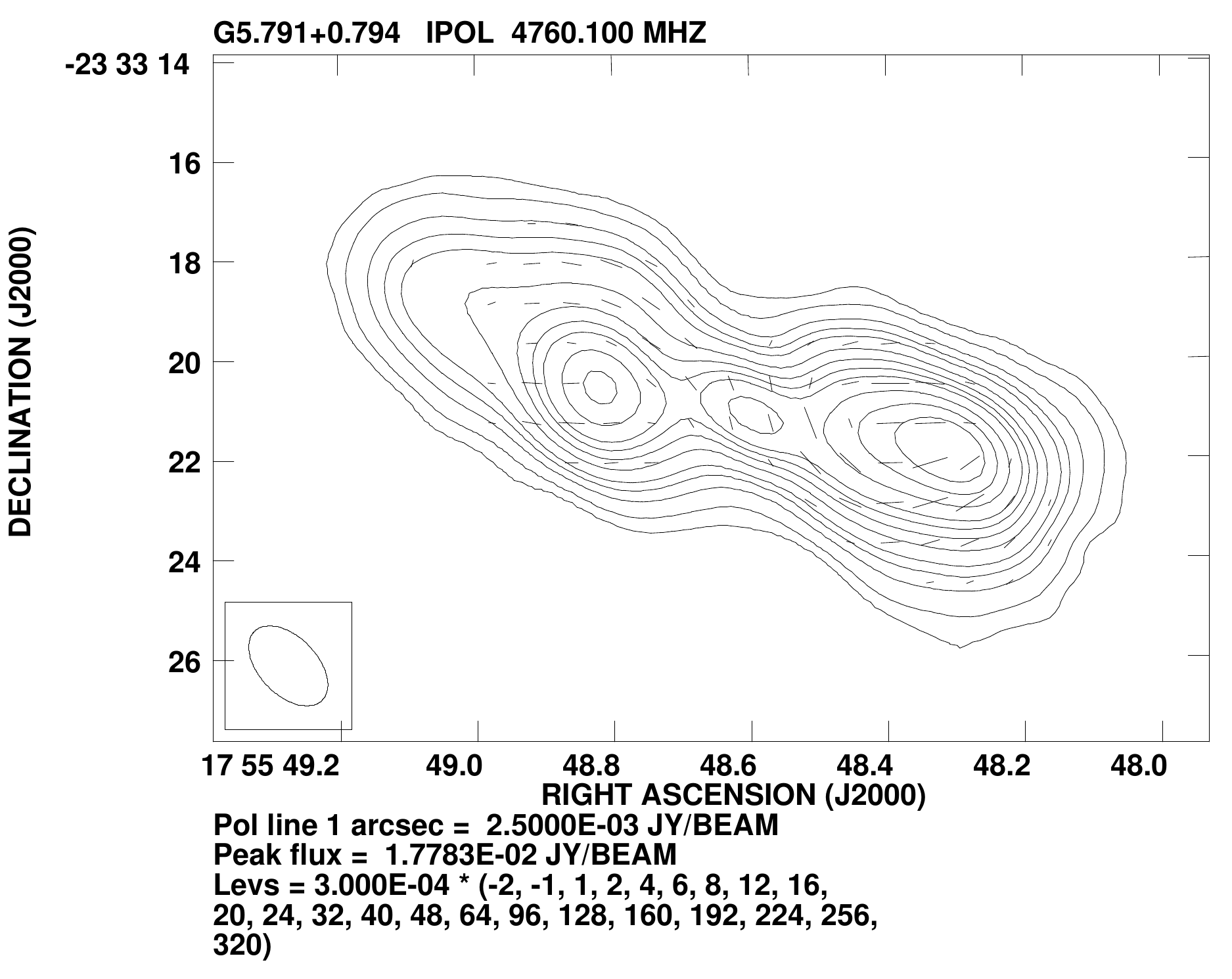}
\includegraphics[width=4.2cm, angle=0,clip=]{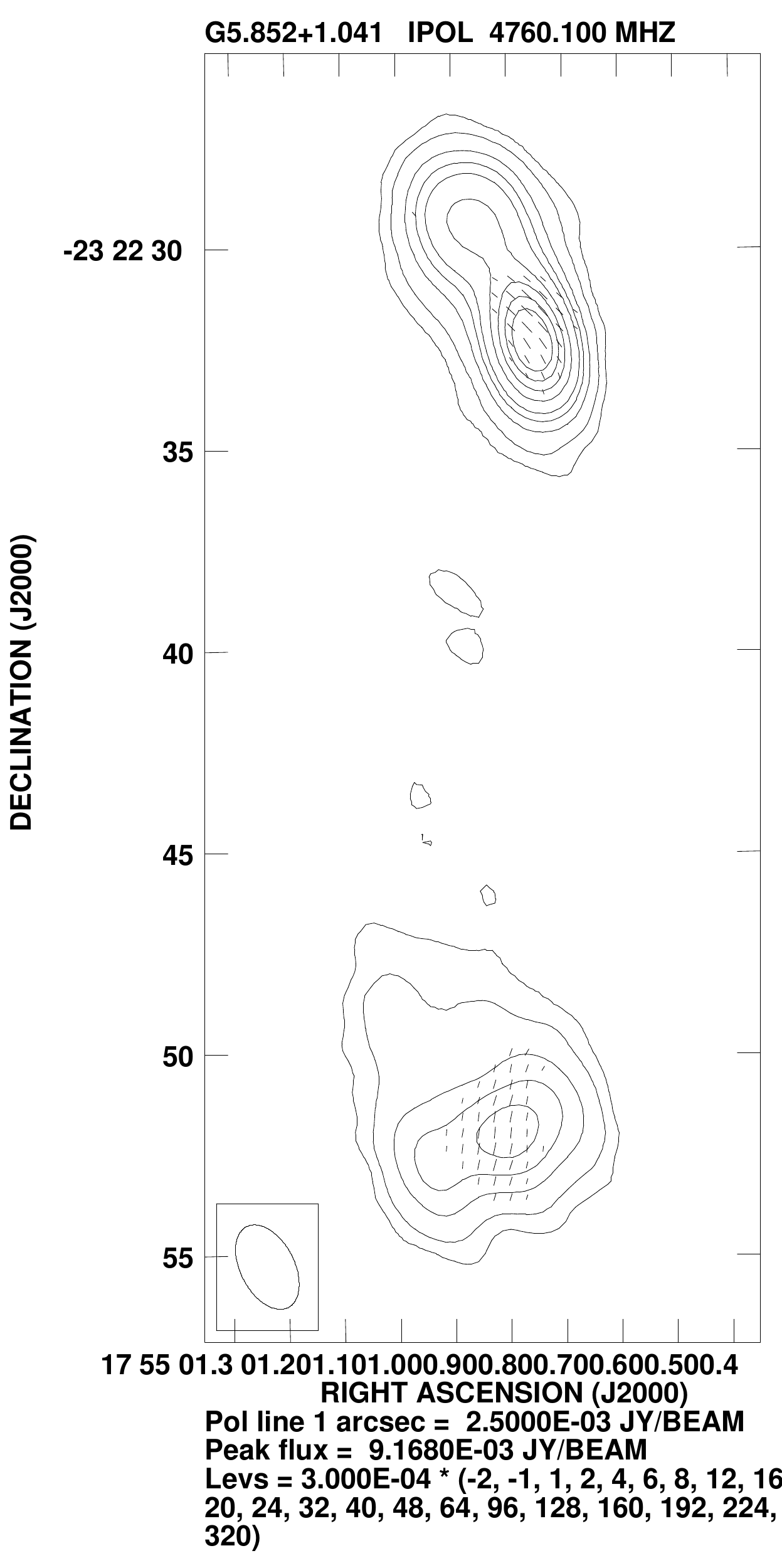}
}
\hbox{
\includegraphics[width=6.6cm, angle=0,clip=]{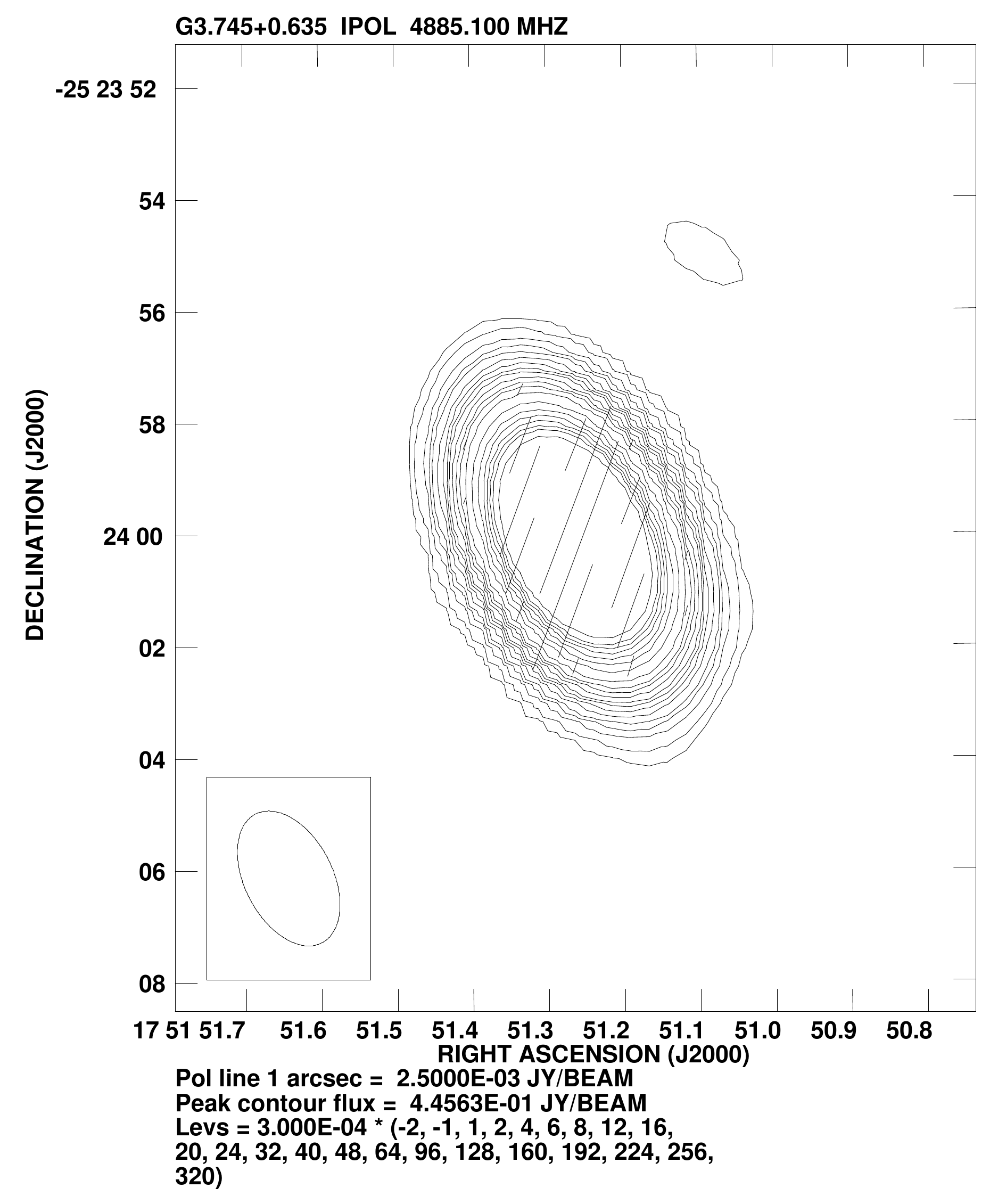}
\includegraphics[width=4.5cm, angle=0,clip=]{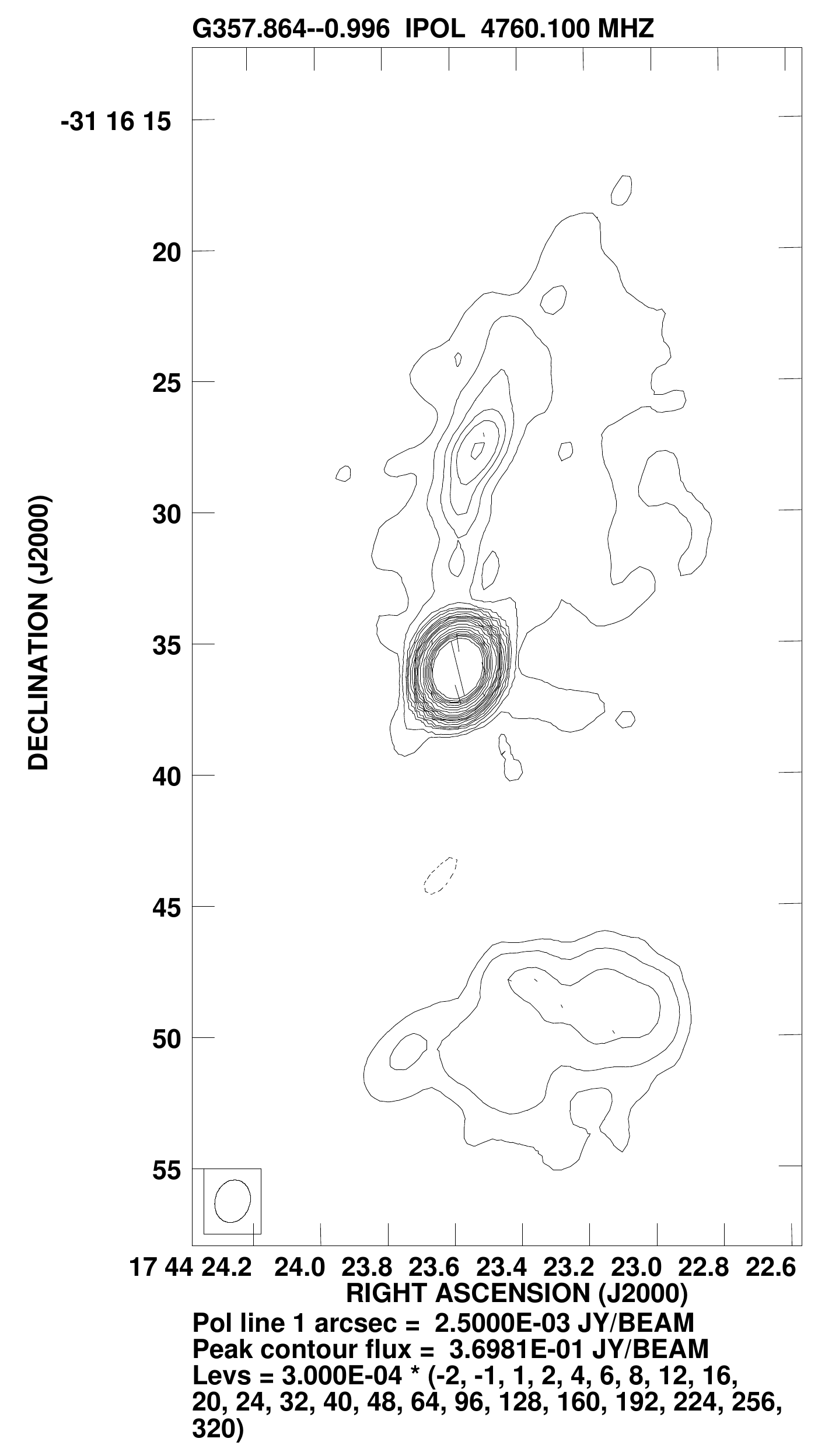}
}
}
\caption{Continued}
\end{figure*}
\vspace{-0.4cm}

\begin{figure*}
\centering
\vbox{
\hbox{
\includegraphics[width=4.9cm, angle=0,clip=]{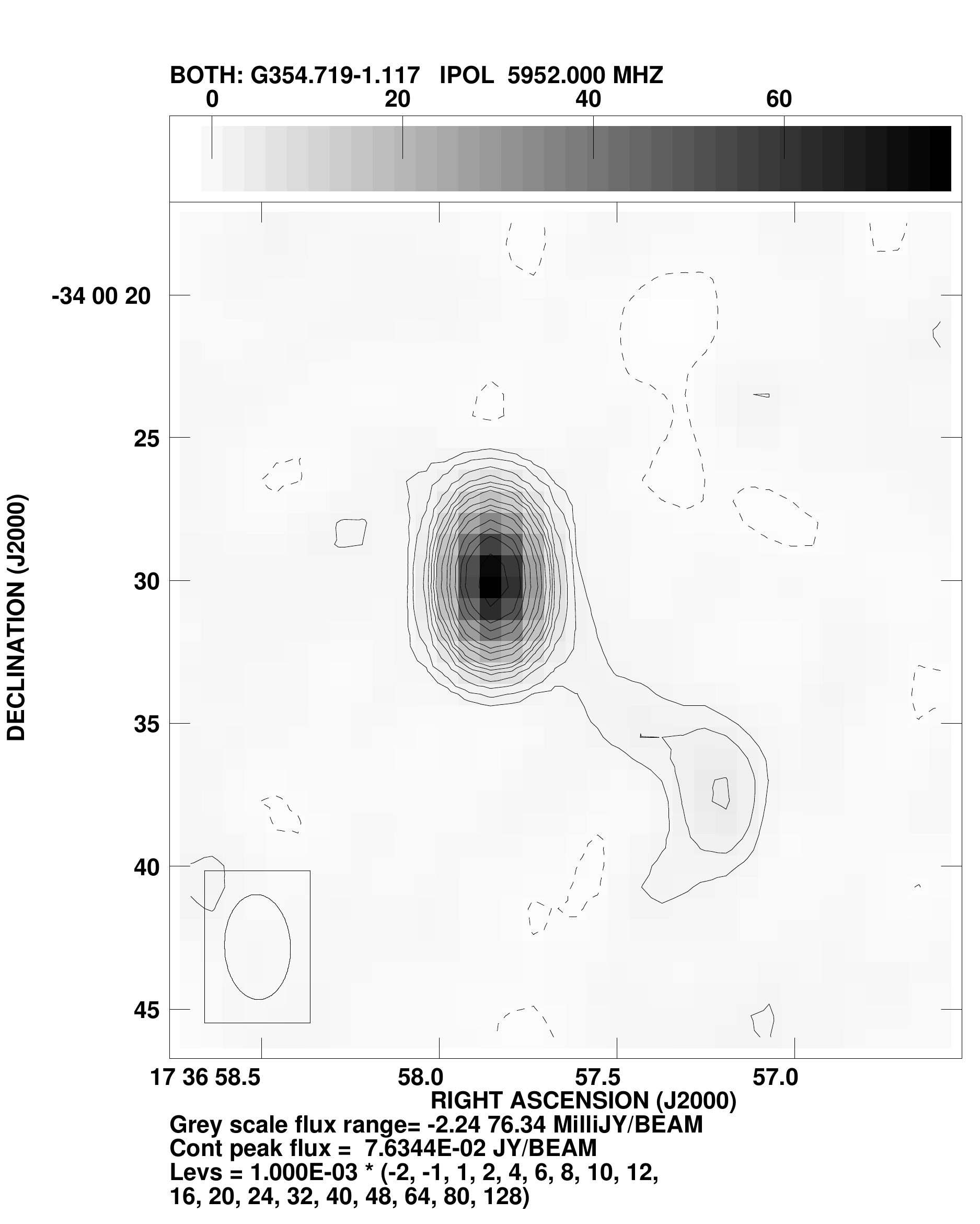}
\includegraphics[width=4.5cm, angle=0,clip=]{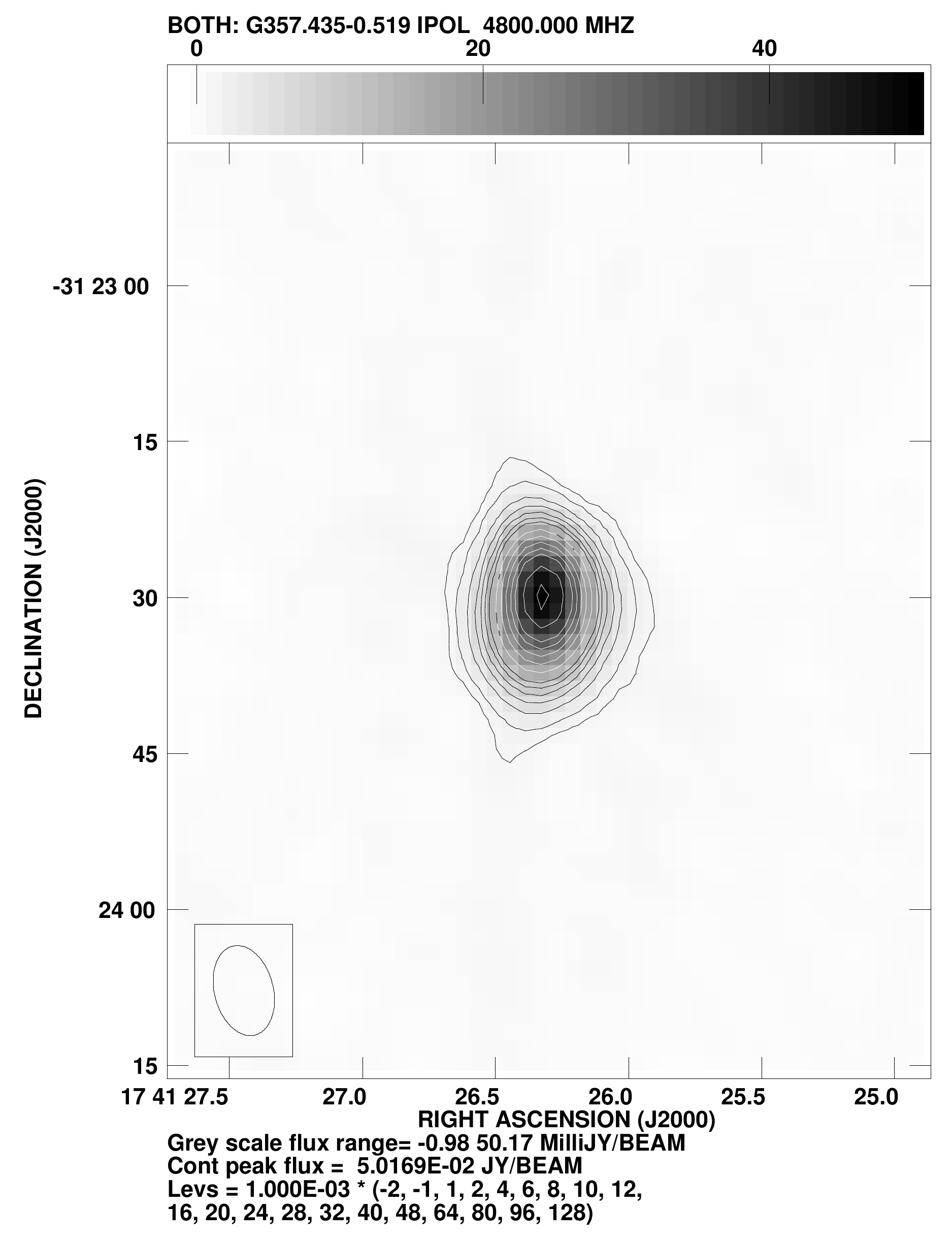}
\includegraphics[width=3.6cm, angle=0,clip=]{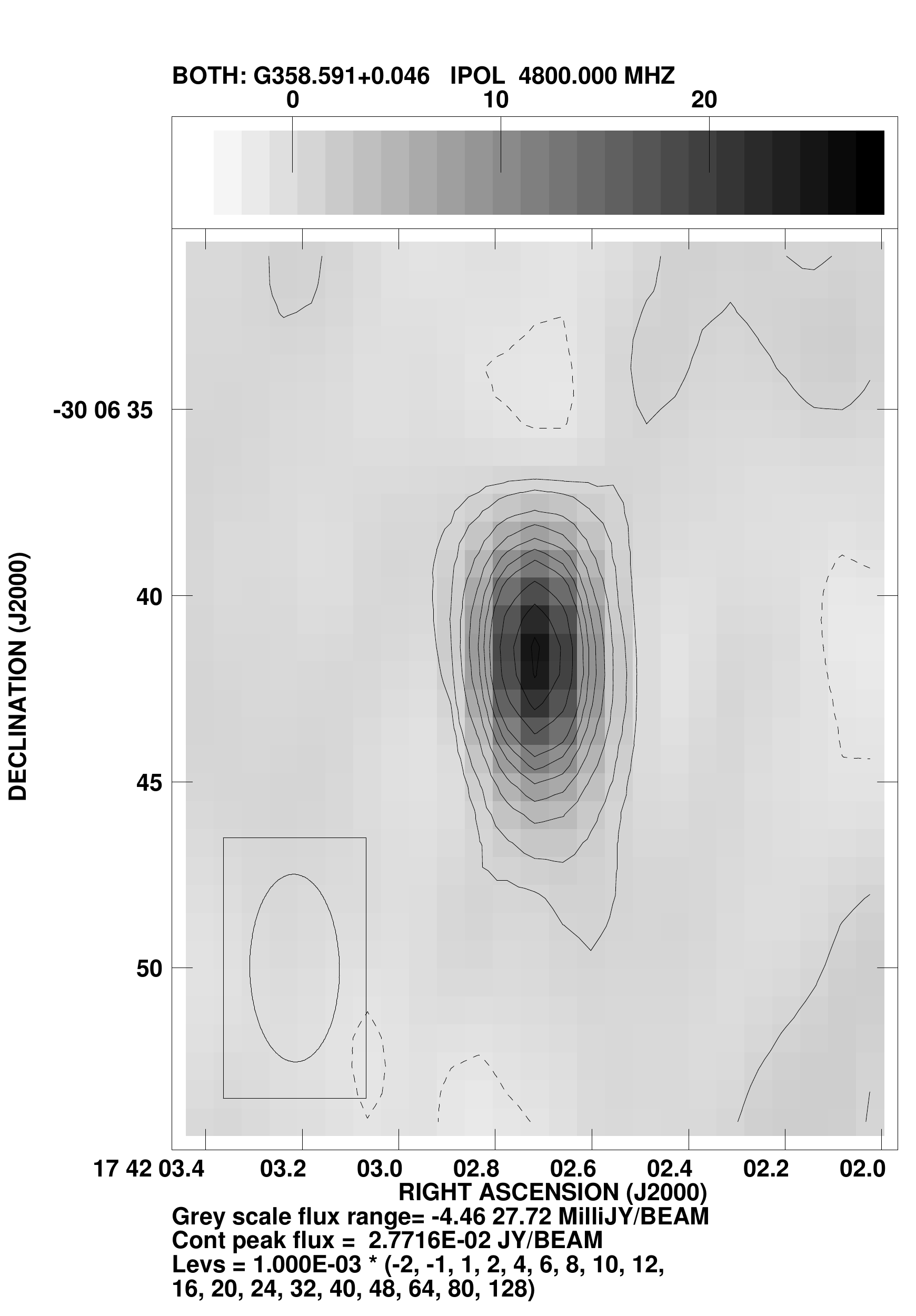}
\includegraphics[width=4.7cm, angle=0,clip=]{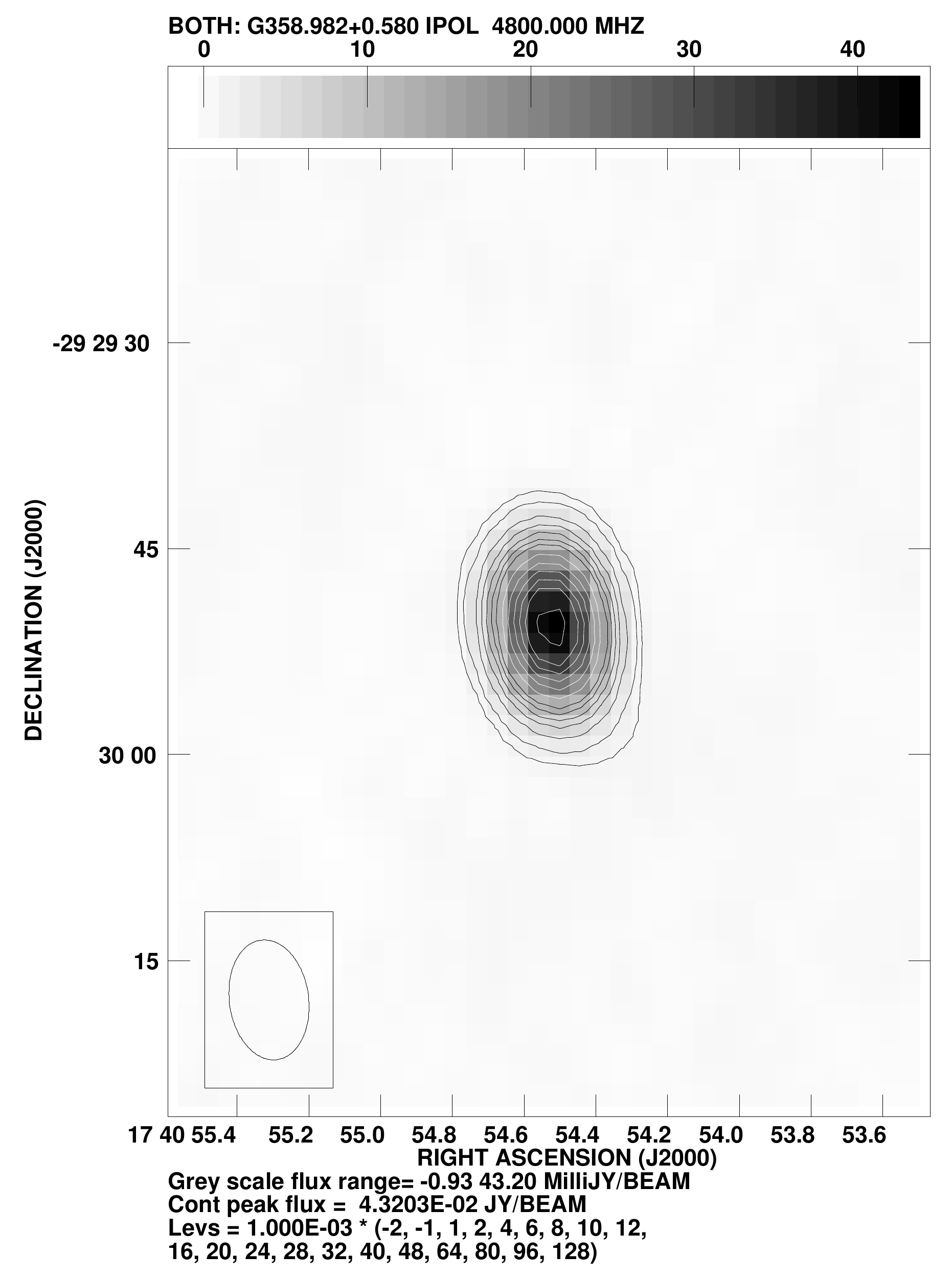}
}
\hbox{
\includegraphics[width=5.5cm, angle=0,clip=]{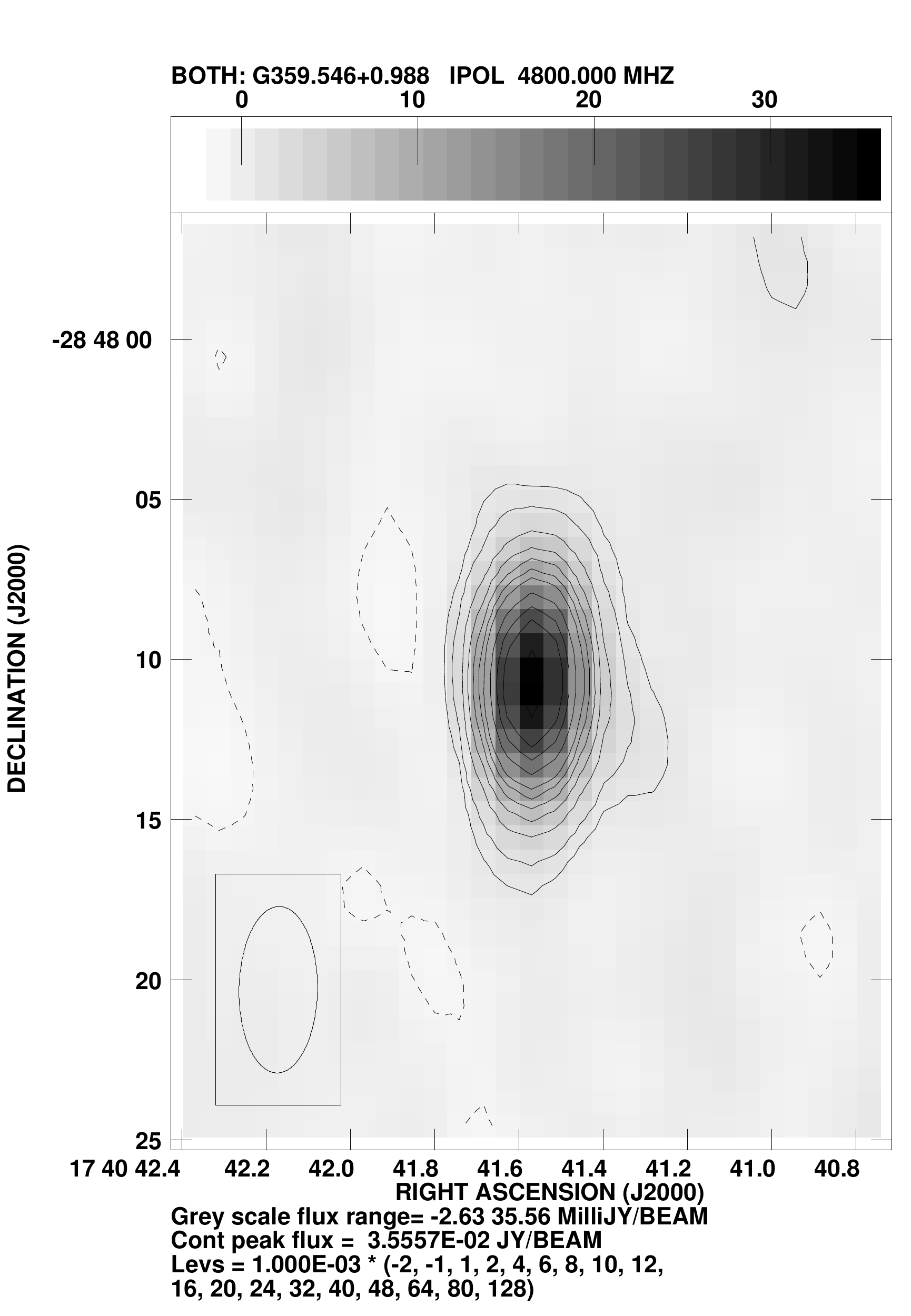}
\includegraphics[width=5.5cm, angle=0,clip=]{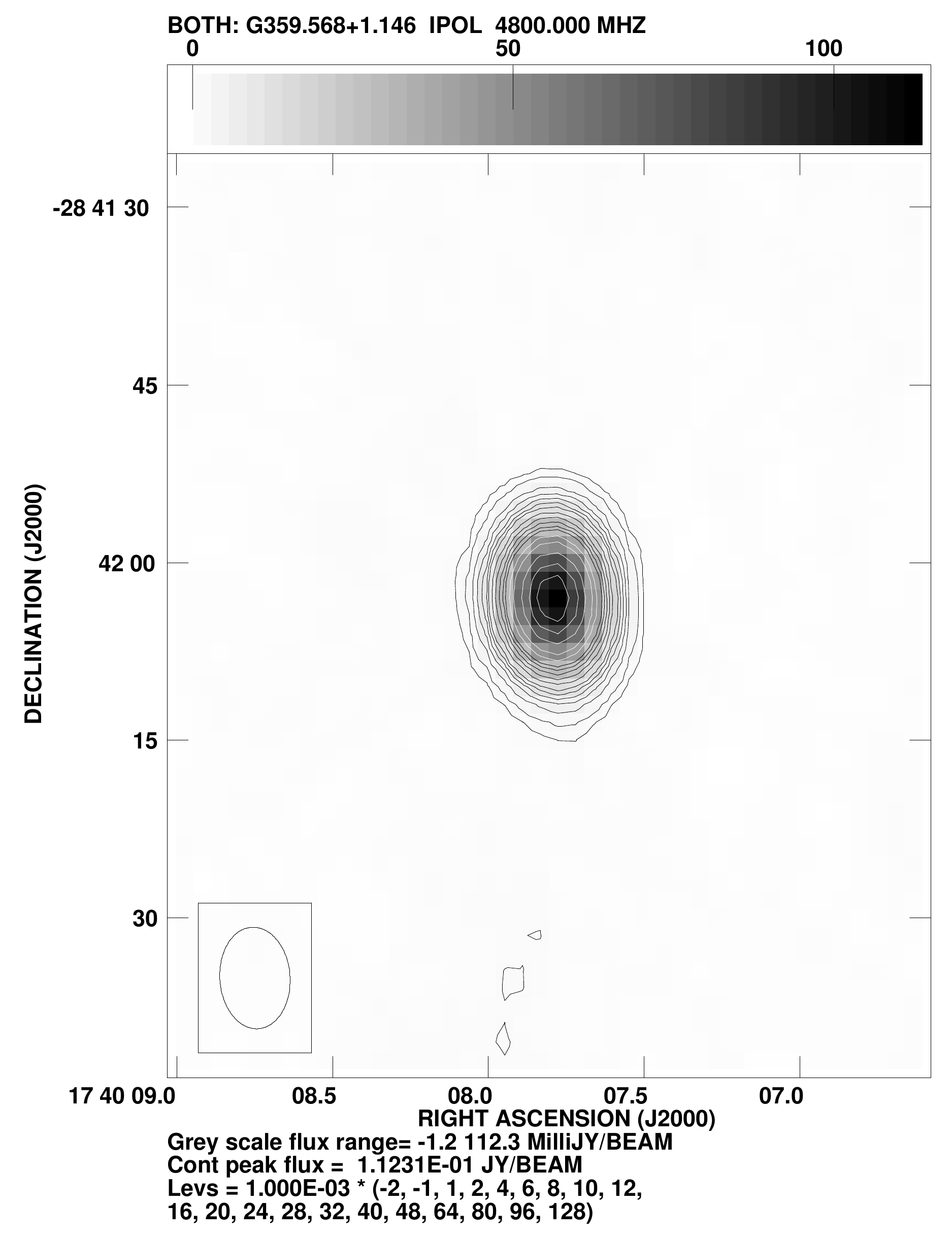}
\includegraphics[width=5.8cm, angle=0,clip=]{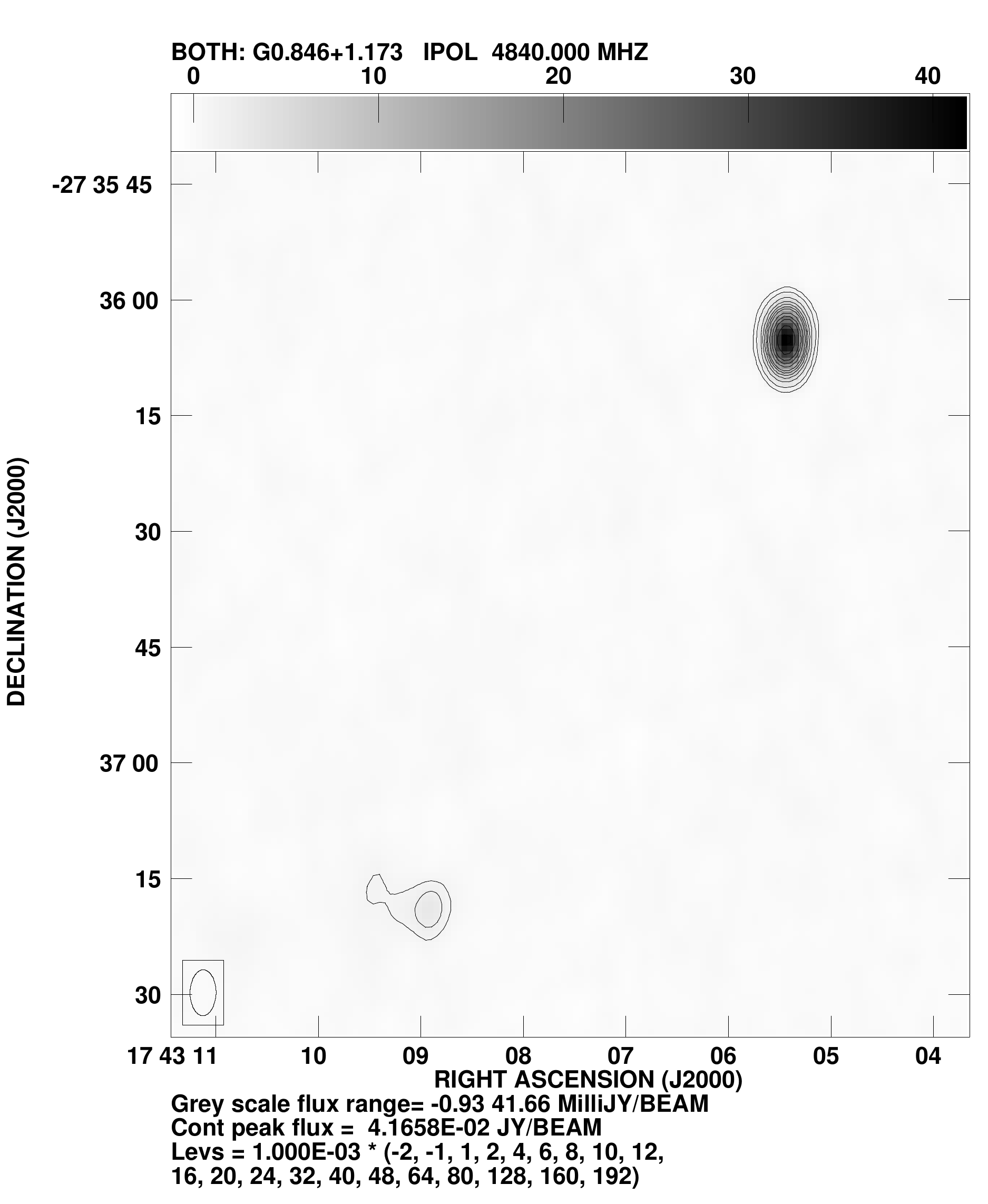}
}
\hbox{
\includegraphics[width=5.2cm, angle=0,clip=]{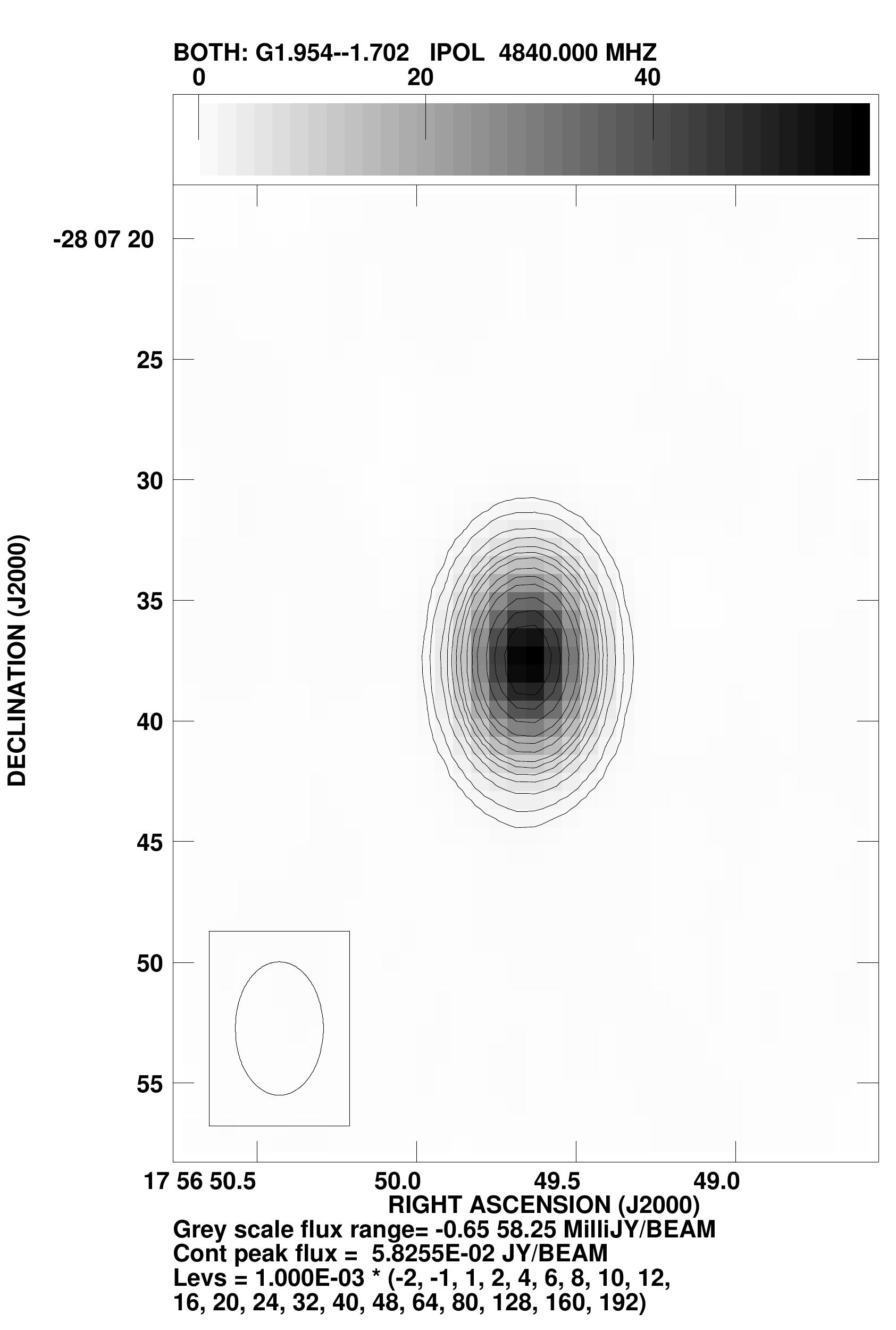}
\includegraphics[width=5.5cm, angle=0,clip=]{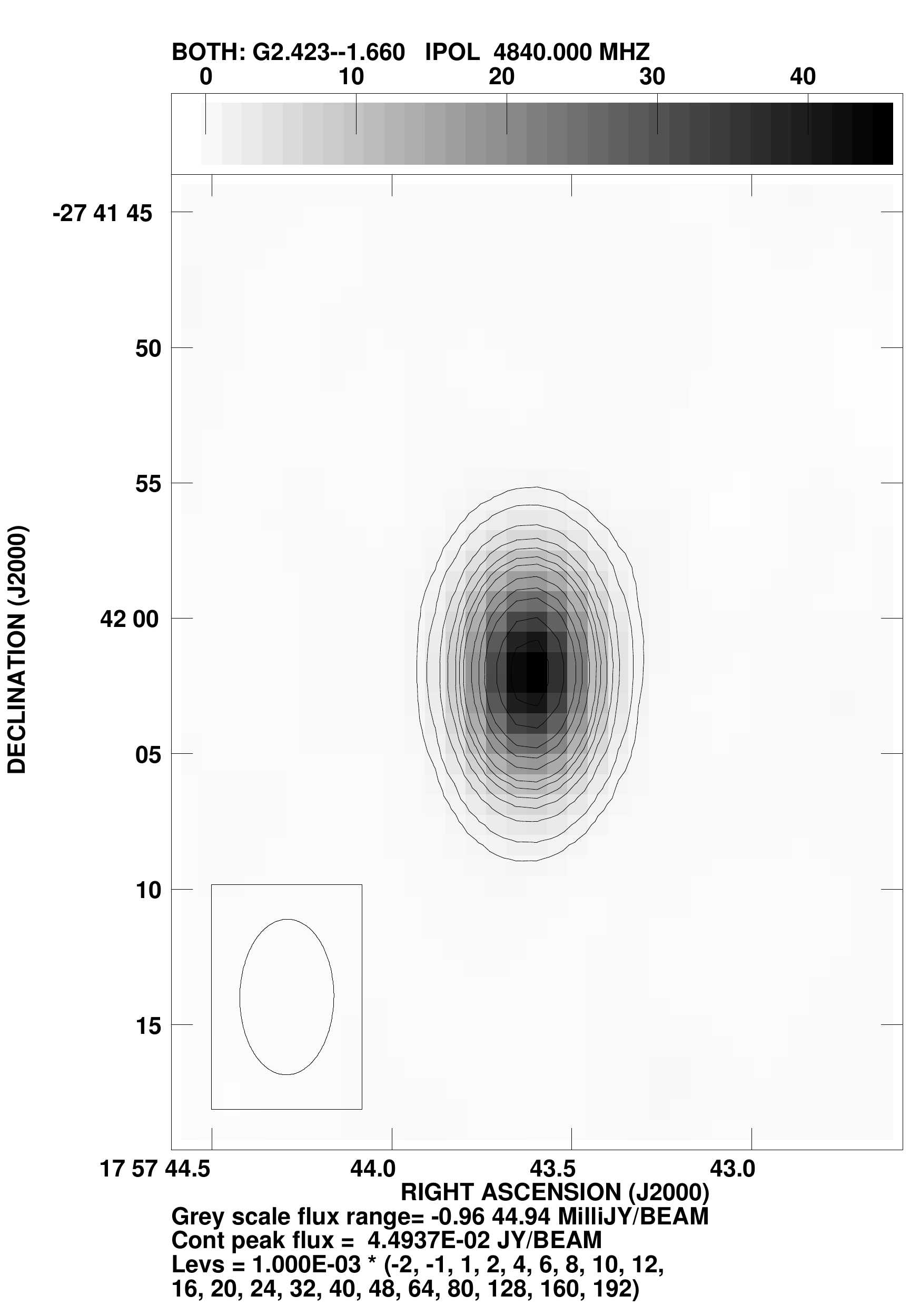}
\includegraphics[width=5.5cm, angle=0,clip=]{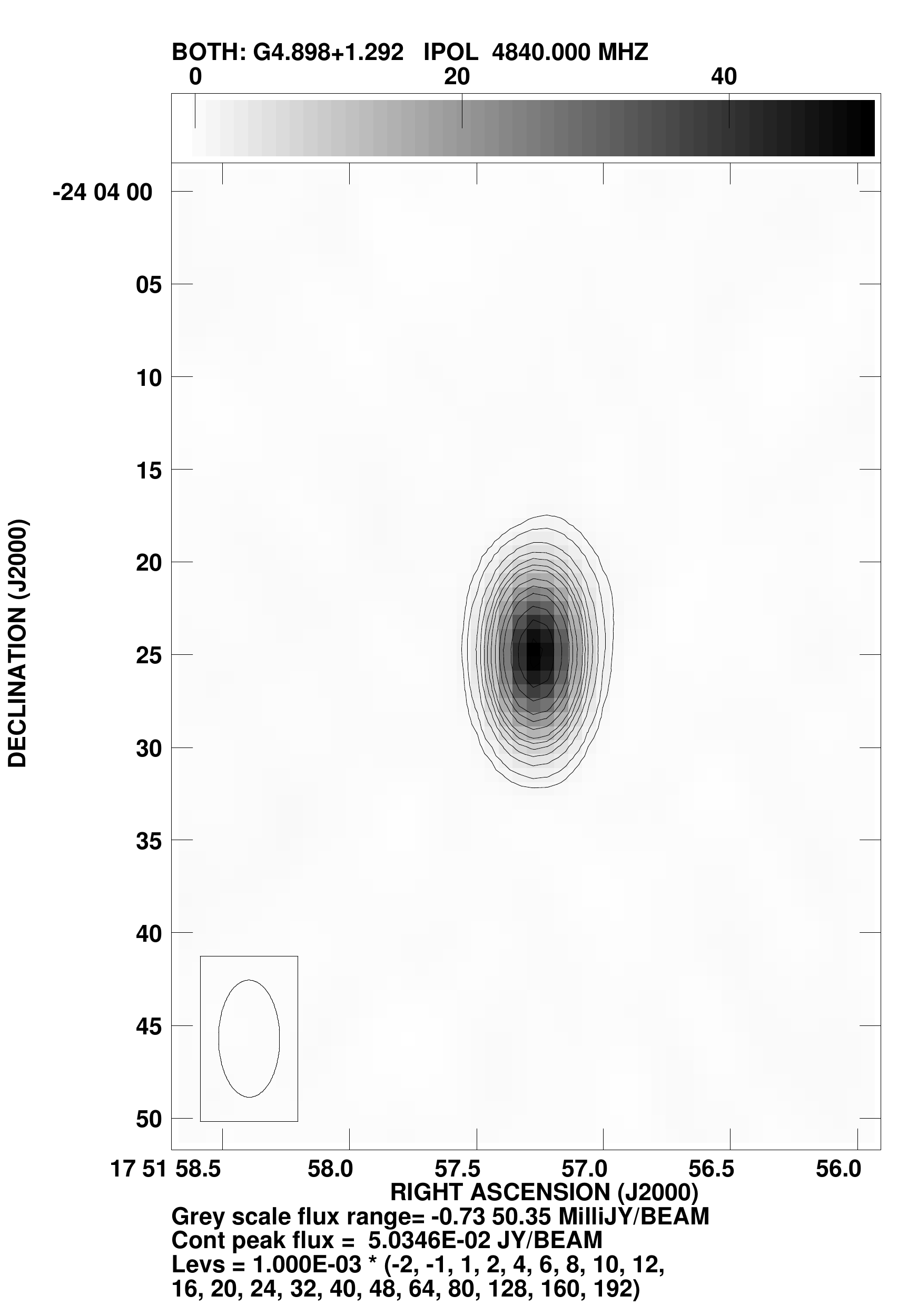}
}
}
\caption{4.8 GHz continuum images of the unpolarised sources observed with the
ATCA. Typical rms noise in Stokes I is about 0.23 mJy/beam, and beamsize
$\approx$6$^{''} \times 2^{''}$.} 
\label{5ghz.atca.unpol.src.maps}
\end{figure*}
\vspace{-0.4cm}

\newpage

\begin{figure*}
\centering
\vbox{
\hbox{
\includegraphics[width=5.5cm, angle=0,clip=]{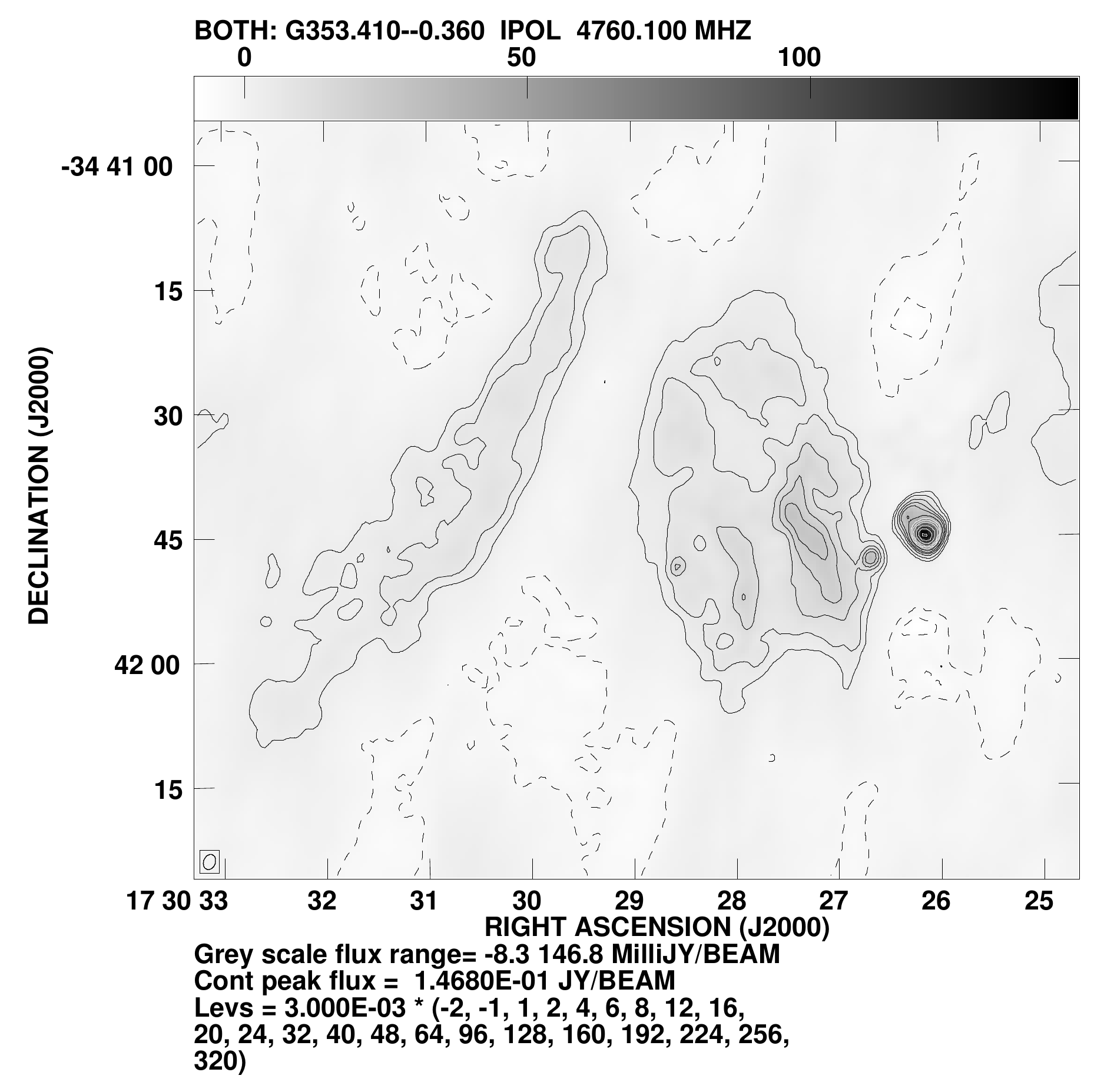}
\includegraphics[width=5.5cm, angle=0,clip=]{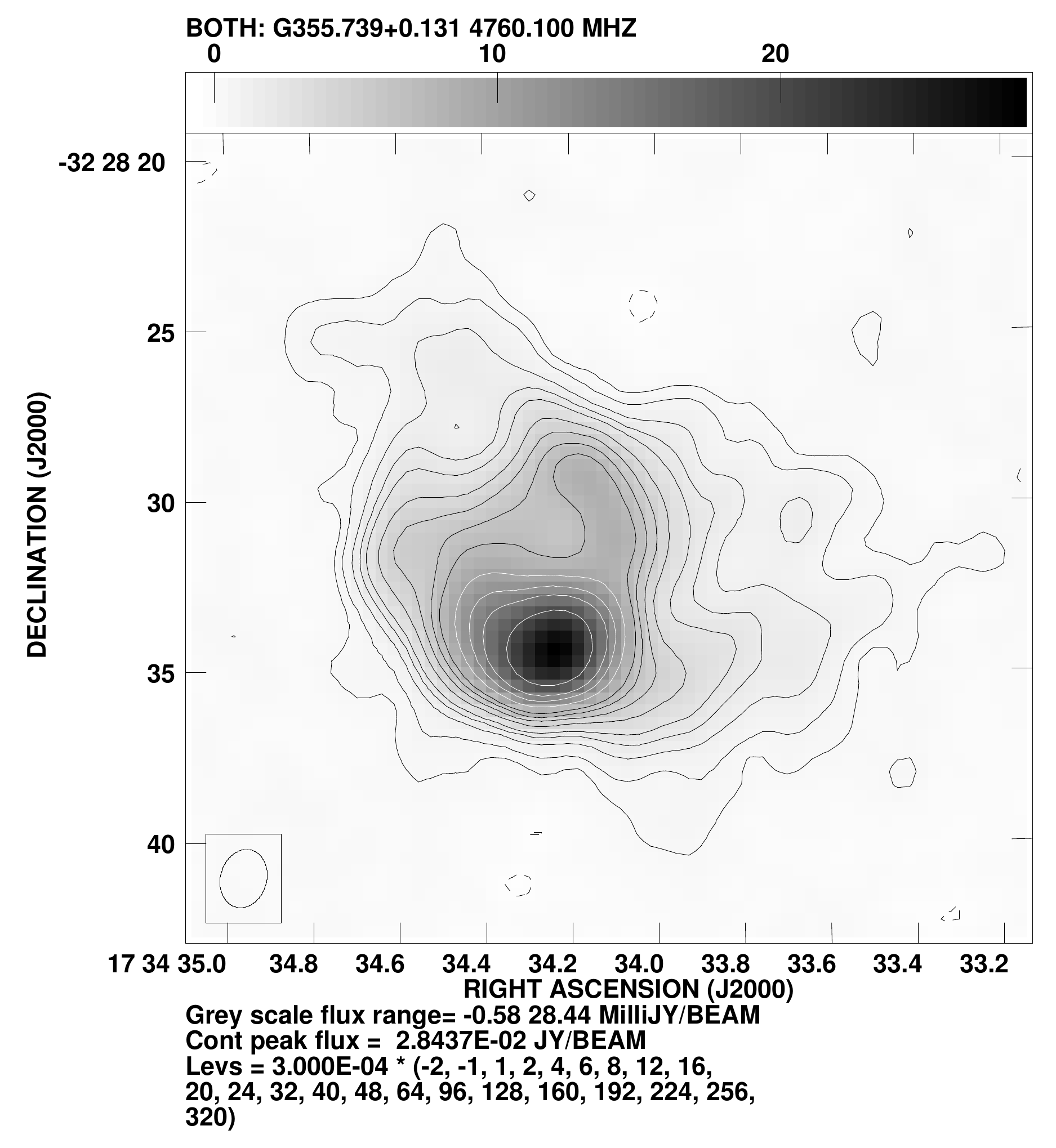}
\includegraphics[width=5.5cm, angle=0,clip=]{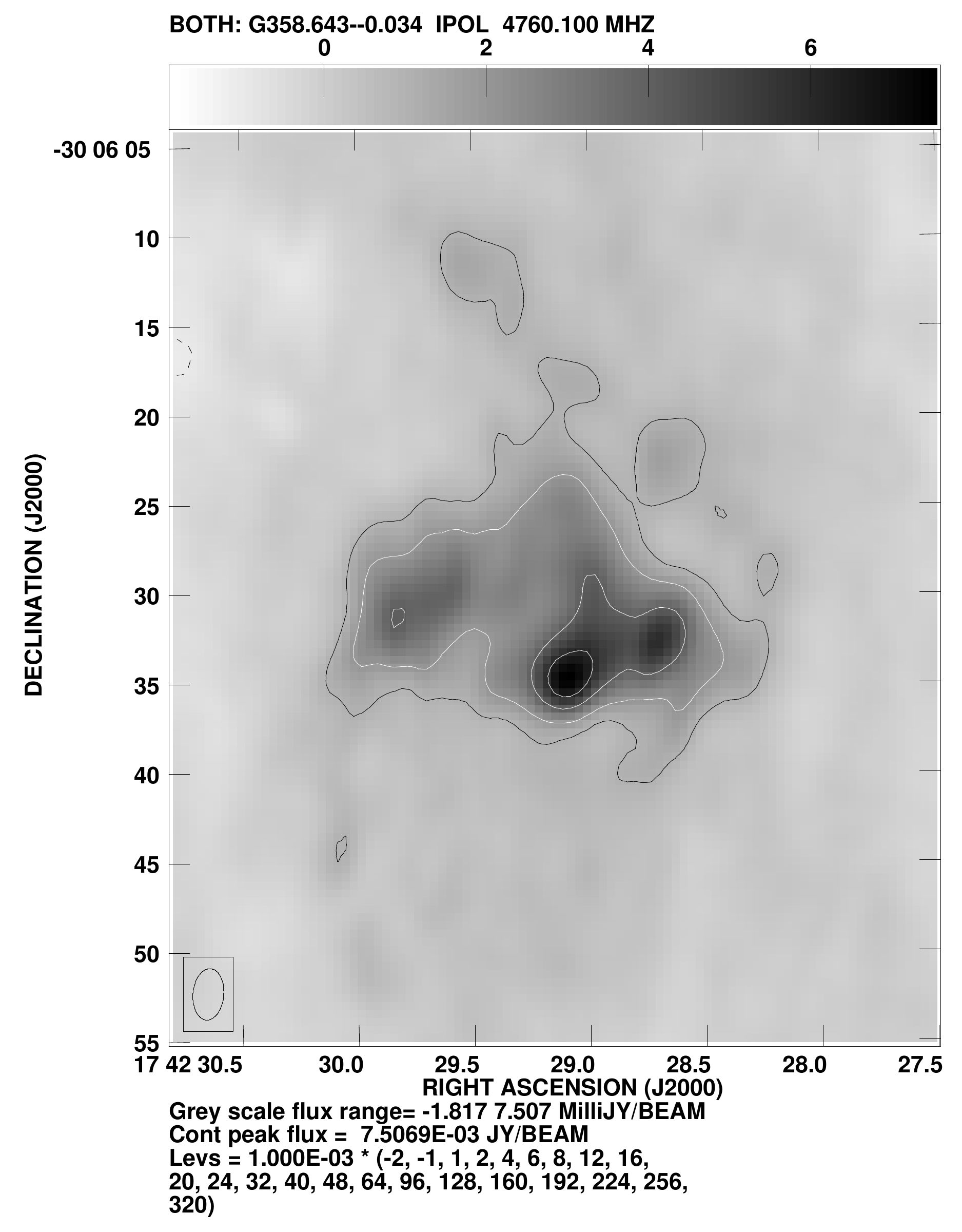}
}
\hbox{
\includegraphics[width=5.5cm, angle=0,clip=]{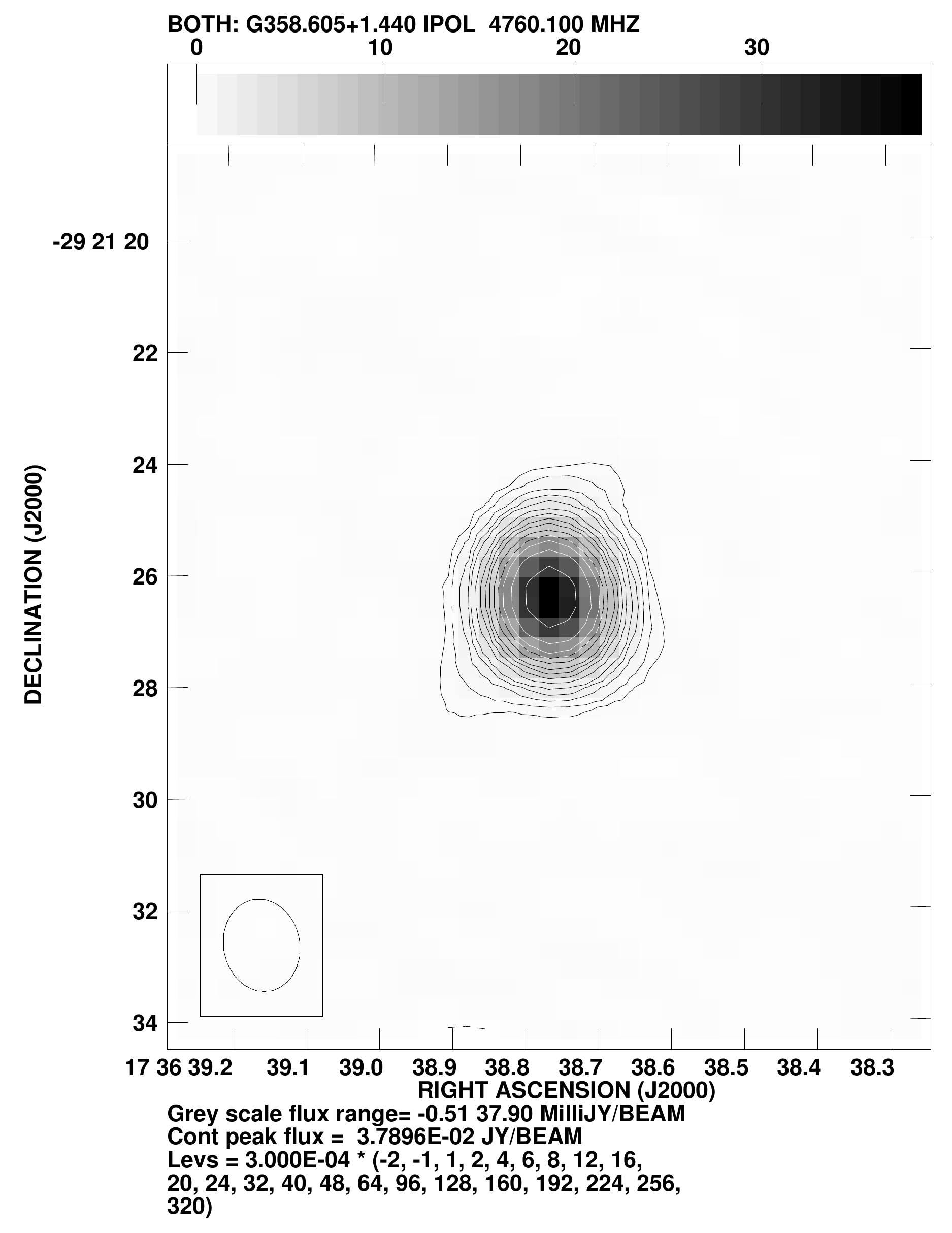}
\includegraphics[width=4.5cm, angle=0,clip=]{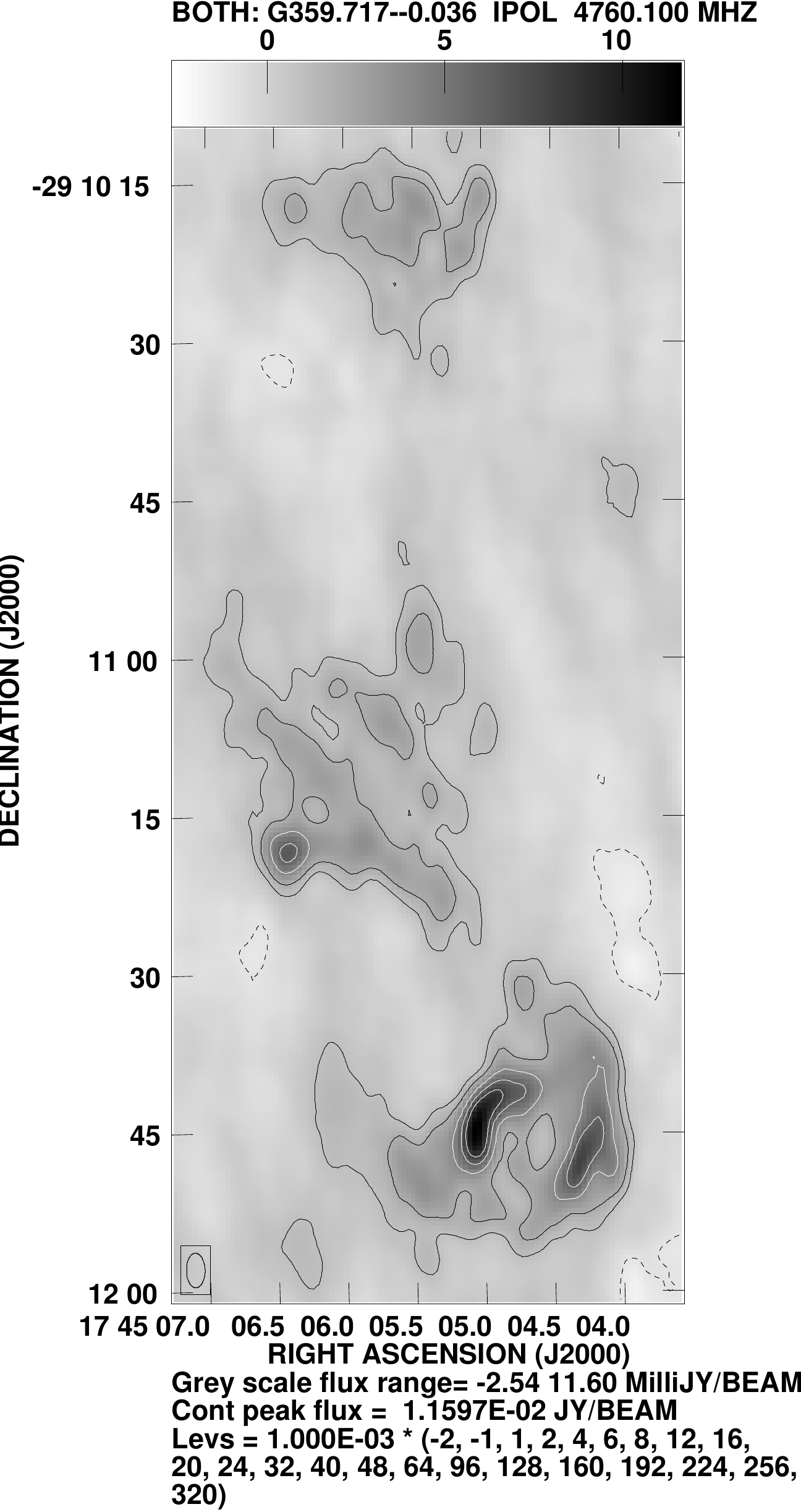}
\includegraphics[width=5.8cm, angle=0,clip=]{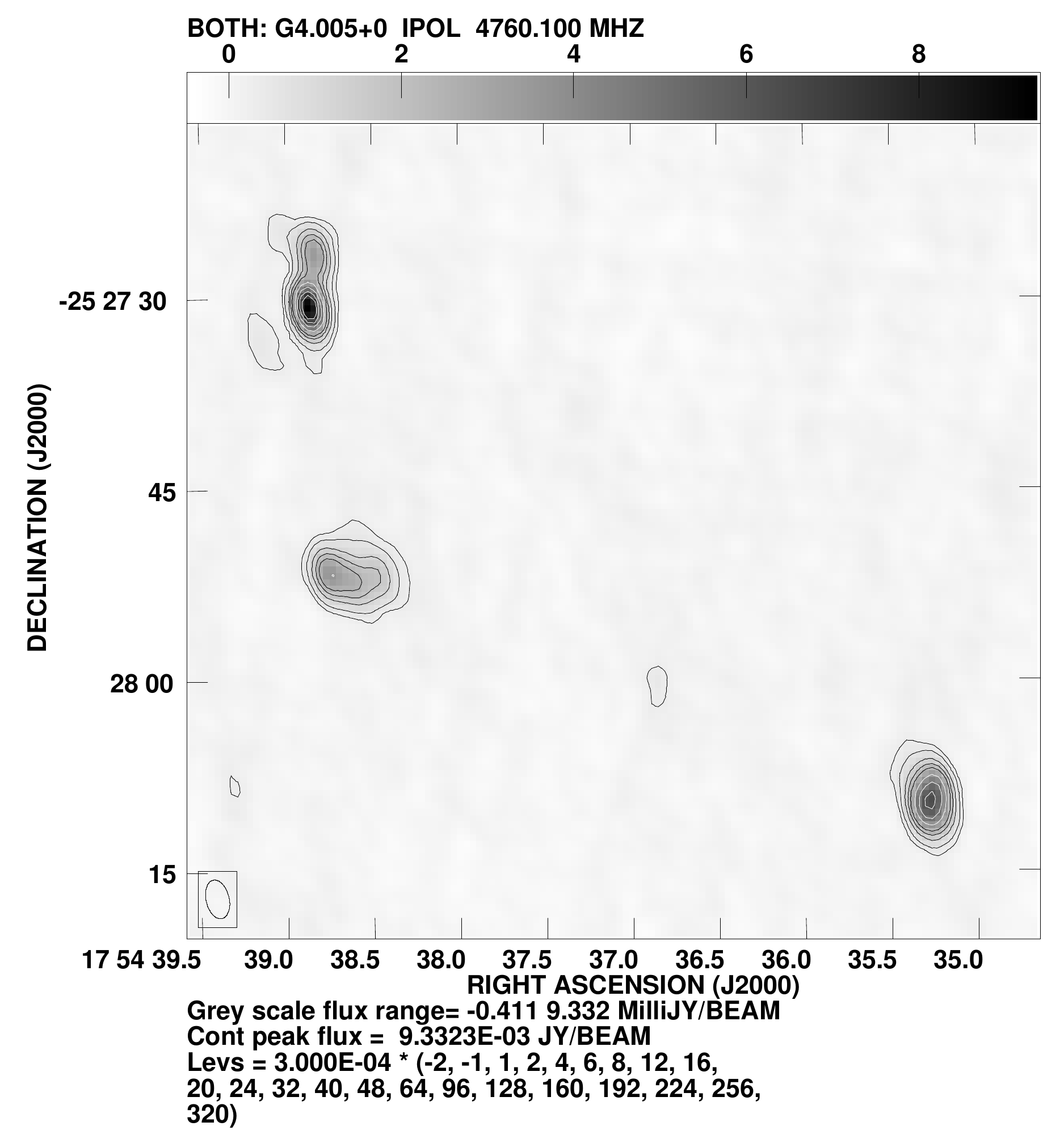}
}
\hbox{
\includegraphics[width=6.0cm, angle=0,clip=]{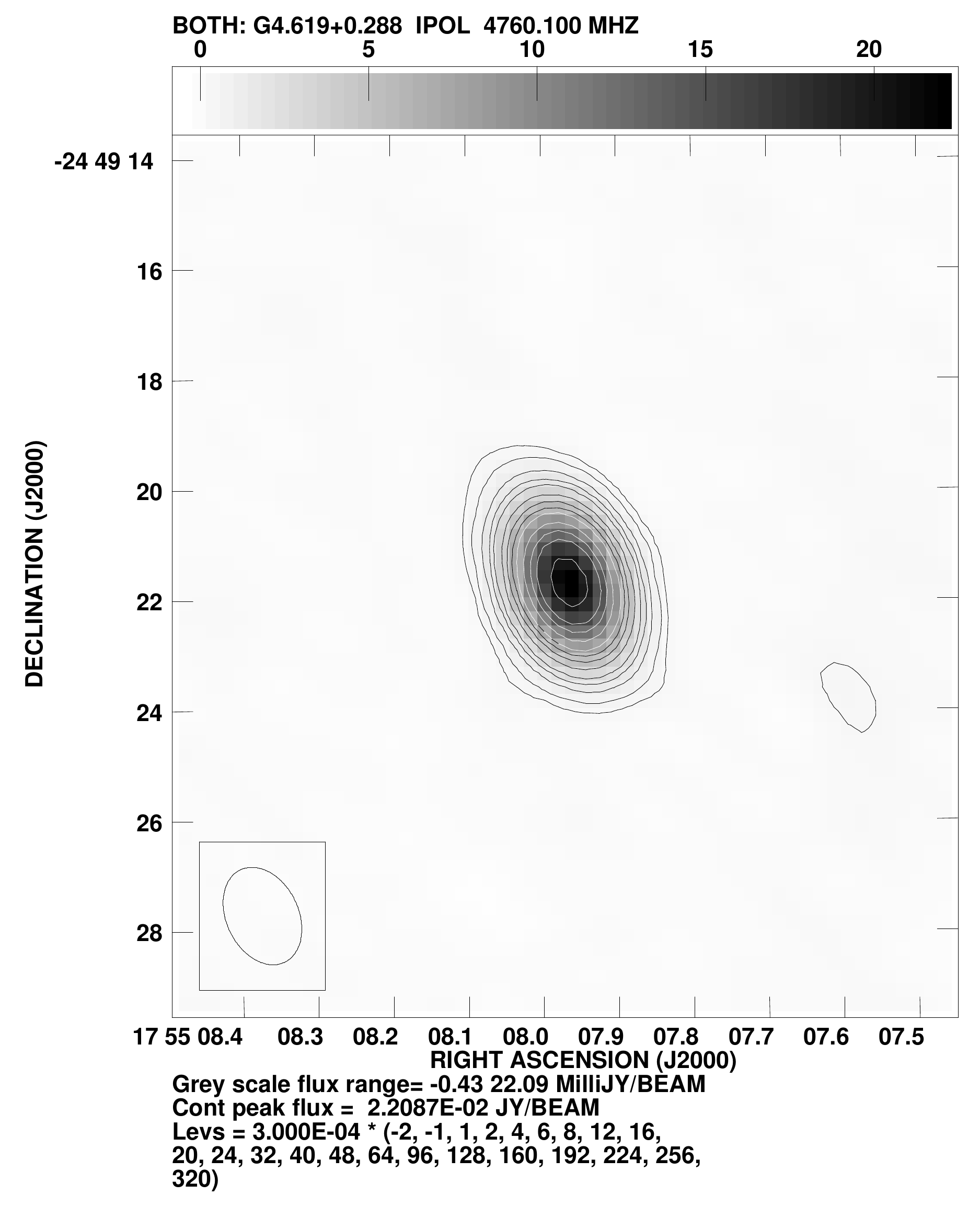}
}
}
\caption{4.8 GHz continuum image of the unpolarised sources observed with the
VLA. Typical rms noise is about 90 $\mu$Jy/beam, and beamsize of
$\approx$2$^{''} \times 1.5^{''}$.}
\label{5ghz.vla.unpol.src.maps}
\end{figure*}

\small

\onecolumn
\begin{landscape}
\begin{longtable}{c c c c c c c c c c l}
\caption{Estimated RM towards the background sources towards the GC region} \\
\label{gc.rm.list} \\
\hline
Source Name & RA & DEC  & RM & Error & D & $\Delta$(D) & $\theta$ & $\Delta\theta$ & $\chi^2$ & Measured polarisation angles \\
      & (J2000)&(J2000) &    & on    &  & (\%)   &          &                &          & at different frequencies  \\
      &        &        &    & RM    &  &        &          &                &          &                        \\
\hline
\endfirsthead

\hline
Source Name & RA & DEC  & RM & Error & D & $\Delta$(D) & $\theta$ & $\Delta\theta$ & $\chi^2$ & Measured polarisation angles \\
      & (J2000)&(J2000) &    & on    &   &  (\%)   &          &                &          & at different frequencies \\
      &        &       &     & RM    &   &         &          &                &          &   \\
\hline
\endhead

353.462-0.691 &  17 31 55.40   &  $-$34  49 59.9   & 1011   & 30   &  0.5   & 14  &  123  &   5.7 &  2.3   &  8540 $-$83$\pm$5.4,  8484 $-$67$\pm$5.5,  4885 74$\pm$4.8,  4635 94$\pm$ 3.7 \\                                          
              &                &                   &        &      &        &     &       &       &        & \\
354.740+0.138 &  17  31  56.96 &  $-$33  18 33.72  &   1193 &   32 &  0.7 & 15  &   49   &   5.8  &  4.6  &   8512 39$\pm$4.1,  5952 $-$46$\pm$3.8,  5312 04 $\pm$3.1,  4804 41$\pm$2.9 \\
              &  17  31  57.00 &  $-$33  18 42.12  &    858 &   45 &  0.8 & 25  &  122   &   8.1  &  2.7  &   8512 $-$85$\pm$6.2,  5952 $-$33$\pm$4.6,  5312 07 $\pm$3.8,  4804 44$\pm$4.8,  4764 52$\pm$5.7 \\
              &                &                   &        &      &      &     &        &        &        &  \\
354.815+0.775 &  17  29  36.653 &  $-$32  53 51.88 & $-$99 &   16 &  0.8   & 10  &  153  & 2.7  &  0.1 &     8685 57$\pm$2.6,  8335 55$\pm$2.4,  4885 42$\pm$2.6,  4635 39$\pm$2.5  \\
              &                &                   &        &      &        &     &       &      &      &    \\
355.424-0.809 &  17  37  32.514 &  $-$33  14 36.88 &    611 &   17 &  0.9   & 11  &  167  & 3.5  &  4.0 &     8685 $-$58$\pm$3.1,  8335 $-$60$\pm$3.9,  4885 25$\pm$2.1,  4635 47$\pm$1.9  \\
              &  17  37  31.413 &  $-$33  15 09.56 &    574 &   23 &  0.9   & 17  &   37  & 4.6  &  1.1 &     8685 $-$20$\pm$5.4,  8335 $-$07$\pm$3.8,  4885 70$\pm$2.8,  4635 85$\pm$2.4  \\
              &                 &                 &         &      &        &     &       &      &      &     \\
356.000+0.023 &  17  35  40.37 &  $-$32  18 59.03  &    328 &   25 &  1.1 & 15  &   79   &    4.6 &   0.5  &  8512 11$\pm$4.0,  5952 39$\pm$2.0,  5312 49$\pm$1.8,  4804 63$\pm$1.8  \\
              &  17  35  40.05 &  $-$32  18 55.83  &    439 &   29 &  0.9 & 15  &  112   &    5.1 &   3.0  &  8512 54$\pm$4.0,  5952 83$\pm$2.3,  5312 $-$73$\pm$2.5,  4804 $-$62$\pm$2.5  \\
              &  17  35  39.74 &  $-$32  18 54.23  &    720 &   30 &  1.1 & 17  &   73   &    5.6 &   0.2  &  8512 33$\pm$4.8,  5952 89$\pm$2.6,  5312 $-$66$\pm$2.3,  4804 $-$36$\pm$2.3  \\
              &                &                   &        &      &      &     &        &        &        &   \\
356.161+1.635 &  17  29  43.86 &  $-$31  18  04.77  &    679 &   27 &  1.0 & 16 &   13   &    4.9 &   3.9  &  8512 39$\pm$4.1,  5952 $-$46$\pm$3.8,  5312 04$\pm$3.1,  4804 41$\pm$2.9 \\
              &  17  29  43.64 &  $-$31  18  04.77  &    681 &   23 &  1.4 & 20 &  109   &    4.4 &   2.5  &  8512 61$\pm$4.7,  5952 $-$58$\pm$1.8,  5312 $-$36$\pm$1.6,  4804 $-$08$\pm$1.4 \\
              &  17  29  42.71 &  $-$31  18  02.37  &    629 &   34 &  0.9 & 18 &  119   &    6.4 &   3.8  &  8512 68$\pm$5.0,  5952 $-$58$\pm$3.9,  5312 $-$29$\pm$3.4,  4804 $-$14$\pm$2.6 \\
              &                &                    &        &      &      &    &        &        &        &     \\
356.567+0.869 &  17  33  45.73 &  $-$31  22 50.38   &    960 &   35 &  1.1 & 12 &  163   &    6.5 &   0.2  &  8512 $-$37$\pm$3.7,  5952 31$\pm$2.4,  5312 70$\pm$2.1,  4804 $-$72$\pm$1.8 \\
              &                &                    &        &      &      &    &        &        &        &    \\
356.719-1.220 &  17  42  27.90 &  $-$32  22 14.04 &   $-$511 &   37 &  1.1 & 23 &   51   &    6.7 &   1.8  &  8512 $-$83$\pm$6.6,  5952 70$\pm$2.5,  5312 46$\pm$2.4,  4804 27$\pm$2.6 \\
              &  17  42  27.52 &  $-$32  22 09.64 &   $-$464 &   41 &  1.2 & 23 &   27   &    7.6 &   3.3  &  8512 82$\pm$6.7,  5952 56$\pm$3.6,  5312 27$\pm$3.2,  4804 16$\pm$2.8 \\
              &                &                  &          &      &      &    &        &        &        &    \\
357.865-0.996 &  17  44  23.580 & $-$31  16 35.970 &  1883 &  2.9 &   0.8   & 1.3 &  151  & 0.4  &  5.5 &     8512 14.7$\pm$0.3,  5952 $-$24.3$\pm$0.3,  5312 45.8$\pm$0.3,  4804 $-$59.3$\pm$0.4\\
 (1741-312)   &                &                   &       &      &         &     &       &      &      &        \\
              &                &                   &       &      &         &     &       &      &      &        \\
358.002-0.636 &  17  43  17.87 &  $-$30  58 18.73 &   869 &   13 &    1.4 & 09 &   76   &    2.0 &   17   &  8685 50$\pm$1.7,  8335 51$\pm$2.2,  5988 $-$79$\pm$2.6,  5920 $-$83$\pm$2.6,  4836 $-$01$\pm$2.5,  4764 06$\pm$1.9 \\
              &                &                  &        &      &        &    &        &        &        &     \\
358.149-1.675 &  17  47  50.144 &  $-$31  23 03.86 &$-$22 &   13 &  0.9 & 09  &  142  & 2.6  &  4.0 &       8685 46$\pm$2.5,  8335 56$\pm$2.5,  4885 48$\pm$1.7,  4635 47$\pm$1.6 \\
              &  17  47  45.300 &  $-$31  23 36.21 &   276  &   18  &  1.0 & 12  &  156  & 3.6  &  0.4 &      8685 85$\pm$3.7,  8335 88$\pm$3.3,  4885 $-$55$\pm$1.9,  4635 $-$46$\pm$1.9 \\
              &  17  47  45.411 &  $-$31  23 38.40 &    346 &   24 &  1.4   & 18  &  102  & 5.1  &  0.3 &     8685 34$\pm$4.9,  8335 41$\pm$5,  4885 87$\pm$2.1,  4635 $-$84$\pm$2.1 \\
              &                 &                  &        &      &        &     &       &      &      &      \\
358.156+0.028 &  17  41  02.909 &  $-$30  29 22.11 &    533 &    7 &  0.9   & 05  &  112  & 1.3  &  2.4 &     8685 57$\pm$1.3,  8335 62$\pm$1.2,  4885 $-$42$\pm$0.8,  4635 $-$31$\pm$0.8 \\
              &  17  41  02.668 &  $-$30  29 36.53 &    469 &   28 &  1.1   & 17  &  117  & 5.1  &  0.5 &     8685 58$\pm$4.8,  8335 62$\pm$5.3,  4885 $-$50$\pm$3.1,  4635 $-$44$\pm$4.5 \\
              &                 &                  &        &      &        &     &       &      &       &      \\
358.917+0.072 &  17  42  43.99 &  $-$29  49 16.03 &    4768 &   88 &  0.3  & 10 &   95   &    3.9 &   1.3  &  8512 $-$17$\pm$2.3,  5988 $-$37$\pm$3.3,  5952 $-$26$\pm$3.3,  5920 $-$20$\pm$4.1,  5312 $-$3.6$\pm$3.9,  4804 $-$12$\pm$4.7 \\
              &                &                  &         &      &       &    &        &        &        &     \\
358.930-1.197 &  17  47  46.474 &  $-$30  28 17.39 & $-$905 &   26 &  1.0   & 17  &   71  & 4.9  &  9.4 &     8685 $-$69$\pm$4.4,  8335 77$\pm$5.1,  4885 $-$31$\pm$4.4,  4635 $-$57$\pm$3.1 \\
              &                 &                  &        &      &        &     &       &      &       &       \\
359.2-0.8     &  17 47 15.78   &  $-$29 58 01.0    &   $-$05  &   18 & 1.0  & 06  &   94  & 3.4  &  1.0 &     8685 $-$7$\pm$2.9,  8335 $-$2$\pm$4.0,  4885 $-$8.2$\pm$2.3,  4635 $-$5$\pm$2.4 \\
( Mouse)       &                &                   &          &      &      &     &       &      &      &        \\
              &                &                   &          &      &      &     &       &      &      &       \\
359.388+0.460 &  17  42  21.47 &  $-$29  12 59.73 &    $-$568 &   29 &  0.4 & 10 &  36   &    4.2 &   0.9  &  8512 85$\pm$2.7,  5988 54$\pm$4,  5920 42$\pm$4.0,  5660 36.6$\pm$5.6,  5312 24$\pm$4.1,  4804 $-$1.5$\pm$4.5 \\
              &                &                  &           &      &      &    &       &        &        &     \\
359.392+1.272 &  17  39  13.384 &  $-$28  46 54.08 &   $-$318 &   27 &  1.2 & 20  &   53  & 5.4  &  1.7 &     8685 $-$55$\pm$5.4,  8335 $-$63$\pm$5.0,  4885 70$\pm$3.3,  4635 70$\pm$3.3 \\
              &  17  39  12.505 &  $-$28  47 01.35 &   $-$271 &   30 &  1.1 & 20  &   66  & 5.5  &  0.2 &     8685 $-$40$\pm$5.3,  8335 $-$46$\pm$5.3,  4885 $-$83$\pm$3.8,  4635 $-$88$\pm$4.4 \\
              &                 &                  &          &      &      &     &       &      &      &       \\
359.604+0.306 &  17  43  29.266 &  $-$29  06 46.736 &   $-$502 &  34 &  -- & --   &  162  & 7.2  &  0.3 &     8685 38$\pm$5.0,  5056 $-$32$\pm$4.7,  4885 $-$35$\pm$5.2,  4635 $-$48$\pm$3.0 \\
              &                 &                   &          &     &     &     &       &      &       &        \\
359.710-0.904 &  17  48  29.373 &  $-$29  39 08.83 &   1423 &   93 &   --  &  --  &  117  & 19   &  2.2  &    5696 79$\pm$6.1,  5056 $-$50$\pm$4.1,  4885 $-$31$\pm$5.0,  4635 14$\pm$4.7 \\
              &  17  48  27.686 &  $-$29  39 09.76 &   1475 &   52 &   --  &  --  &   28  & 11   &  1.8  &    5696 $-$07$\pm$3.0,  5056 57$\pm$3.1,  4885 72$\pm$2.9,  4635 $-$65$\pm$3.3 \\
              &                &                   &        &      &       &      &       &      &       &       \\
359.844-1.843 &  17  52  30.88 &  $-$30  01 06.59 &    $-$58 &   47 &   --  & -- &  96   &    9.4 &   0.3  &  5988 $-$03$\pm$3.8,  5920 00$\pm$4.4,  4836 $-$7.7$\pm$3.4,  4764 $-$6$\pm$3.4\\
              &                &                  &          &      &       &    &       &        &        &     \\
359.871+0.179 &  17  44  37.069 &  $-$28  57 09.27 &    772 &   14 &   0.9 & 10   &  106  & 2.7  &  0.2  &    8685 68$\pm$2.8,  8335 75$\pm$2.7,  4885 03$\pm$1.4,  4635 22$\pm$1.5 \\
              &                 &                  &        &      &       &      &       &      &       &       \\
359.911-1.813 &  17  52  32.97 &  $-$29  56 44.28 &    165   &   47 &   --  & -- &  116  &    9.0 &   0.3  &  5952 49$\pm$4.6,  5696 52$\pm$2.7,  5056 61$\pm$2.8,  4804 62$\pm$2.3 \\
              &                &                  &          &      &       &    &       &        &        &     \\
359.993+1.591 &  17  39  26.94 &  $-$28  06 12.65 &     16   &   41 &  0.7  & 20  &  172  &   7.4 &   1.3   & 8512 80$\pm$5.3,  5952 $-$89$\pm$5.4,  5312 89$\pm$5.3,  4804 84$\pm$3.5 \\                                              
              &                &                  &          &      &       &     &       &       &         &    \\
0.313+1.645   &  17  40   0.21 &  $-$27  48 10.37 &   $-$121 &   22 &  1.1  & 12  &   00  &   4.0 &   0.5   & 8512 82$\pm$3.2,  5952 74$\pm$2.2,  5312 67$\pm$1.7,  4804 64$\pm$1.7 \\                                                 
              &                &                  &          &      &       &     &       &       &         &    \\
0.404+1.061   &  17  42  27.934 &  $-$28  02 08.49 & $-$176 &   16 &  0.8  & 12   &   21  & 3.2  &  0.4  &    8685 $-$82$\pm$3.8,  8335 $-$82$\pm$2.8,  4885 71$\pm$1.9,  4635 69$\pm$1.7 \\
              &                &                   &        &      &       &      &       &      &       &       \\
0.537+0.263   &  17  45  52.50 &  $-$28  20 26.64 &  $-$1176 &   26 &  0.7  & 15  &   56  &   3.5 &   2.7   & 8512 63$\pm$1.9,  5952 $-$30$\pm$3.1,  5344 $-$59$\pm$3.8,  5284 $-$60$\pm$5.6,  4836 62$\pm$6.7,  4764 48$\pm$6.0 \\            
              &                &                  &          &      &       &     &       &       &         &    \\
1.028-1.110   &  17  52  22.81 &  $-$28  37 41.31 &    $-$65 &   23 &  0.7  & 12  &   22  &   4.0 &   0.7   & 8512 $-$75$\pm$3.1,  5952 $-$76$\pm$2.2,  5312 $-$80$\pm$2.0,  4804 $-$84$\pm$2.0 \\                                     
              &  17  52  22.51 &  $-$28  37 32.51 &    218   &   43 &  0.7  & 20  &   83  &   7.5 &   0.7   & 8512 10$\pm$6.0,  5952 25$\pm$3.4,  5312 29$\pm$3.7,  4804 44$\pm$3.8\\                                             
              &                &                  &          &      &       &     &       &       &         &     \\
1.035+1.559   &  17  42   1.90 &  $-$27  13 9.94  &   $-$136 &    9 &   --  & --  &   67  &   1.6 &   0.4   & 5988 $-$41.7$\pm$0.8,  5920 $-$43.2$\pm$0.8,  4836 $-$52.6$\pm$0.7,  4764 $-$53.6$\pm$0.7 \\
              &                &                  &          &      &       &     &       &       &         &      \\
1.505-1.231   &  17  53  55.37 &  $-$28  15 53.77 &    385 &   12 &   1.0   & 07  &  95   &   2.2 &  1.6    & 8512 33$\pm$1.9,  5952 62$\pm$1.0,  5312 74$\pm$1.1,  4804 $-$88$\pm$0.9 \\                                              
              &  17  53  59.21 &  $-$28  17 21.35 &    689 &   24 &   1.4   & 15  &  00   &   5.0 &  1.3    & 8512 $-$36$\pm$4.1,  5952 09$\pm$1.8,  5312 37$\pm$1.8,  4804 65$\pm$1.6 \\                                              
              &                &                  &        &      &         &     &       &       &         &       \\
1.826+1.070   &  17  45  47.185 &  $-$26  49 21.48 &    806 &  9.5 &   1.1  & 07 &  100  & 2.0  &  1.3  &     8685 64$\pm$1.8,  8335 72$\pm$2.8,  4885 3.6$\pm$1.1,  4635 24.5$\pm$1.1 \\
              &  17  45  47.702 &  $-$26  49 06.24 &    638 &   40 &   0.6  & 23 &  152  & 6.9  &  3.9  &     8685 $-$63$\pm$6.4,  8335 $-$83$\pm$6.3,  4885 14$\pm$7.0,  4635 38$\pm$6.2 \\
              &                 &                  &        &      &        &    &       &      &       &        \\
2.143+1.772   &  17  43  51.24 &  $-$26  10 59.24 &    158 &   50 &  --     & --  &  18   &   9.8 &  0.01  &  5952 $-$49$\pm$3.8,  5696 $-$47$\pm$3.4,  5056 $-$40$\pm$2.5,  4804 $-$37$\pm$2.8 \\ 
              &                &                  &        &      &         &     &       &       &        &      \\
2.922+1.028   &  17  48  29.121 &  $-$25  54 17.97 &   1563 &   15 &   1.0  & 11 &   78  & 3.1  &  0.3  &     8685 $-$83$\pm$3.0,  8335 $-$77$\pm$2.9,  4885 $-$34$\pm$1.6,  4635 04$\pm$1.6 \\
              &  17  48  29.150 &  $-$25  54 15.97 &   1510 &   21 &   0.7  & 13 &   85  & 3.8  &  1.4  &     8685 $-$79$\pm$4.5,  8335 $-$75$\pm$3.1,  4885 $-$35$\pm$3.4,  4635 $-$05$\pm$2.9 \\
              &                 &                  &        &      &        &     &       &      &       &         \\
3.347-0.327   &  17  54  38.331 &  $-$26  13 50.76 &  $-$68 &   21 &  0.8   & 15  &  166  & 4.1  &  0.3  &    8685 71$\pm$3.2,  8335 74$\pm$5.5,  4885 60$\pm$3.2,  4635 61$\pm$2.7 \\
              &                 &                  &        &      &        &     &       &      &       &          \\
3.745+0.635   &  17  51  51.26 &  $-$25  24 0.05   &   1295 &    2 &  0.7   &     & 115.3 & 0.4  &  3.8  &    8512 $-$62.1$\pm$0.3,  5952 33.4$\pm$0.2,  5312 81.8$\pm$0.2,  4804 $-$45.1$\pm$0.2 \\
(1748-253)    &                &                   &        &      &        &     &       &      &      &           \\
              &                &                   &        &      &        &     &       &      &      &           \\
3.748-1.221   &  17  58  58.649 &  $-$26  20 01.31 &   1247 &   18 &  0.8   & 12  &  167  & 3.5  &  0.9 &     8685 $-$17$\pm$3.4,  8335 $-$11$\pm$3.6,  4885 $-$16$\pm$2.1,  4635 18$\pm$2.2 \\
              &                 &                  &        &      &        &     &       &      &      &           \\
3.928+0.253   &  17  53  43.371 &  $-$25  26 10.26 &   1457 &    7 &  1.4   & 04  &  80   & 1.2  &  3.7 &     8685 88$\pm$1.1,  8335 $-$80$\pm$1.2,  4885 $-$54$\pm$1.0,  4635 $-$21$\pm$1.0 \\
              &                 &                  &        &      &        &     &       &      &      &           \\
4.005+1.403   &  17  49  31.68 &  $-$24  46 57.70 &    490 &   28 &   0.8   & 10  &  68   &   5.0 &  3.4   &  8512 06$\pm$3.8,  5952 54$\pm$2.5,  5312 70$\pm$2.8,  5284 69$\pm$4.8,  4836 75$\pm$5.3,  4764 83$\pm$4.3 \\
              &                &                  &        &      &         &     &       &       &        &        \\
4.188-1.680   &  18  01  44.34 &  $-$26  10 43.63 &   $-$625 &   15 &  1.0  & 08  &  12   &   2.8 &  1.0   &  8512 55$\pm$2.3,  5952 11.5$\pm$1.4,  5312 $-$12$\pm$1.3,  4804 $-$39$\pm$1.1 \\
              &                &                  &          &      &       &     &       &       &        &        \\
4.256-0.726   &  17  58  12.97 &  $-$25  38 44.29 &   1788  &   22 &  0.7   & 12  &  73   &   3.9 &  3.6   &  8512 $-$74$\pm$3.2,  5952 64$\pm$1.8,  5312 $-$46$\pm$2.4,  4804 20$\pm$1.7 \\
              &                &                  &         &      &        &     &       &       &        &        \\
4.752+0.255   &  17  55  33.409 &  $-$24  43 29.77 &    191 &   18 &  0.9   & 13  &  110  & 3.7  &  3.2 &     8685 27$\pm$3.6,  8335 40$\pm$3.5,  4885 63$\pm$2.1,  4635 65$\pm$1.9 \\
              &  17  55  33.424 &  $-$24  43 31.16 &    112 &   24 &  0.7   & 17  &  65   & 4.8  &  0.2 &     8685 $-$19$\pm$5.3,  8335 $-$16$\pm$4.1,  4885 01$\pm$3.7,  4635 01$\pm$2.6 \\
              &                 &                  &        &      &        &     &       &      &      &           \\
5.260-0.754   &  18  00  31.38 &  $-$24  47 21.73 &   1145  &   12 &  0.7   & 06  &  126  &   2.1 &  0.7   &  8512 $-$62$\pm$1.6,  5952 22$\pm$1.3,  5312 64$\pm$1.3,  4804 $-$67.5$\pm$1.1 \\
              &                &                  &         &      &        &     &       &       &        &        \\
5.358+0.899   &  17  54  27.81 &  $-$23  52 36.17 &    851 &   29  &  0.8   & 15  &  29   &   5.3 &  2.0   &  8512 $-$04$\pm$4.2,  5952 68$\pm$2.8,  5312 $-$86$\pm$2.2,  4804 $-$51$\pm$2.4 \\
              &                &                  &        &       &        &     &       &       &        &        \\
5.511-1.515   &  18   3  59.28 &  $-$24  56 43.63 &    319 &   20 &   0.8   & 09  &  01   &   3.5 &  0.5   &  8512 $-$65$\pm$2.6,  5952 $-$44$\pm$1.9,  5312 $-$31$\pm$2.1,  4804 $-$17$\pm$2.0 \\
              &  18   3  59.25 &  $-$24  56 48.03 &    396 &   32 &   1.2   & 17  &  50   &   6.1 &  1.4   &  8512 $-$13$\pm$4.6,  5952 23$\pm$4.4,  5312 29$\pm$3.0,  4804 49$\pm$2.3 \\
              &                &                  &        &      &         &     &       &       &        &         \\
5.791+0.794   &  17  55  48.323 &  $-$23  33 22.71 &   1203 &   13 &  0.9   & 11  &  91   & 2.8  &  2.8 &     8685 79$\pm$2.8,  8335 $-$85$\pm$2.7,  4885 82$\pm$1.3,  4635 $-$70$\pm$1.2 \\
              &  17  55  48.526 &  $-$23  33 21.09 &   1294 &   17 &  1.1   & 13  &  155  & 3.5  &  0.2 &     8685 $-$25$\pm$3.1,  8335 $-$21$\pm$4.3,  4885 $-$15$\pm$1.8,  4635 16$\pm$1.7 \\
              &  17  55  48.802 &  $-$23  33 18.67 &   1144 &   24 &  1.4   & 17  &  66   & 4.9  &  0.8 &     8685 55$\pm$5.6,  8335 59$\pm$4.4,  4885 45$\pm$2.1,  4635 69$\pm$2.1 \\
              &                 &                  &        &      &        &     &       &      &      &           \\
5.852+1.041   &  17  55  00.749 &  $-$23  22 31.56 &    581 &   26 &  1.0   & 18  &  172  & 5.3  &  2.7 &     8685 $-$58$\pm$4.5,  8335 $-$52$\pm$6.3,  4885 22$\pm$3.0,  4635 45$\pm$2.6 \\
              &  17  55  00.800 &  $-$23  22 52.16 &    623 &   33 &  0.8   & 23  &  113  & 6.7  &  0.1 &     8685 66$\pm$6.2,  8335 69$\pm$6.9,  4885 $-$21$\pm$3.6,  4635 $-$08$\pm$3.6 \\
              &                 &                  &        &      &        &     &       &      &      &           \\
6.183-1.480   &  18   5  18.04 &  $-$24  20 45.50 &   $-$876 &   24 &  0.9  & 12  &  10   &   4.4 &  3.8   & 8512 44$\pm$3.5, 8952 $-$27$\pm$2.0, 5348 $-$63$\pm$2.4, 5312 $-$64$\pm$1.7, 5284 $-$66$\pm$3.1,  4764 $-$95$\pm$2.3 \\
\hline
\end{longtable}
\twocolumn
\end{landscape}
\normalsize

\normalsize

\section{Discussion}
\subsection{Galactic HII regions}

Extended unpolarised structures, typical of HII regions, are seen in the VLA
images of the sources G353.410$-$0.360, G355.739+0.131, G358.643$-$0.034 and
G359.717$-$0.036. A small diameter source is seen in the field of
G353.410$-$0.360, which has a flat spectrum between 8.5 and 4.8 GHz. It is well
resolved in our BnA array image and is unlikely to be the flat spectrum core of
a background extragalactic source. We believe that it is a Galactic HII region.

\begin{figure}
\centering
\includegraphics[width=8.0cm, angle=0,clip=]{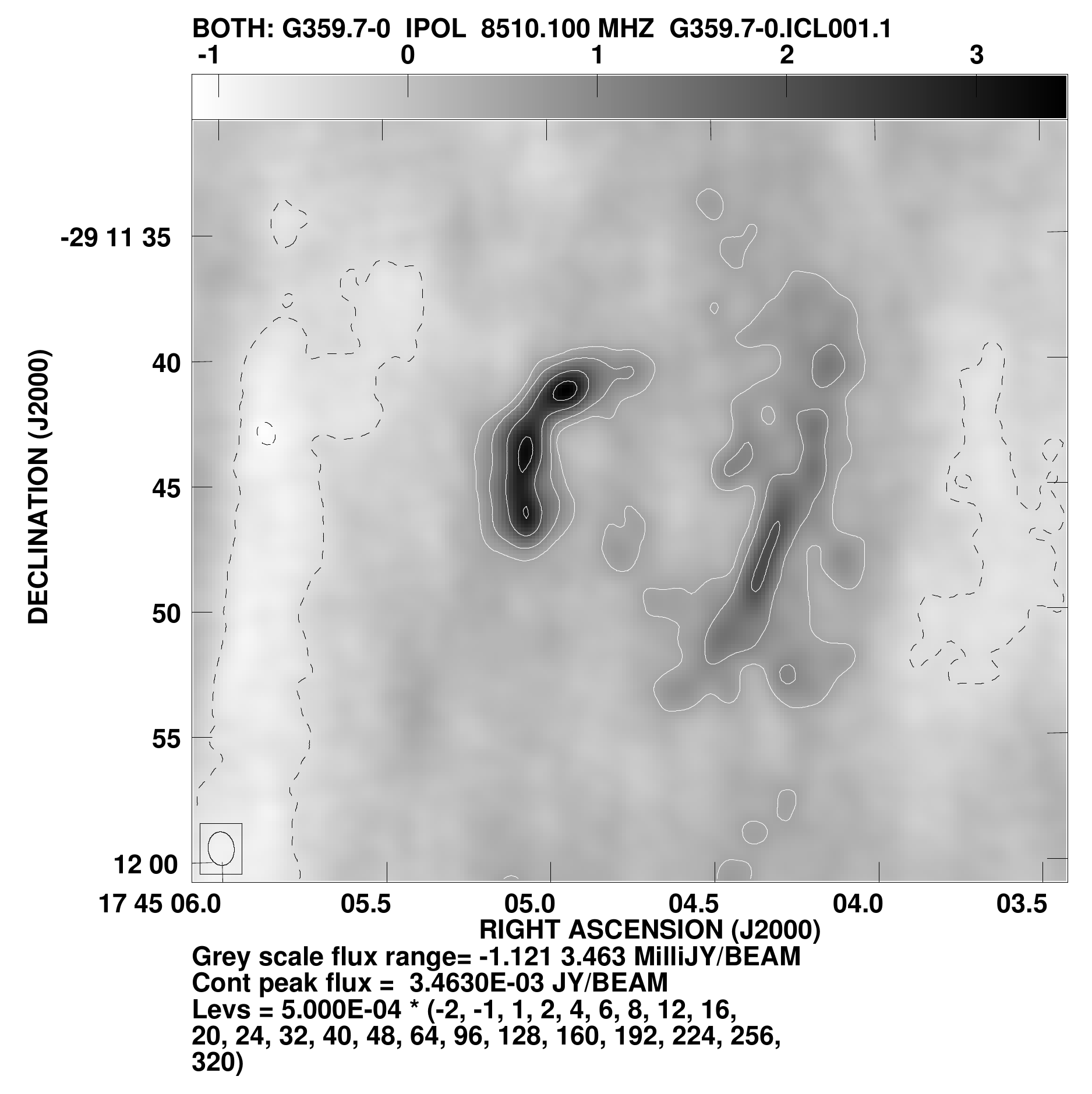}
\caption{8.5 GHz continuum image of the object G359.717$-$0.036 with a partial
shell morphology}
\label{8ghz.G359.7$-$0.SNR.candidate}
\end{figure}

The 8.5 GHz VLA BnA array image of G359.717$-$0.036
(Fig.~\ref{8ghz.G359.7$-$0.SNR.candidate}) shows a partial shell-like
structure.  Extended emission can also be seen around this object. Since either
HII regions or shell type SNRs can have shell like morphology, we have tried to
measure its spectral index accurately. We made an image of this object at 8.4
GHz using the archival VLA CD array data (Project Code AY68 made by Farhad
Yusef Zadeh). We also imaged it at 1.4 GHz using the archival Galactic Centre
data acquired and presented by \citet{PEDLAR1989} using the B, C and D array of
the VLA during 1981--1984. From the low resolution 8.4 and 1.4 GHz image of
this source, the measured spectrum of this component is flat, indicating it to
be an HII region.

The image of G358.643$-$0.034 at 8.5 GHz suffers from missing short spacings
and, consequently, the measured steep spectral index between 8.5 and 4.8 GHz
(Table~\ref{unpol.src.prop.vla}) is underestimated. However, its spectrum
between 4.8 and 1.4 GHz is inverted and the source is not detected in the 330
MHz GC image. The extended structure and inverted spectrum at low frequencies
suggest that it is an HII region.

G355.739+0.131 has a flat spectrum between 8.5 and 1.4 GHz and is well
resolved. This source is also not detected at 0.3 GHz, which suggests that its
spectrum has turned over due to significant free-free absorption at low radio
frequencies: these suggest that the source is an HII region.

\subsection{Are the rest of the sources extragalactic ?}

Excluding the sources described above, the rest of the sources observed are
either polarised or the deconvolved diameters imply a brightness
temperature greater than 10$^3$ K at 4.8 GHz. Except for a few small diameter
sources, all the other sources have steep spectral indices indicating the
emission from these sources is non-thermal. The small diameter objects
have flat spectra ($\alpha \ge -0.3$) down to 0.3 GHz. A thermal source with
brightness temperature of 10$^3$ K at 4.8 GHz should show self absorption
(i.e., positive spectral index) at 0.3 GHz (otherwise, the required brightness
temperature exceeds 10$^5$ K). Therefore, these objects with flat spectra at
0.3 GHz indicate that the emission from these small diameter sources are also
non-thermal. Among the non-thermal sources, the object G359.2$-$0.8 (Mouse)
(Table~\ref{pol.src.prop.vla}) \citep{YUSEF-ZADEH1987d, UCHIDA1992} is known to
be a Galactic non-thermal source located within 5 kpc from the Sun.  We review
below non-thermal emission from Galactic objects with the aim of rejecting
any of the remaining sources in the sample, which might admit such a
classification.

\subsubsection{Non thermal emission from Galactic sources}

There are several types of Galactic sources that emit non-thermal emission.\\
(i) Supernova Remnants (SNRs): These are the remnants of supernova explosions.
The electrons in these objects are accelerated to high energies near the
expanding shock front. The SNRs are usually spherical in shape, and when
projected on the sky appear like a ring for the shell type SNRs, while they
appear to be filled with emission in the case of plerion-type SNRs. None of
the non-thermal objects in Tables~\ref{pol.src.prop.at}, \ref{pol.src.prop.vla},
~\ref{unpol.src.prop.at} \& \ref{unpol.src.prop.vla} have such a morphology,
indicating that there is no resolved supernova remnant among these sources.
However, if an SNR is young, its angular size will be small and will appear
like an unresolved source.  Assuming an initial expansion velocity of 3000
\kms, an SNR at a distance of 10 kpc will expand to an angular size of 6$^{''}$
in about 100 years after the explosion. An object of angular size of 6$^{''}$
will be well resolved in our images and should have a ring or filled centre
morphology: the absence of these rule out any SNR older than 100 years in our
sample.  Since the expected number of supernova explosions in our Galaxy is
believed to be about 1 in 50 years, the probability of finding an SNR of age
less than 100 years in the central 12\dg $\times$ 4\dg of the Galaxy is
less than 0.1, which suggests that there is no young SNR in our sample.

(ii) Radio Pulsars: Pulsars typically have a very steep spectrum, with spectral
indices of $-1.5$ to $-4$ at cm wavelengths.  They would be unresolved and
almost all would be undetectable at 5 GHz owing to their steep spectrum.  Since
none of the unresolved sources in our sample have a very steep spectrum, we
believe that there is no pulsar in our list of sources.

(iii) Radio Stars: Most stars are weak radio emitters; however, some are
detected in non-thermal emission.  \citet{BECKER1994} show that the radio flux
densities from these stars are $\le$1 mJy at 5 GHz. Since the observed flux
densities of our sources are much higher than the upper limit for these stars,
such objects can be ruled out from our list of sources.

(iv) Transient sources: These are highly variable and transient radio sources
which include radio counterparts of X-ray sources. The sources in our sample
are not only detected at 4.8 and 8.5 GHz, but also detected at 1.4 GHz in
the GPS and the NVSS. These sources are also detected at 0.3 GHz, either in the
Texas survey or in the 330 MHz GC observations. These observations were
separated by days to years and the measured spectral indices determined between
any two bands are quite close to the mean value (differences are less than
0.6). Therefore, the flux densities of these sources have not changed by more
than a factor of two and this rules out the possibility that there are
transient sources in our catalogue.

(iv) Galactic Microquasar: These are stellar-mass black holes in our Galaxy
that mimic, on a smaller scale, many of the phenomena seen in quasars (see
\citet{MIRABEL1999}, and the references therein).  For a black hole accreting
at the Eddington limit, the characteristic black body temperature at the last
stable orbit in the surrounding accretion disk is given by T~$\sim2 \times 10^7
M^{-1/4}$ (Rees 1984). Therefore, compared to the AGNs, the emission from
microquasars are shifted towards higher frequencies and the microquasars
are usually identified by their X-ray properties \citep{MIRABEL1999}.

Though many of the already known microquasars are highly variable, two of
these sources, 1E1740.7$-$2942 \citep{MIRABEL1999} and GRS~1758$-$258
\citep{MARTI2002}, are persistent sources of both X-rays and
relativistic jets. At radio wavelengths, these two sources are
morphologically similar to typical radio galaxies, which have a central compact
component and two extended lobes.  Therefore, based on morphology,
microquasars cannot be separated from the distant radio galaxies. However, as
mentioned above, microquasars are believed to have X-ray counterparts.  We have
searched the ROSAT PSPC all sky survey \citep{VOGES1999} and a
catalogue of soft X-ray sources ($|l| \le$ 1.5\dg, $|b| \le $ 2.0\dg) in the GC
region \citep{SIDOLI2001}.  Twenty of our sources are also located within the
boundary of the ASCA survey of the GC region ($|l| \le$ 2.5\dg, $|b| \le $
2.5\dg) \citep{SAKANO2002}. However, none of the radio sources in our sample 
were found to have any counterpart in these catalogues.  Therefore, it is
unlikely that any of the sources we have observed
(Tables~\ref{pol.src.prop.at}, \ref{pol.src.prop.vla}, ~\ref{unpol.src.prop.at}
\& \ref{unpol.src.prop.vla}) is a microquasar.

\subsection{Extragalactic source counts}

The expected number of extragalactic sources (N) in 1 square arc minute of the
sky at 5 GHz and with a flux density limit of S mJy is N($>$S)= $0.032 \times$
S$^{-1.13}$ \citep{LEDDEN1980}. Therefore, the expected number of extragalactic
sources seen through the central $l \times b$ = 12\dg $\times$ 3.6\dg\ region
of the Galaxy above a flux density limit of 20 mJy at 5 GHz is 168.  However,
as we selected sources with steep spectral indices ($\alpha <-0.4$), typical
sizes $\le$10$^{''}$ and excessive confusion prevails in the region, we could
identify only 59 extragalactic sources, which indicates that about two thirds
of the extragalactic sources in this region are yet to be identified.  The
median angular size of these 59 sources is 7.6$^{''}$, and the median flux
density at 1.4 GHz is 160 mJy. 

\section{Summary}

We have observed 64 sources towards the central $-$6\dg~$~<~l~<~$6\dg,
$-$2\dg~$~<~b~<~$2\dg\ of the Galaxy using the 6 and 3.6 cm band of the ATCA
and the VLA. Based on our work described herein, 59 of these sources are
classified to be extragalactic. This increases the number of known
extragalactic radio sources towards this unique region by almost an order of
magnitude and provides the first systematic study of the polarisation
properties of the background sources in the region. We provide 4.8 GHz images
of all the observed sources and measure the angular sizes and the spectral
indices of these sources.  
Based on the morphology, spectral characteristics and polarisation properties,
we identify 4 Galactic HII regions in the sample. 

\section*{Acknowledgements}
S.R. thanks D. J. Saikia for useful discussions. The Australia Telescope is
funded by the Commonwealth of Australia for operation as a National Facility
managed by CSIRO. The National Radio Astronomy Observatory is a facility of the
National Science Foundation operated under cooperative agreement by Associated
Universities, Inc.

\bibliographystyle{mn2e}
\bibliography{eg.src.pap}
\label{lastpage}
\end{document}